%Final 0106
%Final form sent to CPAM, texas, xxx
\magnification \magstep 1
\input amstex
\input amsppt.sty
%\def\newpage{\vfill\eject}
%\def\em{\it}
%\def\RR{{\bf R}}
%\input nume2.tex
%\magnification=\magstephalf
%\magnification=\magstep1
%\baselineskip 20pt

\def\text{\hbox}
%\input amstex
%\def\cal{\Cal}
%\def\Cal{\cal}

%%%%%%%%%%%%%%%%% EQUAZIONI CON NOMI SIMBOLICI
%%%
%%% Per assegnare un nome simbolico ad una equazione basta
%%% scrivere \Eq(...) o, in \eqalignno, \eq(...) o,
%%% nelle appendici, \Eqa(...) o \eqa(...);
%%% dentro le parentesi e al posto di ... si puo' scrivere qualsiasi commento;
%%% per avere i nomi simbolici segnati a sinistra delle formule si deve
%%% dichiarare il documento come bozza, iniziando il testo con
%%% \BOZZA. Sinonimi: \Eq,\EQ,\EQS; \eq,\eqs; \Eqa,\Eqas;\eqa,\eqas.
%%% All' inizio di ogni paragrafo si devono definire il
%%% numero del paragrafo e della prima formula dichiarando
%%% \numsec=... \numfor=...  (brevetto Eckmannn).
%%% Si possono citare formule seguenti; le corrispondenze fra nomi
%%% simbolici e numeri effettivi sono memorizzate nel file \jobname.aux, che
%%% viene letto all'inizio, se gia' presente. E' possibile citare anche
%%% formule che appaiono in altri file, purche' sia presente il
%%% corrispondente file .aux; basta includere all'inizio l'istruzione
%%%           \include{nomefile}
%%%
%%%%%%%%%%%%%%%%%%%%%%%%%%%%%%%%%%%%%%%%%%%%%%%%%%%%%%%%%%%%%%%

\global\newcount\numsec
\global\newcount\numfor
\global\newcount\numtheo
\global\advance\numtheo by 1

\def\senondefinito#1{\expandafter\ifx\csname#1\endcsname\relax}

\def\SIA #1,#2,#3 {\senondefinito{#1#2}%
\expandafter\xdef\csname #1#2\endcsname{#3}\else
\write16{???? ma #1,#2 e' gia' stato definito !!!!} \fi}

\def\etichetta(#1){(\veroparagrafo.\veraformula)%
\SIA e,#1,(\veroparagrafo.\veraformula) %
\global\advance\numfor by 1%
\write15{\string\FU (#1){\equ(#1)}}%
\write16{ EQ #1 ==> \equ(#1) }}

\def\letichetta(#1){\veroparagrafo.\verotheo
\SIA e,#1,{\veroparagrafo.\verotheo}
\global\advance\numtheo by 1
 \write15{\string\FU (#1){\equ(#1)}}
 \write16{ Sta \equ(#1) == #1 }}

\def\tetichetta(#1){\veroparagrafo.\veraformula %%%%copy four lines
\SIA e,#1,{(\veroparagrafo.\veraformula)}
\global\advance\numfor by 1
\write15{\string\FU (#1){\equ(#1)}}
\write16{ tag #1 ==> \equ(#1)}}

\def\FU(#1)#2{\SIA fu,#1,#2 }

\def\etichettaa(#1){(A\veroparagrafo.\veraformula)%
\SIA e,#1,(A\veroparagrafo.\veraformula) %
\global\advance\numfor by 1%
\write15{\string\FU (#1){\equ(#1)}}%
\write16{ EQ #1 ==> \equ(#1) }}

\def\BOZZA{
\def\alato(##1){%
 {\rlap{\kern-\hsize\kern-1.4truecm{$\scriptstyle##1$}}}}%
\def\aolado(##1){%
 {%\vtop to \profonditastruttura
{%\baselineskip
 %\profonditastruttura\vss
 \rlap{\kern-1.4truecm{$\scriptstyle##1$}}}}}
}

\def\alato(#1){}
\def\aolado(#1){}

\def\veroparagrafo{\number\numsec}
\def\veraformula{\number\numfor}
\def\verotheo{\number\numtheo}

\def\Eq(#1){\eqno{\etichetta(#1)\alato(#1)}}
\def\eq(#1){\etichetta(#1)\alato(#1)}
\def\teq(#1){\tag{\aolado(#1)\tetichetta(#1)\alato(#1)}}%%%%%this line for \tag
\def\Eqa(#1){\eqno{\etichettaa(#1)\alato(#1)}}
\def\eqa(#1){\etichettaa(#1)\alato(#1)}
\def\eqv(#1){\senondefinito{fu#1}$\clubsuit$#1
\write16{#1 non e' (ancora) definito}%
\else\csname fu#1\endcsname\fi}
\def\equ(#1){\senondefinito{e#1}\eqv(#1)\else\csname e#1\endcsname\fi}

%next six lines by paf (no responsibilities taken)
\def\Lemma(#1){\aolado(#1)Lemma \letichetta(#1)}%
\def\Theorem(#1){{\aolado(#1)Theorem \letichetta(#1)}}%
\def\Proposition(#1){\aolado(#1){Proposition \letichetta(#1)}}%
\def\Corollary(#1){{\aolado(#1)Corollary \letichetta(#1)}}%
\def\Remark(#1){{\noindent\aolado(#1){\bf Remark \letichetta(#1).}}}%
\def\Definition(#1){{\noindent\aolado(#1){\bf Definition \letichetta(#1)$\!\!$\hskip-1.6truemm}}}
\def\Example(#1){\aolado(#1) Example \letichetta(#1)$\!\!$\hskip-1.6truemm}

\def\include#1{
\openin13=#1.aux \ifeof13 \relax \else
\input #1.aux \closein13 \fi}

\openin14=\jobname.aux \ifeof14 \relax \else
\input \jobname.aux \closein14 \fi
\openout15=\jobname.aux

\def\today{\rightline{\ifcase\month\or
  January\or February\or March\or April\or May\or June\or
  July\or August\or September\or October\or November\or December\fi
  \space\number\day,\space\number\year}}

%\BOZZA

%
%%%%%%%%%%%%%%%%%%extra definitions already in yau's file%%%%%%%%
%
%\def\qed{\mathchoice\sqr64\sqr64\sqr{2.1}3\sqr{1.5}3} 

\def\NN{N\!\!\!\!\!I\,\,}

\def\11{\hbox{l}\!\!\!\hbox{1}\,}

\newfam\mitbfam
\def\and{ \hbox{ and } }

\def\pa{\parallel}

\def\l{\lambda}

\def\e{\varepsilon}
\def\a{\alpha}
\def\be{\beta}

\def\d{\delta}
\def\g{\gamma}

\def\s{\sigma}
\def\t{\theta}
\def\newpage{\vfill\eject}
\def\to{\rightarrow}

\def\\{\cr}
\overfullrule=0in
\def\s{\sigma}

\def \a {\alpha}

\def\[{\left [ \  }
\def\]{\ \right ]  }
\def\){\  \right ) }
\def\({ \left (  \  }

%%%%%%%%References macros start here%%%%%%%%%%
%
\catcode`\X=12\catcode`\@=11
\def\n@wcount{\alloc@0\count\countdef\insc@unt}
\def\n@wwrite{\alloc@7\write\chardef\sixt@@n}
\def\n@wread{\alloc@6\read\chardef\sixt@@n}
\def\crossrefs#1{\ifx\alltgs#1\let\tr@ce=\alltgs\else\def\tr@ce{#1,}\fi
   \n@wwrite\cit@tionsout\openout\cit@tionsout=\jobname.cit 
   \write\cit@tionsout{\tr@ce}\expandafter\setfl@gs\tr@ce,}
\def\setfl@gs#1,{\def\@{#1}\ifx\@\empty\let\next=\relax
   \else\let\next=\setfl@gs\expandafter\xdef
   \csname#1tr@cetrue\endcsname{}\fi\next}
\newcount\sectno\sectno=0\newcount\subsectno\subsectno=0\def\r@s@t{\relax}
\def\resetall{\global\advance\sectno by 1\subsectno=0
  \gdef\firstpart{\number\sectno}\r@s@t}
\def\resetsub{\global\advance\subsectno by 1
   \gdef\firstpart{\number\sectno.\number\subsectno}\r@s@t}
\def\v@idline{\par}\def\firstpart{\number\sectno}
\def\l@c@l#1X{\firstpart.#1}\def\gl@b@l#1X{#1}\def\t@d@l#1X{{}}
\def\m@ketag#1#2{\expandafter\n@wcount\csname#2tagno\endcsname
     \csname#2tagno\endcsname=0\let\tail=\alltgs\xdef\alltgs{\tail#2,}%
  \ifx#1\l@c@l\let\tail=\r@s@t\xdef\r@s@t{\csname#2tagno\endcsname=0\tail}\fi
   \expandafter\gdef\csname#2cite\endcsname##1{\expandafter
 %the following line was replaced by the subseqent one, DNA 7/6/89
  %  \ifx\csname#2tag##1\endcsname\relax?\else\csname#2tag##1\endcsname\fi
     \ifx\csname#2tag##1\endcsname\relax?\else{\rm\csname#2tag##1\endcsname}\fi
    \expandafter\ifx\csname#2tr@cetrue\endcsname\relax\else
     \write\cit@tionsout{#2tag ##1 cited on page \folio.}\fi}%
   \expandafter\gdef\csname#2page\endcsname##1{\expandafter
     \ifx\csname#2page##1\endcsname\relax?\else\csname#2page##1\endcsname\fi
     \expandafter\ifx\csname#2tr@cetrue\endcsname\relax\else
     \write\cit@tionsout{#2tag ##1 cited on page \folio.}\fi}%
   \expandafter\gdef\csname#2tag\endcsname##1{\global\advance
     \csname#2tagno\endcsname by 1%
   \expandafter\ifx\csname#2check##1\endcsname\relax\else%
\fi%      \immediate\write16{Warning: #2tag ##1 used more than once.}\fi
   \expandafter\xdef\csname#2check##1\endcsname{}%
   \expandafter\xdef\csname#2tag##1\endcsname
     {#1\number\csname#2tagno\endcsnameX}%
   \write\t@gsout{#2tag ##1 assigned number \csname#2tag##1\endcsname\space
      on page \number\count0.}%
   \csname#2tag##1\endcsname}}%
\def\m@kecs #1tag #2 assigned number #3 on page #4.%
   {\expandafter\gdef\csname#1tag#2\endcsname{#3}
   \expandafter\gdef\csname#1page#2\endcsname{#4}}
\def\re@der{\ifeof\t@gsin\let\next=\relax\else
    \read\t@gsin to\t@gline\ifx\t@gline\v@idline\else
    \expandafter\m@kecs \t@gline\fi\let \next=\re@der\fi\next}
\def\t@gs#1{\def\alltgs{}\m@ketag#1e\m@ketag#1s\m@ketag\t@d@l p
    \m@ketag\gl@b@l r \n@wread\t@gsin\openin\t@gsin=\jobname.tgs \re@der
    \closein\t@gsin\n@wwrite\t@gsout\openout\t@gsout=\jobname.tgs }
\outer\def\localtags{\t@gs\l@c@l}
\outer\def\globaltags{\t@gs\gl@b@l}
\outer\def\newlocaltag#1{\m@ketag\l@c@l{#1}}
\outer\def\newglobaltag#1{\m@ketag\gl@b@l{#1}}

\def\t@gsoff#1,{\def\@{#1}\ifx\@\empty\let\next=\relax\else\let\next=\t@gsoff
   \expandafter\gdef\csname#1cite\endcsname{\relax}
   \expandafter\gdef\csname#1page\endcsname##1{?}
   \expandafter\gdef\csname#1tag\endcsname{\relax}\fi\next}
\def\verbatimtags{\let\ift@gs=\iffalse\ifx\alltgs\relax\else
   \expandafter\t@gsoff\alltgs,\fi}
\catcode`\X=11 \catcode`\@=\active
\localtags
%%%%%%%%%%%%%%%references macro end here%%%%%%%%%%%%
%%%%%%%%%%%%%%%%%end of macros%%%%%%%%%%%
%

\input epsf
\loadbold
\loadmsbm
\voffset -6mm
\hsize 17 truecm
\vsize 23 truecm
\baselineskip 6mm

\def\label{\Eq}
\def\be{$$}
\def\ee{$$}
\def\ref{\equ}

\global\newcount\numsec
\global\newcount\numfor
\global\newcount\numtheo
\global\advance\numtheo by 1

\def\senondefinito#1{\expandafter\ifx\csname#1\endcsname\relax}

\def\SIA #1,#2,#3 {\senondefinito{#1#2}%
\expandafter\xdef\csname #1#2\endcsname{#3}\else
\write16{???? ma #1,#2 e' gia' stato definito !!!!} \fi}

\def\etichetta(#1){(\veroparagrafo.\veraformula)%
\SIA e,#1,(\veroparagrafo.\veraformula) %
\global\advance\numfor by 1%
\write15{\string\FU (#1){\equ(#1)}}%
\write16{ EQ #1 ==> \equ(#1) }}

\def\letichetta(#1){\veroparagrafo.\verotheo
\SIA e,#1,{\veroparagrafo.\verotheo}
\global\advance\numtheo by 1
 \write15{\string\FU (#1){\equ(#1)}}
 \write16{ Sta \equ(#1) == #1 }}

\def\tetichetta(#1){\veroparagrafo.\veraformula %%%%copy four lines
\SIA e,#1,{(\veroparagrafo.\veraformula)}
\global\advance\numfor by 1
\write15{\string\FU (#1){\equ(#1)}}
\write16{ tag #1 ==> \equ(#1)}}

\def\FU(#1)#2{\SIA fu,#1,#2 }

\def\etichettaa(#1){(A\veroparagrafo.\veraformula)%
\SIA e,#1,(A\veroparagrafo.\veraformula) %
\global\advance\numfor by 1%
\write15{\string\FU (#1){\equ(#1)}}%
\write16{ EQ #1 ==> \equ(#1) }}

\def\BOZZA{
\def\alato(##1){%
 {\rlap{\kern-\hsize\kern-1.4truecm{$\scriptstyle##1$}}}}%
\def\aolado(##1){%
 {%\vtop to \profonditastruttura
{%\baselineskip
 %\profonditastruttura\vss
 \rlap{\kern-1.4truecm{$\scriptstyle##1$}}}}}
}

\def\alato(#1){}
\def\aolado(#1){}

\def\veroparagrafo{\number\numsec}
\def\veraformula{\number\numfor}
\def\verotheo{\number\numtheo}

\def\Eq(#1){\eqno{\etichetta(#1)\alato(#1)}}
\def\eq(#1){\etichetta(#1)\alato(#1)}
\def\teq(#1){\tag{\aolado(#1)\tetichetta(#1)\alato(#1)}}%%%%%this line for \tag
\def\Eqa(#1){\eqno{\etichettaa(#1)\alato(#1)}}
\def\eqa(#1){\etichettaa(#1)\alato(#1)}
\def\eqv(#1){\senondefinito{fu#1}$\clubsuit$#1
\write16{#1 non e' (ancora) definito}%
\else\csname fu#1\endcsname\fi}
\def\equ(#1){\senondefinito{e#1}\eqv(#1)\else\csname e#1\endcsname\fi}

%next six lines by paf (no responsibilities taken)
\def\Lemma(#1){\aolado(#1)Lemma \letichetta(#1)}%
\def\Theorem(#1){{\aolado(#1)Theorem \letichetta(#1)}}%
\def\Proposition(#1){\aolado(#1){Proposition \letichetta(#1)}}%
\def\Corollary(#1){{\aolado(#1)Corollary \letichetta(#1)}}%
\def\Remark(#1){{\noindent\aolado(#1){\bf Remark \letichetta(#1).}}}%
\def\Definition(#1){{\noindent\aolado(#1){\bf Definition \letichetta(#1)$\!\!$\hskip-1.6truemm}}}
\def\Example(#1){\aolado(#1) Example \letichetta(#1)$\!\!$\hskip-1.6truemm}

\def\include#1{
\openin13=#1.aux \ifeof13 \relax \else
\input #1.aux \closein13 \fi}

\openin14=\jobname.aux \ifeof14 \relax \else
\input \jobname.aux \closein14 \fi
\openout15=\jobname.aux

\def\today{\rightline{\ifcase\month\or
  January\or February\or March\or April\or May\or June\or
  July\or August\or September\or October\or November\or December\fi
  \space\number\day,\space\number\year}}

%\BOZZA

%%%%%%%%References macros start here%%%%%%%%%%
%
\catcode`\X=12\catcode`\@=11
\def\n@wcount{\alloc@0\count\countdef\insc@unt}
\def\n@wwrite{\alloc@7\write\chardef\sixt@@n}
\def\n@wread{\alloc@6\read\chardef\sixt@@n}
\def\crossrefs#1{\ifx\alltgs#1\let\tr@ce=\alltgs\else\def\tr@ce{#1,}\fi
   \n@wwrite\cit@tionsout\openout\cit@tionsout=\jobname.cit 
   \write\cit@tionsout{\tr@ce}\expandafter\setfl@gs\tr@ce,}
\def\setfl@gs#1,{\def\@{#1}\ifx\@\empty\let\next=\relax
   \else\let\next=\setfl@gs\expandafter\xdef
   \csname#1tr@cetrue\endcsname{}\fi\next}
\newcount\sectno\sectno=0\newcount\subsectno\subsectno=0\def\r@s@t{\relax}
\def\resetall{\global\advance\sectno by 1\subsectno=0
  \gdef\firstpart{\number\sectno}\r@s@t}
\def\resetsub{\global\advance\subsectno by 1
   \gdef\firstpart{\number\sectno.\number\subsectno}\r@s@t}
\def\v@idline{\par}\def\firstpart{\number\sectno}
\def\l@c@l#1X{\firstpart.#1}\def\gl@b@l#1X{#1}\def\t@d@l#1X{{}}
\def\m@ketag#1#2{\expandafter\n@wcount\csname#2tagno\endcsname
     \csname#2tagno\endcsname=0\let\tail=\alltgs\xdef\alltgs{\tail#2,}%
  \ifx#1\l@c@l\let\tail=\r@s@t\xdef\r@s@t{\csname#2tagno\endcsname=0\tail}\fi
   \expandafter\gdef\csname#2cite\endcsname##1{\expandafter
 %the following line was replaced by the subseqent one, DNA 7/6/89
  %  \ifx\csname#2tag##1\endcsname\relax?\else\csname#2tag##1\endcsname\fi
     \ifx\csname#2tag##1\endcsname\relax?\else{\rm\csname#2tag##1\endcsname}\fi
    \expandafter\ifx\csname#2tr@cetrue\endcsname\relax\else
     \write\cit@tionsout{#2tag ##1 cited on page \folio.}\fi}%
   \expandafter\gdef\csname#2page\endcsname##1{\expandafter
     \ifx\csname#2page##1\endcsname\relax?\else\csname#2page##1\endcsname\fi
     \expandafter\ifx\csname#2tr@cetrue\endcsname\relax\else
     \write\cit@tionsout{#2tag ##1 cited on page \folio.}\fi}%
   \expandafter\gdef\csname#2tag\endcsname##1{\global\advance
     \csname#2tagno\endcsname by 1%
   \expandafter\ifx\csname#2check##1\endcsname\relax\else%
\fi%      \immediate\write16{Warning: #2tag ##1 used more than once.}\fi
   \expandafter\xdef\csname#2check##1\endcsname{}%
   \expandafter\xdef\csname#2tag##1\endcsname
     {#1\number\csname#2tagno\endcsnameX}%
   \write\t@gsout{#2tag ##1 assigned number \csname#2tag##1\endcsname\space
      on page \number\count0.}%
   \csname#2tag##1\endcsname}}%
\def\m@kecs #1tag #2 assigned number #3 on page #4.%
   {\expandafter\gdef\csname#1tag#2\endcsname{#3}
   \expandafter\gdef\csname#1page#2\endcsname{#4}}
\def\re@der{\ifeof\t@gsin\let\next=\relax\else
    \read\t@gsin to\t@gline\ifx\t@gline\v@idline\else
    \expandafter\m@kecs \t@gline\fi\let \next=\re@der\fi\next}
\def\t@gs#1{\def\alltgs{}\m@ketag#1e\m@ketag#1s\m@ketag\t@d@l p
    \m@ketag\gl@b@l r \n@wread\t@gsin\openin\t@gsin=\jobname.tgs \re@der
    \closein\t@gsin\n@wwrite\t@gsout\openout\t@gsout=\jobname.tgs }
\outer\def\localtags{\t@gs\l@c@l}
\outer\def\globaltags{\t@gs\gl@b@l}
\outer\def\newlocaltag#1{\m@ketag\l@c@l{#1}}
\outer\def\newglobaltag#1{\m@ketag\gl@b@l{#1}}

\def\t@gsoff#1,{\def\@{#1}\ifx\@\empty\let\next=\relax\else\let\next=\t@gsoff
   \expandafter\gdef\csname#1cite\endcsname{\relax}
   \expandafter\gdef\csname#1page\endcsname##1{?}
   \expandafter\gdef\csname#1tag\endcsname{\relax}\fi\next}
\def\verbatimtags{\let\ift@gs=\iffalse\ifx\alltgs\relax\else
   \expandafter\t@gsoff\alltgs,\fi}
\catcode`\X=11 \catcode`\@=\active
\localtags
%%%%%%%%%%%%%%%references macro end here%%%%%%%%%%%%
%%%%%%%%%%%%%%%%%end of macros%%%%%%%%%%%
%

\def\l{\lambda}

\def\a{\alpha}

\def\d{\delta}
\def\e{\varepsilon}
\def\g{\gamma}

\def\s{\sigma}
\def\text{\hbox}

\def\\{\cr}
\def\ref{}

\def\NN{{\bold N}}

\overfullrule=0in
\def\({\left(}
\def\[{ \left[}
\def\){ \right)}
\def\]{ \right]}
\def\s{\sigma}

\def\ov{\overline}
\def\ba{{\bold a}}

\def\bk{{\bold k}}

\def\balpha{{\boldsymbol \alpha}}

\def\rd{d}
\def\bE{{\bold E}}

\def\cB{{\Cal B}}

\def\cD{{\Cal D}}

\def\cF{{\Cal F}}

\def\cS{{\Cal S}}

\def\bb{{\bold b}}
\def\ba{{\bold a}}
\def\bR{{\bold R}}

\def\l{\langle}

\def\fp{{\bold p}}
\def\fq{{\bold q}}
\def\fr{{\bold r}}

\def\fv{{\bold v}}
\def\hV{\widehat V}

\def\btau{{\boldsymbol \tau}}

\def\cO{{\Cal O}}
\def\fw{{\bold w}}

\def\({\left(}
\def\[{ \left[}
\def\){ \right)}
\def\]{ \right]}

\def\fp{{\bold p}}
\def\bk{{\bold k}}
\def\rd{d}

\def\l{\lambda}
\def\l{\lambda}

\def\e{\varepsilon}
\def\a{\alpha}

\def\d{\delta}
\def\g{\gamma}

\def\s{\sigma}
\def\t{\theta}
%\let\rcite=\cite

%%%%%%%%%%%%%%%%%end of macros%%%%%%%%%%%

\vskip -2cm 
Jan 06 1999 \hfill
\vskip 2cm

\topmatter

\title 
Linear Boltzmann Equation as the Weak Coupling Limit of 
a Random Schr\"odinger Equation
\endtitle

\rightheadtext{Weak coupling limit of 
random Schr\"odinger equation}

\author L\'aszl\'o Erd\H os, Horng-Tzer Yau \endauthor

\affil{  School of Mathematics, Georgiatech \\
Courant Institute, New York University}  \endaffil

\address{L\'aszl\'o Erd\H os: School of Mathematics, Georgiatech,
ATLANTA, GA-30332-0160, USA } \endaddress
\email  lerdos\@math.gatech.edu \endemail

\address{Horng-Tzer Yau: Courant Institute, New York University,
251 Mercer Street, New York
NY 10012, USA. } \endaddress 
\email yau\@cims.nyu.edu \endemail

\thanks
Partially 
supported by U.S. National Science Foundation grants DMS-9403462, 9703752.
\endthanks

\abstract
We study  the  time evolution of a quantum particle in a 
Gaussian random environment. We show that in the weak coupling limit
the Wigner distribution of the wave function converges
to a solution of a linear Boltzmann equation globally in
time. The Boltzmann collision kernel is given by the Born
approximation of the quantum differential scattering  cross section.
\endabstract

\endtopmatter

\subheading{ 1. Introduction}
\numsec=1\numfor=1\numtheo=1

\bigskip\bigskip

Schr\"odinger equations with random potentials describe the motion
of  quantum particles in a random environment
such as electrons in a semiconductor. The random potential is modeled
by a Gaussian random field $V(x)= V_\omega(x)$ 
in $\bR^d$, $d\ge 2$, with $\bE V(x)=0$
and  covariance $G(x-y):=\bE V(x) V(y)$. We assume that
$G$ is spherically symmetric, smooth  and decays fast at infinity.
Let $\psi_t = \psi_{t,\omega}
\in L^2(\bR^d)$
solve the random Schr\"odinger equation
\be
	i \partial_t \psi_{t,\omega}(x)
	= \Bigg(-{1\over 2}\Delta_x + \lambda
	V_\omega(x)\Bigg)\psi_{t,\omega}(x)
\label(scheq)
\ee
with a coupling constant $0< \lambda \ll 1$.
Here $x, t$ are atomic (microscopic) space and time coordinates.

Our goal is to understand the long time behavior of the random 
Schr\"odinger
equation \equ(scheq). The first scaling yielding 
a nontrivial (non-free) evolution  is the weak coupling limit 
which we now describe. 
Let $\e$ be the scaling parameter which determines the ratio 
of  the microscopic and
macroscopic scales, and let 
$$
	(X, T): = (x\e, t\e)
\Eq(es)
$$
be the macroscopic coordinates. 
Note that the velocity is not rescaled, $V=v$. The coupling constant is chosen 
to scale as  
$$
\lambda = \sqrt{\e} \; .
\Eq(wl)
$$
This is called the 
{\it weak coupling limit }.

The Gaussian random potential can be replaced with a potential generated
by random point obstacles: 
$$
V_\omega (x) = \sum_{\alpha} V_0(x-x_\alpha (\omega))
$$
where $\{ x_\alpha (\omega)\}$ is a random point process 
(say Poisson) with
a  density $\varrho $. In this setting, we can consider 
both the weak coupling limit $ \rho =1, \lambda = \sqrt{\e} $
and the low density limit
$$
\rho = \e \; , \lambda = 1 \; .
$$
The model we just described in the low density limit is the quantum 
analogue of  a  {\it classical Lorentz gas}. 
It  is proved by 
G.~Gallavotti [\rcite{Ga}],  H.~Spohn [\rcite{Sp2}] and 
Boldrighini, Bunimovich and Sinai 
[\rcite{BBS}] that  the evolution of the
 phase space density of a classical Lorentz 
gas converges to a linear Boltzmann equation.
For the weak coupling limit, the velocity 
field under the classical dynamics converges to a {\it Brownian motion} on the 
energy surface, see Kesten and Papanicolaou [\rcite {KP}] and  
Durr, Goldstein and Lebowitz 
[\rcite{DGL}].

The picture changes for the quantum dynamics. It was first 
proved by H. Spohn [\rcite{S}] that for sufficiently small $T$ 
the weak coupling limit of 
the  quantum dynamics \equ(scheq) is governed by 
a linear Boltzmann equation.  His result was extended 
to higher order correlation functions by Ho, Landau and Wilkins 
[\rcite{HLW}] under the same assumption. Notice that, 
instead of converging
to a Brownian motion, it converges  now to a Boltzmann equation. 
For the low density limit, the quantum Lorentz gas still converges 
to a Boltzmann equation  [\rcite{EY1}].  
In this model the Boltzmann collision
kernel $\sigma(U,V)$ is the full quantum scattering cross section. 
Hence  the full quantum collision mechanism is preserved
in the scaling limit. In the weak coupling limit only
the first Born approximation of the quantum scattering cross section
appears in the limiting Boltzmann equation.

The main objective of this paper is to prove the convergence of 
the quantum dynamics \equ(scheq) to a Boltzmann equation 
in the weak coupling limit globally
in time. An outline of our approach was given in 
[\rcite{EY1}] focusing on point obstacles in the low density limit. 
This approach was very robust and the point obstacles 
can be replaced by random fields with certain modifications 
in the estimates. 
We choose in this paper  the Gaussian law for the random potential 
only to fix the convention. 
The approach of  [\rcite{S},  \rcite{HLW}] 
relies on the convergence
of the Duhamel expansion and 
is  valid only for 
the Gaussian law in short time. 
Our approach, though still based on Duhamel formula,
estimates directly the error terms in the Duhamel formula (rather
than expand them to infinity) 
and the precise 
tail behavior of the Gaussian law is  not used.  We shall consider
 the case $d \ge 2$ in this paper.
 The case of dimension $d=1$ 
is different. We do not know the limit equation for the choice 
of weak coupling limit \equ(es) and 
\equ(wl). Some logarithmic  scaling was considered in a recent preprint
[\rcite{EP}], obtaining  a modified Boltzmann equation in the case the 
potential  of the  obstacle is a delta function.

Recall that the linear Boltzmann equation with
collision kernel $\sigma(U, V)$ is 
$$\eqalign{ 
	\partial_T F_T(X,V) + V\cdot\nabla_X F_T(X, V)
	& = \int \Big[ \sigma(U, V) F_T(X, U) - \sigma(V, U)F_T(X, V)
	\Big] dU \cr
	& =  \int  \sigma(U, V) F_T(X, U)dU - \sigma_0(V) \, F_T(X, V)
}
\Eq(Beq)
$$
with the total cross section $\sigma_0(V) : = \int \sigma(V, U)dU$.
In case of elastic collision, $\sigma(U, V)$ is supported on the
set $\{ (U, V)\, : \, U^2 = V^2\}$ (onshell condition).
If, in addition, the collision mechanism is
 spherically symmetric, then $\sigma(U, V)
= \sigma (U-V)$ and $\sigma_0(V)$ depends only on $|V|$.

\bigskip

\subheading{ 1.2  Notations}

\bigskip

For convenience, we fix a convention to avoid problems with
the $(2\pi)$'s in the Fourier transform. We define $dx$ to be the
Lebesgue measure on $\bR^d$ {\it divided by} $(2\pi)^{d/2}$, i.e.
$$
        \int dx = {1\over (2\pi)^{d/2}}\int d^* x \; ,
$$
where the notation $d^*x$ is reserved for the genuine $d$-dimensional
Lebesgue measure, which  will rarely  be used.
This convention will apply to any space or momentum variables
(like $x,y, p, q, r, v$ etc.)
in $d$-dimensions, but {\bf not} to  one dimensional integrations
(like time variables $s$, $t$ and their conjugate variables $\a, \beta$
 etc.), where these symbols
still mean the Lebesgue measure. With this convention, the
$d$-dimensional Fourier transform (usually denoted by hat) is
$$
        \widehat f(p) = \cF f(p): = \int f(x) e^{-ipx} dx
$$
and its inverse
$$
        f(x) = \cF^{-1} \widehat f(x) = \int \widehat f(p) e^{ipx} dp \; .
$$
{F}rom this notation is follows that
$$
        \int e^{ipx} dx = \delta(p)
$$
(where $\delta(p)$ is understood with respect to the $dp$ measure),
while in one dimension
$$
        \int_{-\infty}^\infty e^{i\a t} d\a = 2\pi\delta (t) \; .
$$
To avoid confusion, the notation $\int$ is reserved for $\int_{\bR^d}$,
and the limits of integration
 are always indicated for one dimensional integrals (with Lebesgue measure).

\bigskip

The {\it Wigner transform} of a function $\psi\in L^2(\bR^d)$ is
defined as
$$
        W_\psi(x,v): = \int e^{i\eta x}\overline{\widehat \psi\Big(v-{\eta\over
        2}\Big)}
        \widehat\psi\Big(v+{\eta\over 2}\Big) d\eta \; ; 
\Eq(wd)
$$
in position space it is
$$
       W_\psi(x,v) = \int e^{ivy} \overline{\psi\Big(x +{y\over 2}\Big)}
        \psi\Big( x-{y\over 2}\Big) dy \; .
$$
Define the rescaled Wigner transform as
$$
        W^\e_\psi (X, V) : = \e^{-d}W_\psi\Big( {X\over \e}, V\Big).
$$
Let $J(X, V)$ be a function in the Schwarz space 
$\cS(\bR^d\times\bR^d)$ . We need to compute
the expectation value of the quantity
$$
       \cO^J_\e:= \langle J, W^\e_{\psi_t}\rangle
        = \int \overline{J(X, V)} W^\e_{\psi_t}(X, V) dX dV \; .
$$
Recall that $X= x\e$, $T= t\e$ and $V=v$. 
Let  $J_\e (x, v) : =  J(x\e, v)$ and  $\widehat J_\e (\xi, v)
= \e^{-d}\widehat J(\xi\e^{-1}, v)$ denote the 
Fourier transform
in the first variable. Notice that 
$J_\e(x, v) \not = \e^d J(\e x, v)$ and
$\int J_\e(x, v) dx dv = \e^{-d}$ with our convention. 
We can rewrite $\cO^J_\e $ as 
$$
        \cO^J_\e 
        =\int \overline{J(x\e, v)} W_{\psi_t}(x,v) dx dv
        = \int \overline{J_\e(x, v)} W_{\psi_t}(x,v) dx dv \; .
$$
In the Fourier space, it is 
$$
     \cO^J_\e    =  \int\overline{\widehat J_\e(\xi, v)}
      \widehat W_{\psi_t}(\xi, v) d\xi dv
        =  \e^{-d}\int \overline{\widehat J(\xi \e^{-1}, v)}
        \overline{\widehat \psi_t\Big(v-{\xi\over 2}\Big)}
        \widehat \psi_t\Big( v+ {\xi\over 2}\Big) d\xi dv
\Eq(Obs)
$$
where we have used the definition \equ(wd) that 
$$
	 \widehat W_{\psi_t}(\xi, v) =	
	 \overline{\widehat \psi_t\Big(v-{\xi\over 2}\Big)}
        \widehat \psi_t\Big( v+ {\xi\over 2}\Big) \; .
\Eq(wignerfour)
$$
Following the usual convention, we use 
$$
        <x>: = (1+x^2)^{1/2}.
$$
Define the following norm for  functions on $\bR^d$
$$
        \pa f \pa_{{\a, \beta}}\; : \; = 
	\; \pa  <x>^\a <\nabla_x>^\beta f(x)   \pa_2 .
\Eq(normdef)
$$

\bigskip

\subheading{ 1.3 Main Result}

\bigskip
\noindent
{\it Assumption on the potential and the initial function}:

\medskip

We consider a centered  real Gaussian field $V(x)$ with 
covariance $G(x-y):= \bE V(x)V(y)$. Since $G$ is real, 
$\widehat G(p) = \overline{\widehat G(-p)}$. From the 
property of the covariance, 
$\widehat G(p) \ge 0$. Thus we can define $R(p): = \widehat G(p)^{1/2}$.
The covariance of $V$ in the Fourier space can be written as 
$$
        \bE \hV(p) \hV(q) = \delta(p+q)\widehat G(p) =
        \delta(p+q) R(p)R(q)
$$
and
$$
        \bE \overline{\hV(p)}\hV(q) = \delta(p-q) \widehat G(p)
        =\delta(p-q)R(p)R(q) \; .
$$ 
We assume $R$ to be  radially symmetric and smooth. In fact
$$
        \| R(p) \|_{20d, 20d} < \infty \; 
\Eq(assump)
$$
is enough, but for simplicity we assume $R\in \cS(\bR^d)$.

We shall choose the initial data 
$$
        \psi_0 (x) =\psi_0^\e (x) : = \e^{d/2} h(\e x) e^{i S(\e x)/\e}
$$
with some $h, S\in \cS(\bR^d)$. We never use
the full smoothness and decay properties of the Schwarz space and 
some weaker conditions such as $\| h\|_{20d, 20d}<\infty$
and $\| S\|_{20d, 20d}<\infty$ are sufficient. 
Clearly, $\psi_0^\e$ satisfies 
$$
        \Big\|<p>^{20d}\widehat\psi_0^\e(p)\Big\|_2\leq C
\Eq(assumpi)
$$
with some constant $C$ uniformly in $\e$.
 It is well known that 
\be
	W^\e_{\psi_0^\e} (X, V) \to |h(X)|^2 \delta (V- \nabla S(X))
	: = F_0(X, V) 
\label(initwig)
\ee
weakly in $\cS(\bR^{2d})$ as $\e\to0$.

\proclaim{\Theorem (mainth)}
Let $\lambda= \sqrt{\e}$, $d\ge 2$ and let $\psi_{t,\omega}^\e$
solve the Schr\"odinger equation \equ(scheq) with
initial condition
\be
	\psi_0^\e(x) : = \e^{d/2}h(\e x) \exp{\Big({iS(x\e)\over \e}\Big)}.
\label(init)
\ee
Assume that $h(X), S(Y)$ are in the Schwarz class.
In $d=2$ we assume in addition that the set $\{ (X, V) \; : \; V=0\}$
has zero measure with respect to the probability measure $F_0(X, V)
dX dV$ (cf. \equ(initwig)).
Then for any $T>0$
$$
	\bE W^\e_{\psi_{T/\e, \omega}^\e} (X, V) 
	\to F_T(X, V) 
$$
weakly in $\cS(\bR^{2d})$
as $\e\to0$ , and $F_T$ satisfies the linear Boltzmann
equation \equ(Beq) with initial condition $F_0(X,V)$ and
collision kernel $\sigma (U, V) = 4\pi\widehat G(U-V)\delta(U^2-V^2)$.
\endproclaim

\smallskip

\noindent {\it Convention 1.}  Throughout the paper, $C$ denotes various
constants which depend on $d$, $T$, $h$, $S$ 
 and $R$ but does
not depend on $\e$.

\smallskip

\noindent {\it Convention 2.} The random parameter $\omega$
will be omitted from the notation in most cases.

\bigskip

\subheading{ 2. Outline of the Proof}
\numsec=2\numfor=1\numtheo=1

\medskip

\subheading{ 2.1 The Duhamel formula}

\bigskip

The Duhamel formula states that for any $n_0 \ge 1$
$$
        \psi_t = e^{-itH}\psi_0 = \sum_{n=0}^{n_0-1} \psi_n (t)
	 + \Psi_{n_0}(t) \; ,
\Eq(decomp)
$$
where $H_0= -{1\over 2}\Delta$,
$$
        \psi_n(t) : = (-i)^n 
	\lambda^n\int_0^t\ldots \int_0^t \Big(\prod_{j=0}^n ds_j\Big)
        \delta\Big( t- \sum_{j=0}^n s_j\Big) 
	e^{-is_0H_0} V e^{-is_1H_0} V\ldots
        V e^{-is_n H_0}\psi_0
$$
and
$$
        \Psi_{n_0} (t): = (-i)\lambda \int_0^t ds \, e^{-i(t-s)H} 
	V \psi_{n_0-1}(s)
$$
is the  error term.  
Sometimes we shall use the notations 
$$
        \widetilde \psi_n (t) : = \lambda V\psi_{n-1}(t)
$$
and
$$
        \psi_{<n_0}(t) = \sum_{n=0}^{n_0-1} \psi_n(t) \; .
$$
We shall choose  
$$
       n_0 =  n_0(\e) =   {\gamma|\log \e |\over\log |\log \e| }
\Eq(n0)
$$
with some small fixed $\gamma >0$.
Notice that 
$$
      \e^{\alpha}  |\log \e|^{n_0(\e)} \to 0 \qquad
	\hbox{and} \qquad \e^\a \Big[ n_0(\e)\Big]! \to 0
\Eq(n0p)
$$
as $\e\to0$ for any $\alpha > \gamma$.

For any subset $I \subset \NN$, we use the convention that
$$
        \fp_I: = \{ p_i \, : \, i\in I\} \; .
$$
We also introduce the infinite sequence 
$$
        \fp: = \{ p_0, p_1, \ldots \} \; .
$$
In the special case 
$$
        I_n: = \{ 0, 1, \ldots, n\}
$$
we denote  $\fp_n: = \fp_{I_n}$. Denote  
$\fp_{n, \widehat 0} : = \fp_{I_n\setminus \{0\}}$.
Define the kernel for the free evolution in the Fourier
(momentum) space by 
$$
        K(t; \fp, I): = K(t; \fp_I) = (-i)^{|I|-1}\int_0^t \Bigl( \prod_{j\in I} \rd s_j\Bigr)
        \delta(t-\sum_{j\in I}s_j)\prod_{j\in I} e^{-is_jp_j^2/2} \; .
\Eq(Kkernel)
$$
Again, we denote the special case $I= I_n$ by 
$$
	  K(t; \fp, n): = K(t; \fp, {I_n}) \; .
$$
Denote the contribution from the potential term by
$$
        L(\fp, n):  = \prod_{j=0}^{n-1} \widehat V( p_j- p_{j+1}) \; .
\Eq(L)
$$
Then we can rewrite the $n-$th order wave function as
$$
        \widehat\psi_n(t, p_0) = \lambda^n \int d\fp_{n, \widehat 0}
         K(t; \fp, n) L(\fp, n) \widehat\psi_0(p_n) \; .
\Eq(psiexp)
$$

The decomposition \equ(decomp) has the disadvantage that
the threshold $n_0$ depends on the scaling parameter. Hence, we
rewrite it as
$$
        \psi_t = \sum_{n=0}^{M-1}\psi_n(t) + \sum_{n=M}^{n_0-1}
        \psi_n(t) + \Psi_{n_0}(t)
        = \psi^{main}_t + \psi^{error}_t \; ,
$$
where
$$
        \psi^{main}_t = \psi^{main}_{t, M} : = \sum_{n=0}^{M-1}\psi_n(t) 
$$
and
$$
        \psi^{error}_t = \psi^{error}_{t, M} : =
        \sum_{n=M}^{n_0-1} \psi_n (t) +\Psi_{n_0}(t) \; .
$$
We shall show in Sections 3. and 5. that for any $T>0$
$$
        \lim_{M\to \infty}\lim_{\e\to 0} \bE \| \psi^{error}_{T/\e, M}
        \|^2 = 0 \; .
\Eq(errors)
$$
{F}rom the unitarity of the Schr\"odinger evolution, 
  $\|\widehat\psi_t\|^2=1$ for all $t$. Thus 
$\|\widehat \psi^{main}\|^2$ 
is  bounded from the inequality 
$$
        \|\widehat \psi^{main}_t\|^2\leq 2\|\widehat\psi_t\|^2 + 2\|\widehat
        \psi^{error}_t\|^2 \; .
$$

We shall consider test functions $\Gamma =\Gamma_\e$ in Fourier space
 such that 
$\Gamma_\e(\xi, v)\in L^1_\xi (\bR^d, L^\infty_v(\bR^d))$
uniformly in $\e$, i.e.
$$
\sup_{0<\e\leq 1}\int\sup_v  | \Gamma_\e(\xi, v)|d\xi < \infty \; .
\Eq(supsup)
$$
In particular $\Gamma_\e =\widehat J_\e$ satisfies \equ(supsup).
The expected value of the 
observable generated by a test function in  this class is bounded 
in the sense that
$$
\Bigg|  \bE \int \overline{\Gamma(\xi, v)}  \widehat
        W_{\psi}(\xi, v)  d\xi dv \Bigg|
 \leq C \Big(\int\sup_v  |\Gamma(\xi, v)|d\xi\Big)
          \bE \|\psi\|^2  \; .
\Eq(wb)
$$
To prove this, we have from the  Schwarz inequality and \equ(wignerfour)
$$\eqalign{ 
\Bigg|  \bE \int \overline{\Gamma(\xi, v)}  \widehat
        W_{\psi}(\xi, v)  d\xi dv \Bigg|
 & \leq     \bE \int  d\xi dv |\Gamma(\xi, v)|
 \( \Big|\widehat\psi \Big(v-{\xi\over 2}\Big)
        \Big|^2 +
        \Big|\widehat\psi \Big(v+{\xi\over 2}\Big)\Big|^2 \) \cr
&         \leq C \Big(\int\sup_v  |\Gamma(\xi, v)|d\xi\Big)
        \bE \|\widehat\psi\|^2 \; .
}
$$
We can generalize this inequality to prove 
 the continuity of the expected value of the observables
with respect to the decomposition of the wave function in $L^2$.  
Suppose that we have a decomposition of a wave function
$\psi= \psi_1+\psi_2$.  Then
$$
         \Bigg| \bE\cO^J_\e - \bE\int
         \overline{\widehat J_\e(\xi, v) }\widehat
        W_{\psi_{1}} (\xi, v)d\xi dv\Bigg| 
= \Bigg| \bE\int \overline{\widehat J_\e (\xi, v)} \[ \widehat
        W_{\psi}(\xi, v) -\widehat
        W_{\psi_{1}}(\xi,v) \] d\xi dv\Bigg|
$$
{F}rom the Schwarz inequality and  \equ(wignerfour)
$$
\eqalign{
	&\Big|  \widehat  W_{\psi}(\xi, v) 
       - \widehat  W_{\psi_1}(\xi, v) \Big|\cr
	&\leq C\Bigg( \a  \Big|\widehat\psi_1 \Big(v-{\xi\over 2}\Big)
        \Big|^2 + \a  \Big|\widehat\psi_1 \Big(v+{\xi\over 2}\Big)\Big|^2+
	\a^{-1} \Big|\widehat\psi_2 \Big(v-{\xi\over 2}\Big)
        \Big|^2+ \a^{-1} \Big|\widehat\psi_2
	 \Big(v+{\xi\over 2}\Big)\Big|^2 \Bigg)\; .
}
$$
Optimizing $\a$, we have 
$$
 \Bigg| \bE\cO^J_\e - \bE\int
         \overline{\widehat J_\e(\xi, v) }\widehat
        W_{\psi_{1}} (\xi, v)d\xi dv\Bigg|
 \leq C \Big(\sup_{\e\leq 1}\int\sup_v  |\widehat J_\e(\xi, v)|d\xi\Big)
        \sqrt {  \bE \|\psi_1\|^2  \; \bE \|\psi_2\|^2} \; .
\Eq(optim)
$$
We use this for $\psi_1 = \psi_t^{main}$, $\psi_2 = \psi_t^{error}$.
We already know from \equ(errors) that the $L^2$ norm of the error
terms vanishes. Therefore, 
$$
	\lim_{\e\to 0} \bE {\Cal O}_\e^J =
	 \lim_{M\to\infty}\lim_{\e\to0}      
	 \bE\int  \overline{\widehat J_\e (\xi , v)} \widehat
        W_{\psi^{main}_{T/\e, M}}
        (\xi, v) d\xi dv  \; .
\Eq(wm)
$$
So we only have
to compute the Wigner distribution of the main term  in the limit.
This  will be done in Section 4.

\bigskip 

\subheading{ 2.2 Pairings}

\bigskip

We shall compute 
the expectation of the Wigner transform.  Define
$$
        U_{n,n'}^{\Gamma}(t):=\bE \int
        \overline{\Gamma( \xi, v)}
        \overline{\widehat{\psi}_{n}\Big(t, v-{\xi\over 2}\Big)}
        \widehat\psi_{n'}
        \Big(t, v+{\xi\over 2}\Big) dv d\xi
\Eq(UG)
$$
for any  $n,n'\ge 0$ and any testfunction $\Gamma(\xi, v)\in L^1_\xi (\bR^d, 
L^\infty_v(\bR^d))$ (i.e.
 $\int \sup_v |\Gamma(\xi, v)| d\xi <\infty$).
{F}rom \equ(Obs) we have formally
$$
      \lim_{\e\to0} \bE \cO^J_\e \sim \sum_{n,n'=0}^\infty
        \lim_{\e\to0} U_{n,n'}^{\widehat J_\e}(T/\e) \; .
$$
Note that if $\Gamma( \xi, v) = \d(\xi)$ then 
$$
        U_{n,n}(t) := U_{n,n}^{\Gamma=\delta(\xi)}(t)= 
\bE \|\psi_n(t)\|^2 \; 
$$
is simply the expectation of the $L^2$ norm of $\psi_n$.  
Recall $J_\e (x, v) : =  J(x\e, v)$, $\widehat J_\e (\xi, v)
= \e^{-d}\widehat J(\xi\e^{-1}, v)$ and certainly
$\widehat J_\e (\xi, v) \in L^1_\xi
 (\bR^d, L^\infty_v(\bR^d))$ uniformly in $\e$.

By definition of $\widehat \psi_n$ \equ(psiexp),
 $U_{n,n'}^{\Gamma}$  has 
two sets of potentials with the following arguments:
$$
        \lambda^n\overline{\widehat V(p_0-p_1)}\overline{\widehat V(p_1-p_2)}
         \ldots \overline{\widehat V(p_{n-1}-p_n)}
$$
$$
        \lambda^{n'} \widehat V(p_0'-p_1') \widehat V(p_1'-p_2')\ldots
        \widehat V(p_{n'-1}'-p_{n'}')
$$
with $p_0:= v-\xi/2$, $p_0':=v+\xi/2$.
Since $V$ is Gaussian, we can compute the expectation 
of $U_{n,n'}^{\Gamma}$ explicitly. 
This introduces pairings of potentials
and yields delta
functions among the momenta. 
Clearly, $U_{n,n'}^{\Gamma}=0$ if $n+n'$ is odd. 
We divide the delta functions into the following three
types:
$$
\eqalign{
        Type \,\, I:\,\, &\qquad
	 \delta (-p_i+p_{i+1} - p_j+p_{j+1}) \qquad i<j\cr
        Type \,\,  I':\,\, &\qquad
	\delta (p_i'-p_{i+1}' +p_j'-p_{j+1}') \qquad i<j\cr
        Type \,\,  II:\,\, & \qquad
	\delta( -p_i +p_{i+1} + p_j' - p_{j+1}') \; .
}
$$
In all these delta functions, each $p_i$ and $p_i'$ with $0 < i < n$, 
$0< i'<n'$
appear exactly twice with opposite signs, 
 $p_0'$ and $p_n$  once with
plus sign, and $p_0$ and $p_{n'}'$  once with a minus sign.
Finally, the relation $p_0= v-\xi/2$, $p_0'=v+\xi/2$ is expressed by
$\delta(p_0-p_0'+\xi)$, i.e. we add
$$
        Type \,\, III(\xi): \,\, \delta (p_0 - p_0' +\xi)
$$
to our list. In particular, $p_0=p_0'$ if $\xi=0$, which is needed for 
computing the $L^2$-norm.
 We have exactly ${n+n'\over 2}+1 = \bar n + 1$ delta functions with
$\bar n:= {n+n'\over 2}$.
Notice that if add up the arguments of all
these delta functions, we arrive at a new delta function
$\delta (p_n - p_{n'}'+ \xi)$. We can replace the delta function 
$\delta (p_0 - p_0' +\xi)$ by the new delta funtion $\delta (p_n - p_{n'}'+ \xi)$.

Define
$$
        F(\fp, n): = \prod_{j=0}^{n-1} R(p_j-p_{j+1})
        \widehat\psi_0(p_n) \; .
\Eq(F)
$$
and let
$$
	\Phi(p): = <p>^{20d} |\widehat\psi_0(p)| \; .
\Eq(Phi)
$$
By the property of the initial wave function 
\equ(assumpi), we have $\Phi \in L^2$.

Let $\Pi_{n, n'}$
denote the set of all pairing with $n$ copies of  $\overline{ V} $ and
$n'$ copies of $V$. Denote by $\Delta_\pi(\fp, \fp')$ 
the product of Type I, I', and II delta functions
associated with the pairing $\pi$.  
With these notations, we have 
$$
        U_{n,n'}^\Gamma =\bE \int \int d \xi dv \overline{\Gamma(\xi, v)   }
         \overline{\widehat\psi_n\Big(t, v-{\xi\over 2}\Big)}
         \widehat\psi_{n'}\Big(t, v+{\xi\over 2}\Big)
        = \sum_{\pi \in \Pi_{n, n'}} C_\pi^\Gamma
\Eq(U)
$$
with
$$\eqalign{ 
        C_\pi^\Gamma := & \lambda^{n+n'} \int d \xi dv 
	\overline{\Gamma(\xi, v)} 
	\int d\fp_n d\fp_{n'}'
        \overline{K(t,  \fp,n)}K(t,  \fp', n')\cr
	&\times \Delta_\pi (
        \fp, \fp') \overline{F( \fp,n)}F( \fp',n')
        \delta(p_0-v+\xi/2)\delta(p_0'-v-\xi/2) \; .
}
\Eq(cpigG)
$$
Here $F$ is defined in \equ(F). In the special case that $\Gamma$
is just the delta function in $\xi$, we drop the superscript 
$\Gamma$. This is needed when we estimate the $L^2$ norm. 
We shall use the notation  
$$
        \Delta_\pi^\xi (\fp, \fp'): = \Delta_\pi(\fp, \fp')
        \delta(p_0-p_0'+\xi)= \Delta_\pi(\fp, \fp')
        \delta(p_0-p_{n'}'+\xi)
$$
Notice that 
 the product of all  delta functions in \equ(cpigG)
is $ \Delta_\pi^\xi (\fp, \fp')
\delta(p_0-v+\xi/2)$. When computing $L^2$-norm,
we shall need $\Delta^0_\pi$ i.e. $\xi=0$. Therefore,  
$$
        C_\pi :=  \lambda^{n+n'} \int d\fp_n d\fp_{n'}'
        \overline{K(t,  \fp,n)}K(t,  \fp', n') \Delta_\pi^0 (
        \fp, \fp') \overline{F( \fp,n)}F( \fp',n') \; .
\Eq(cpig)
$$

Next, we describe the structure of $\Delta_\pi^\xi ( \fp, \fp')$.
Sometimes we shall consider  the two sets of momenta in a unified
way, hence we define 
$$
w_j : = p_{n-j}, \quad 0\leq j\leq n \; ; \qquad 
w_{n+j+1} := p_j' , \qquad 0\leq j\leq n'
\Eq(w)
$$
Denote by  $\fw = (w_0, w_1, w_2, \ldots )$.
We shall use the ($\fp_n$, $\fp_{n'}'$) and
 the $\fw_{2\bar n+1}$  notations simultaneously
where $2\bar n = n+ n'$.

\proclaim{\Definition (equiv)}
We consider products of delta functions with arguments
that are linear combinations of the momenta
$\{ w_0, w_1,\ldots , w_{2\bar n +1}\}$.
Two products of such delta functions  are called {\bf equivalent} if
they determine the same affine subspace of $\bR^{2\bar n+2}
= \{ (w_0, w_1, \ldots, w_{2\bar n +1}) \}$.
The {\bf rank} of a product of deltafunctions is the codimension of this
affine subspace.
\endproclaim

One can obtain new delta functions 
from the given ones, by taking linear combinations of their arguments.
In particular, we can obtain  {\it identifications} of
momenta.

\proclaim{\Definition (forced)}
The product of delta functions $\Delta_\pi^\xi(\fw)$ {\bf forces} a new delta
function $\delta (\sum_j a_j w_j)$, if $\sum_j a_j w_j =0$ is an 
identity in the affine subspace determined by 
$\Delta_\pi^\xi(\fw)$.
 If $\Delta_\pi^\xi(\fw)$ {\bf forces} $\delta (w_i-w_j)$,
with some $i, j$, then we say it forces 
the {\bf identification} $w_i=w_j$.
\endproclaim

\proclaim{\Lemma (rank)}
(I)  For every pairing $\pi$, there
exists a partition $A\cup B = \{0, 1, 2, \ldots 2\bar n, 2\bar n+1\}$,
 $A\cap B=\emptyset$ such that 

\noindent (i)  $|A|=|B|=\bar n +1 $, $0\in A$, $ 2\bar n +1\in B$,

\noindent (ii)
 $\Delta_\pi^\xi(\fp, \fp')= \Delta_\pi^\xi( \fw)$ 
is equivalent to the product of the following delta functions:
$$
\eqalign{ &\delta\Big( w_j - \sum_{\ell\in A\atop \ell < j} \pm w_\ell\Big)
	\qquad \forall j\in B, \,\, j\neq 2\bar n +1\cr
   \hbox{ and }  \quad& \delta (w_{0}-w_{2\bar n+1}+ \xi)\; .
}
$$
In other words,  we have
$$
        \fw_B = {\Cal M}_0\fw_{A} + \xi {\bold e}_1
\Eq(AB)
$$
where $\fw_A := ( w_\ell \, : \, \ell\in A )^T$, $\fw_B : = (w_j \, : \,
j\in B)^T$
and ${\bold e}: = (0,0,\ldots, 0,  1 )^T$. Here
${\Cal M}_0 = {\Cal M}_{0}^{(\pi)}$
 is an $(\bar n+1)\times (\bar n+1)$ matrix,
such that all its entries are 0 or $\pm 1$ and 
each row contains one more $+1$ than $-1$. In particular, each row
has an odd number of nonzero elements.
We can order  the elements in $B$ by  $B = \{ b_1 < \ldots <b_\nu <  
\ldots < b_{\bar n+1}\}$,
and similarly for  $A = \{ a_1< \ldots <
a_\mu< \ldots < a_{\bar n+1}\}$.
Then the matrix ${\Cal M}_0= \Big( m_{\nu\mu}\Big)$ 
has the property that  $m_{\nu\mu}=0$ if $a_\mu > b_\nu$.
The rank of  $\Delta_\pi^\xi ( \fw)$
is $\bar n +1$.

(II)
Let $\widetilde \Delta_\pi^\xi (\fp,\fp')$ be the product of a subset
of  delta functions  in $\Delta_\pi^\xi (\fp,\fp')$,
which contains $r\leq \bar n$
 delta functions. We also assume that $\delta (p_n-p_{n'}' + \xi)$
is forced by $\widetilde \Delta_\pi^\xi (\fp,\fp')$.
 Then the sets $A$, $B$ can be chosen exactly with
the properties above, except that $|B|= r$ and $|A|= 2\bar n+2-r$. The
dimension of ${\Cal M}_0$ is $r\times (2\bar n-r+2)$
but all the other properties of ${\Cal M}_0$ are the same.
The rank of $\widetilde \Delta_\pi^\xi (\fp,\fp')$  is $r$.
\endproclaim

\noindent
{\it Remark.} By definition of $B$, 
the last line of \equ(AB)  
is $w_{2\bar n+1} = w_0 + \xi$ (recall $b_{\bar n+1}=2\bar n+1$).
This is the sum of the arguments of all delta functions in $\Delta^\xi_\pi
(\fw)$.

The proof of Lemma \equ(rank) is based upon a simple Gaussian elimination.
{F}rom the coefficients in the arguments
 of the delta functions in $\Delta^\xi_\pi(\fw)$ we form an
$(\bar n+1)\times (2\bar n+1)$ matrix. Each row of
this matrix corresponds to a delta function, and each column corresponds
to a momentum $w_k$, $k=0, 1, \ldots 2\bar n+1$. The  set $B$ consists
of the biggest indices that appear among the arguments in each
delta function. The complementary set is $A$. The corresponding
momenta are called $B-$ and $A-$momenta, respectively.
 Then we perform a Gaussian elimination to
express the $B$-momenta as  linear combinations of the $A$-momenta.
It is easy to show that, after the elimination, the matrix has the
structure $\Big[  {\Cal M}_0 \, \Big| \, -I \Big]$ with a matrix 
${\Cal M}_0$  described in the Lemma.
The details are given in the Appendix.

\bigskip 

\subheading{ 2.3 Graphs}

\bigskip

We classify the pairings according to their combinatorial structure.
This classification and certain properties of the pairings
are easier to illustrate in a graphical representation.
For any pairing $\pi\in \Pi_{n,n'}$
 we define a graph $G=G_\pi$ as follows (Fig. 1). There are  $n+n'$
bullets arranged in two lines. They represent the potentials
in $\psi_n$ and $\psi_{n'}$.
The upper line correspond to $\overline{\widehat\psi_n}$, the lower
line to $\widehat\psi_{n'}$.
  The initial wave function $\overline{\widehat\psi_0(p_n)}\widehat\psi_0
(p_{n'}')$ is denoted
 by a small square on the
right side of the picture. The observable 
$$
	\int d\xi dv 
	\Gamma (\xi, v)\delta(p_0-v+\xi/2)\delta(p_0'
	-v-\xi/2) = \Gamma\Big(p_0'-p_0, (p_0+p_0')/2 \Big)
$$
is denoted  by a small square on the left.
 The bullets are
joined by directed arrows representing momenta; $p_1, \ldots p_{n-1},
p_1', \ldots p_{n'-1}'$.
The squares are connected with the first and last bullets
by arrows corresponding to $p_0, p_0', p_n, p_{n'}'$ as shown on
Fig. 1. The arrows
are called momentum lines, denoted by $e(p_i)$ and $e(p_i')$,
the argument referring to the corresponding momentum.
The bullets and the squares with the arrows form a directed circle
graph, denoted by $C$.

Type I, I' and II delta functions are represented
by undirected dashed lines joining two bullets and we call them pairing lines.
 We draw these lines outside of $C$. For example, Fig 1. represents
a pairing $\pi\in \Pi_{6, 4}$ with
the following deltafunctions: $\delta (-p_0+p_1-p_2+p_3)$,
$\delta(-p_1+p_2-p_3+p_4)$, $\delta(-p_4+p_5 + p_3' - p_4')$,
$\delta (-p_5+p_6 +p_3' - p_2')$ and $\delta(p_0'-p_1' + p_1' - p_2')$.
Notice that the signs of the momenta in the arguments correspond
to the direction of the arrows. At any bullet, an incoming arrow involves
a plus
sign, an outgoing one involves a minus sign in the delta function
pairing that bullet.

\bigskip\bigskip
\centerline{\epsffile{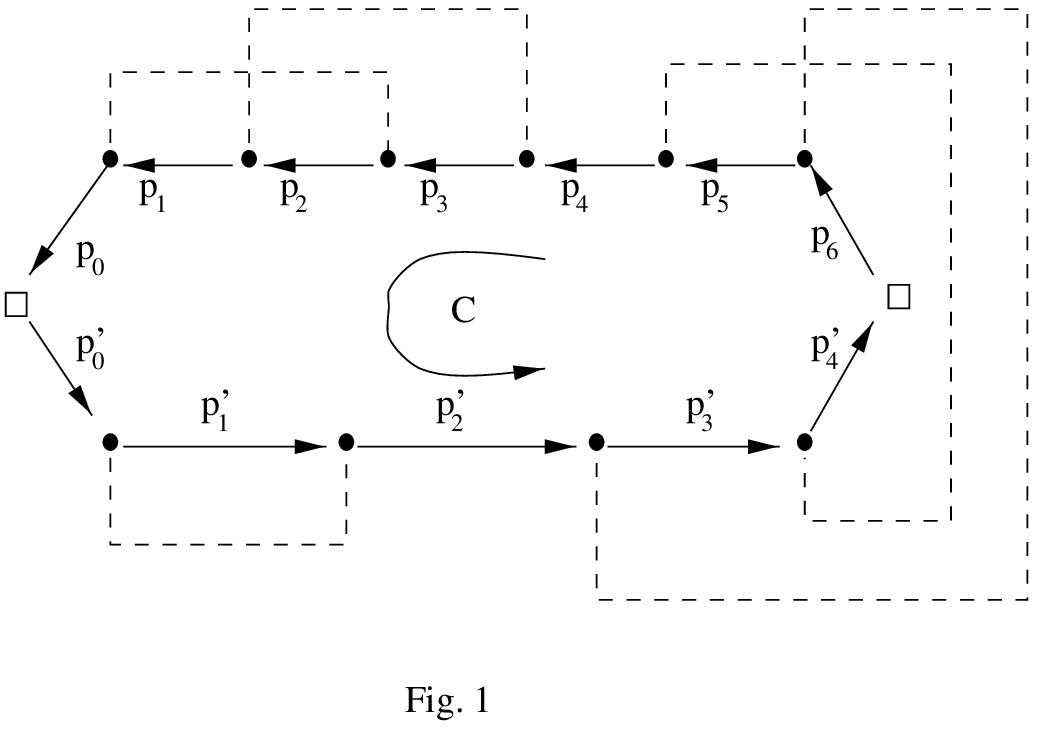}}

Type $III(\xi)$ delta function $\delta (p_0-p_0' + \xi)$
 corresponds to a momentum
transfer $\xi$ at the left square.
Recall that by adding up the arguments of 
all the delta functions, we arrive at
$\delta (p_n-p_{n'}'+\xi)$, implying a momentum transfer $\xi$
at the right square as well. The actual value $\xi$ of these
 momentum transfers will not
play any role in the definitions and theorems related to the
graphical representations. We can set $\xi=0$ for simplicity.
The classification of the pairings
will depend only on Type I, I', II delta functions.

Sometimes we use the notation $w_0, w_1, \ldots, w_{2\bar n +1}$
for labelling the momentum lines.
The momentum and pairing lines together are called the edges
of $G_\pi$, while the bullets and squares are called the vertices
of $G_\pi$. The set of edges of a graph $G$ is $E(G)$, the
set of vertices is $V(G)$.
We introduce the following classes of pairings:

\proclaim{\Definition (crossing)}
The pairing $\pi\in \Pi_{n,n'}$ 
is called {\bf crossing} if there exist
two delta functions among Type I, I' and II; 
$\delta (w_i - w_{i+1} + w_j - w_{j+1})$ and 
$\delta (w_k - w_{k+1} + w_\ell - w_{\ell +1})$
such that $i < k < j < \ell $.
Otherwise, the pairing is called {\bf noncrossing}.
\endproclaim

\proclaim{\Definition (nested)}
A noncrossing pairing is called {\bf nested} if
there are two Type I delta functions $\delta(-p_i+p_{i+1} - p_j+p_{j+1})$,
$\delta (-p_k + p_{k+1} - p_u + p_{u +1})$
with $i<k<u <j$ or there are two Type I' delta functions
$\delta(p_i'-p_{i+1}' + p_j'-p_{j+1}')$,
$\delta (p_k' - p_{k+1}' +p_u' - p_{u +1}')$
with $i'<k'<u' <j'$. A noncrossing and nonnested pairing is
called {\bf simple}.
\endproclaim

\noindent 
{\it Remark.} These notions could be defined directly by using the graph
of a pairing.
Obviously, $\pi$ is crossing if and 
only if it is not possible to draw all the pairing
lines in $G_\pi$  without crossings. Recall that the
pairing lines are always drawn outside  $C$.
Similarly, a noncrossing $\pi$ is nested if and only if
$G_\pi$ contains
two Type I (or two Type I') pairing
 lines such that one is fully inside the other.
These pairing lines are called {\it outer} and {\it inner} deltas.
The inner delta $\delta(-p_k + p_{k+1} - p_u + p_{u+1})$ is called
{\bf minimal} if $u=k+1$.
 Fig. 2
shows a nest with  Type I delta functions.

\bigskip\bigskip
\centerline{\epsffile{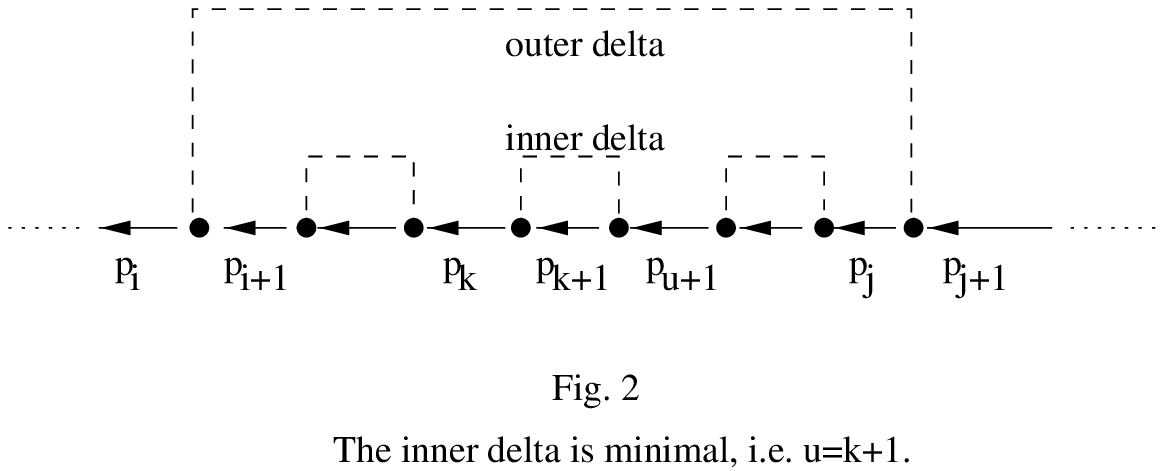}}

\medskip

Recall the Definition \equ(forced) of the forced identification.
The following Lemma gives a characterization of the forced identifications
in terms of the graph:

\proclaim{\Lemma (forceid)}
The product of delta functions $\Delta_\pi^0(\fw)$ forces
$w_i=w_j$, for some $0\leq i, j\leq 2\bar n +1$,
 if and only if the removal of the two momentum lines
corresponding to $w_i, w_j$ makes the graph $G_\pi$ disconnected.
\endproclaim

The proof of Lemma \equ(forceid) is given in the Appendix. Observe that this Lemma properly contains the
identifications $p_0 =p_0'$ and $p_n=p_{n'}'$, as the removal
of one of these pairs of momentum lines
 makes one of the squares disconnected.
Finally, the relation between the graphical and matrix representation of
the crossing pairing is given in the following lemma which
will be proved in the appendix. Recall the
construction of the matrix ${\Cal M}_0^{(\pi)}$ in the proof of
Lemma \equ(rank).

\proclaim{\Lemma (crossmatrix)}
A pairing $\pi\in \Pi_{n,n'}$ is crossing if and only if there is a row in
 the matrix ${\Cal M}_0^{(\pi)}$ which has more than one nonzero element.
\endproclaim

\bigskip

\subheading{3. $L^2$ Estimates of the Duhamel formula}
\numsec=3\numfor=1\numtheo=1
\bigskip 

We first fix some notations. 
The following  property can be  easily checked
$$
        {1 \over <p_n>^{2d}}
         \prod_{j=0}^{n-1}
        {1\over <p_j-p_{j+1}>^{2d}}
        \leq {C^n\over <p_k>^{2d}}
\Eq(iterestim)
$$
for any $0\leq k\leq n$. Hence
$$
        {1 \over <p_n>^{20d}}
         \prod_{j=0}^{n-1}
        {1\over <p_j-p_{j+1}>^{20d}} \leq
        C^n\prod_{i=1}^9{1\over <p_{\ell_i}>^{2d}}
         \prod_{j=0}^{n-1}
        {1\over <p_j-p_{j+1}>^{2d}}
\Eq(normiteration)
$$
for any numbers $\ell_1, \ell_2,\ldots \ell_9$
between 0 and $n$. It is impossible to have the product of all  factors,
$\prod_j < p_j>^{-2 d}$, but one has the freedom
to choose some of them as we did in \equ(normiteration).
By the assumption \equ(assump) on $R$  and the last inequality, 
$F$   satisfies 
$$
        |F(\fp,n)|\leq {C^n \Phi(p_n)\over <p_n>^{20d}}
        \prod_{j=0}^{n-1}
        {1\over <p_j-p_{j+1}>^{20d}}
        \leq  C^n \Phi(p_n) \prod_{i=1}^9
        {1\over <p_{\ell_i}>^{2d}}
         \prod_{j=0}^{n-1}
        {1\over <p_j-p_{j+1}>^{2d}}
\Eq(Festim)
$$
for any $\ell_1, \ell_2, \ldots, \ell_9 \leq n$.

Define the measure
$$
        d\mu(\fp_n) := {|\Phi(p_n)|^2 \; d\fp_n \over <p_n>^{20d}}
        \prod_{j=0}^{n-1}
        {1\over <p_j-p_{j+1}>^{20d }} \; ,
$$
or, more generally for any  set $I:= \{ i_1, i_2, \ldots i_n\}$
$$
        d\mu(\fp_I) : = {|\Phi(p_{i_n})|^2\; d\fp_I \over <p_{i_n}>^{20d}}
        \prod_{j=0}^{n-1}
        {1\over <p_{i_j}-p_{i_{j+1}}>^{20d}} \; .
\Eq(mu)
$$
To estimate this measure we shall recall \equ(normiteration) yielding
$$
	 d\mu(\fp_n) \leq {C^n |\Phi(p_{n})|^2\; d\fp_n \over 
	\prod_{i=1}^9 <p_{\ell_i}>^{2d}}
        \prod_{j=0}^{n-1}
        {1\over <p_{i_j}-p_{i_{j+1}}>^{2d}} \; .
\Eq(muest)
$$	
for any $\ell_1, \ell_2, \ldots \ell_9$.
Moreover, we also have the simple inequality
$$
        \sup_q {1\over <p-q>^{20d}
        <q-r>^{20d}}\leq {C\over <p-r>^{20d}} \; .
\Eq(middlesup)
$$ 

\bigskip
\subheading{ 3.1. K-identity and Estimate}
\bigskip 

\proclaim{\Lemma (K-id2)}
We have the following identity and estimate  for $\eta>0$
$$
        K(t, \fp, I) = i e^{t\eta/2} \int d\alpha e^{-i\alpha t}
        \prod_{j\in I} { 1\over \alpha  -
        p_j^2/2 + i \eta} \; ,
\Eq(Kid)
$$
$$
        |K(t, \fp, I)| \leq C\int d\alpha \prod_{j\in I}
        {1\over |\alpha - p_j^2/2 + it^{-1}|}
\Eq(Kest)
$$
(note that the estimate is finite only for $|I|\ge 2$), and for any $a<1$,
$|I|\ge 2$,
$$
        \int |K(t, \fp, I)|^2 d\mu(\fp_I)
         \leq {\Big(C_a t\Big)^{|I|-1}
        \over \Big[(|I|-1)!\Big]^a} \; .
\Eq(Kestcomb)
$$
\endproclaim

\demo{Proof of Lemma \equ(K-id2)}
Since $s_j\ge 0 $ and $\delta\Big( t-\sum_{j\in I} s_j\Big)$
automatically guarantee that all $s_j\leq t$, 
the upper integration bound of $t$ in the
definition of $K(t, \fp, I)$ \equ(Kkernel) can be replaced with 
$\infty$. Hence
$$
        K(t, \fp, I)  = (-i)^{|I|-1} e^{t\eta/2}
        \int_0^\infty\ldots \int_0^\infty
        \Bigg(\prod_{j\in I} ds_j\Bigg) \prod_{j\in I}
        e^{-is_j(p_j^2/2-i \eta)}
        \delta\Big(t-\sum_{j\in I} s_j \Big)
$$
$$
        = (-i)^{|I|-1} e^{t\eta/2} \int_{-\infty}^\infty
         d\alpha e^{-i\alpha t}
        \prod_{j\in I}\Bigg(\int_0^\infty
         e^{is_j(\alpha - (p_j^2/2-i \eta))}ds_j\Bigg)
$$
$$
        = (-i)(-1)^{|I|-1}  \int_{-\infty}^\infty
         d\alpha e^{-i\alpha t +t \eta/2}
        \prod_{j\in I} {-1\over \alpha  -
        p_j^2/2+ i \eta} \; .
$$
Note that the same identity (and its proof) holds if we replace some of the
$p_j^2$ in $K(t,\fp, I)$ with $p_j^2 - i\mu$ (here $\mu\ge 0$).
Recall that $K(t, \fp, I)$ depends only on the square of the momenta, 
i.e.,  the function
$$
        K^*(t, \fp^2, I): = K(t, \fp, I)
\Eq(Kstardef)
$$
is well defined. The same proof yields the identity
$$
        K^*(t, \ba, I) = i e^{t\eta/2}\int_{-\infty}^\infty  d\alpha
        e^{-i\alpha t }
        \prod_{j\in I} {1\over \alpha  -
        a_j/2+ i \eta}
\Eq(Kstarid)
$$
for any $\ba_I = \{ a_j\, : \, j\in I\}$ with $\hbox{Im}
\; a \leq 0$.
 This remark will be needed in Lemma \equ(partnested).
The proof of the estimate \equ(Kest) is trivial with the choice
of $\eta: = t^{-1}$. In the following proof of \equ(Kestcomb), we shall take $
I = \{0, \cdots, n\}$ for simplicity.

We claim the
following two estimates hold for $K$ with $n: = |I|-1$;
$$ 
        |K(t, \fp, I)| \leq {t^{ n}\over n!}, 
\Eq(Kestfac)
$$
$$
        \int |K(t, \fp, I)|^{2-a} d\mu(\fp_I) \leq
        \Big(C_a t^{1-a}\Big)^{ n } \qquad  0< a<1, \,\,\,  n\ge 1 \; .
\Eq(Kestnorm)
$$
The bound \equ(Kestcomb) follows from these two inequalities:
$$
        \int |K(t, \fp, I)|^{2} d\mu(\fp_I) \leq
\( {t^{ n }\over n!} \)^a
\int |K(t, \fp, I)|^{2-a} d\mu(\fp_I)  \le 
{\Big(C_a t\Big)^{ n }
        \over \Big[n!\Big]^a} \; .
$$

The proof of \equ(Kestfac) is trivial from the original definition 
of $K$ \equ(Kkernel) with multiple time integrals. The denominator 
$n!$ comes from 
the restriction $\sum_j s_j = t$ in the time integration. 
We now prove  \equ(Kestnorm). 
{F}rom the indentity of $K$   \equ(Kid)
$$
        \int |K(t, \fp, I)|^{2-a} d\mu(\fp_I) \le
        C \int d\mu(\fp_I) \Bigg(\int_{-\infty}^\infty
        d\a \prod_{j\in I} {1\over |\a -p_j^2/2 +it^{-1}|}\Bigg)^{2-a} \; .
\Eq(K1)
$$
Denote by 
$$
G (\a, p_0, p_1, \cdots, p_{n-2}) :=  \prod_{j\in I\setminus \{ n-1, n\}}
 {1\over |\a -p_j^2/2 +it^{-1}|} \; .
\Eq(G)
$$
For any fixed $p_{n-1}, p_n$ denote by $\nu$ the probability measure 
$$
    \nu ( d \a):=  R^{-1} (\; \a, \; p_{n-1}\, , \; p_n \; ) \; d \a  \; ,
\Eq(nu)
$$
where 
$$
 R (\; \a, \; p_{n-1}\, , \; p_n \; ) := Z (p_{n-1}, p_n) |\a -A + it^{-1}|\; 
|\a -B + it^{-1}|\; , \quad A:= p_{n-1}^2/2, B:= p_n^2/2
\Eq(R)
$$
and 
$$
	Z (p_{n-1}, p_n) := \int_{-\infty}^\infty 
{d\alpha\over |\a - A + i\eta||\a - B +i\eta|}\; , \quad \eta= t^{-1}\; .
\Eq(Z)
$$
We have the simple estimate for $Z$: 
$$
Z (p_{n-1}, p_n) \leq {C\over |A- B + i\eta|}\Bigg[ 1+ \log_+ \Big|
 {A-B\over \eta}\Big|\Bigg] \; .
\Eq(twodenom)
$$

Rewrite the $d\alpha$  integration in \equ(K1) as
$$
\int_{-\infty}^\infty
        d\a \prod_{j\in I} {1\over |\a -p_j^2/2 +it^{-1}|}
= Z (p_{n-1}, p_n)\int_{-\infty}^\infty  \nu(d\a) 
  G (\a, p_0, \cdots, p_{n-2}) \; .
\Eq(K6)
$$
{F}rom the H\"older inequality,  
$$\eqalign{ 
        \int |K(t, \fp, I)|^{2-a} & d\mu(\fp_I) = 
 \int d\mu(\fp_I)  \[ Z (p_{n-1}, p_n) \int_{-\infty}^\infty  \nu(d\a)  
 G (\a, p_0, \cdots, p_{n-2}) \]^{2-a} \cr & \le 
        \int d\mu(\fp_I)  Z (p_{n-1}, p_n)^{2-a}
  \int_{-\infty}^\infty  \nu(d\a)  
G (\a, p_0, \cdots, p_{n-2})^{2-a} \; .
}
\Eq(K7)
$$
Fix $p_{n-1}$ and $p_n$. Using \equ(normiteration),  the $p_0 \cdots, p_{n-2}$ 
integrations are explicitly estimated by  
$$\eqalign{ 
& \int d\mu(\fp_I)  Z (p_{n-1}, p_n)^{2-a}  G (\a, p_0, \cdots, 
p_{n-2})^{2-a} \cr
& \leq \int {dp_{n-1}dp_n Z (p_{n-1}, p_n)^{2-a}|\Phi(p_n)|^2
         \over 
<p_{n-1}>^{2d}<p_n>^{2d}   } 
 \int  {dp_{n-2}  \over|\a -p_{n-2}^2/2 + it^{-1}|^{2- a}
	<p_{n-2}-p_{n-1}>^{2d}}
  \cr
&   \times     \int {dp_{n-3}\over |\a - p_{n-3}^2/2 + it^{-1}|^{2-a}
      <p_{n-3}-p_{n-2}>^{2d}} \times \cdots \times
\int {dp_0 \over |\a - p_0^2/2 + it^{-1}|^{2-a}
        <p_0-p_1>^{2d}}\; .
}
$$
The following  inequality can be checked easily : 
$$
        \sup_\a\sup_{p_j} \int {dp_{j-1} \over
        |\a - p_{j-1}^2/2 + it^{-1}|^{2-a}
        <p_{j-1}-p_j>^{2d}} \leq C_a t^{1-a} \; .
\Eq(sup)
$$
{F}rom this inequality, 
the integrations of $p_0 \cdots p_{n-2}$ (in this order) 
can be estimated by 
a constant times $t^{1-a}$.
Hence  from \equ(K7)
$$
        \int |K(t, \fp, I)|^{2-a} d\mu(\fp_I) 
        \leq \Big(C_a t^{1-a}\Big)^{ n -1}
 \int  {|\Phi(p_n)|^2 \; dp_{n-1}dp_n
         \over 
<p_{n-1}>^{2d}<p_n>^{2d}   }  Z (p_{n-1}, p_n)^{2-a} \int_{-\infty}^\infty
 \nu(d\a) \; .
\Eq(K2)
$$
Since $\nu$ is a probability measure, $ \int \nu(d\a) =1$.
By the $Z$ estimate \equ(twodenom), 
$$\eqalign{ 
 \int |K(t, & \fp, I)|^{2-a}  d\mu(\fp_I) 
       \cr  &  \leq \Big(C_a t^{1-a}\Big)^{ n -1} 	   
	\int dp_{n-1}dp_n {|\Phi(p_n)|^2 \Big( 1 + \log_+ 
	\Big| 2t  \(p_{n-1}^2-p_n^2 \) \Big|\Big)^{2-a}
         \over <p_{n-1}>^{2d}<p_n>^{2d}
	 |(p_{n-1}^2- p^2_n)/2 + it^{-1}|^{2-a}} \; .
}
\Eq(2.7)
$$
We can now integrate $dp_{n-1}$:
$$
\sup_{p_n} \int dp_{n-1} { \Big( 1 + \log_+ 
	\Big| 2t  \(p_{n-1}^2-p_n^2 \) \Big|\Big)^{2-a}
         \over <p_{n-1}>^{2d}<p_n>^{2d}
	 |(p_{n-1}^2- p^2_n)/2 + it^{-1}|^{2-a}} \le C_a t^{1-a} \; .
$$ 
Since $\Phi \in L^2$, 
the last integral in \equ(2.7) is bounded by $C_a t^{1-a}$.
 We have thus 
proved  \equ(Kestnorm). 
\qed\enddemo

The proof of \equ(Kestnorm) has the following simple extension.
Denote 
$ \bk= (k_0, k_1, \ldots k_m)$ and  $\|\bk\| = \sum k_j$.
Define $Q^*$ (the usage of the star in the notation 
will be clear later on)
$$
        Q^*(t;   \fp;  {\bk}): =
        K\Big( t; \underbrace{p_0, p_0,\ldots p_0}_{k_0+1},
       \ldots,\underbrace{p_{m}, p_{m}\ldots, p_{m}}_{k_{m}+1}\Big) \; .
       \Eq(multKstardef)
$$
Then $| Q^*(t;   \fp;  {\bk})|\leq C\Sigma (t, \fp,  \bk)$ with
$$
	\Sigma (t, \fp,  \bk):=
	\int_{-\infty}^\infty d\a 
	 \prod_{j=0}^m \Bigg| { 1\over \alpha  -
        p_j^2/2 + i t^{-1}} \Bigg|^{k_j+1} \; .
\Eq(Sigma)
$$

\proclaim{\Lemma (Sigmaestlem)}
For any $m\ge 1$, $0\leq a < 1$, we have 
$$
        \int  |Q^*(t; \fp, \bk)|^2 d\mu(\fp_m)
        \leq { (C_a t)^{m + 2\|\bk\|}\over \Big[(m+ \|\bk\|)!\Big]^a}
\Eq(Qestcomb)
$$
and 
$$
	\int  \Big(\Sigma (t, \fp,  \bk)\Big)^{2-a} d\mu(\fp_m)
	\leq \Big( C_a t\Big)^{(1-a)m+ (2-a)\|\bk\|}\; .
\Eq(Sigmaest)
$$
\endproclaim

\medskip

\demo{Proof of Lemma \equ(Sigmaestlem)}
Similar to the proof of \equ(Kestcomb), \equ(Qestcomb)
follows from \equ(Sigmaest) and 
the simple bound
$$
	|Q^*(t; \fp, \bk)|\leq {t^{m+\|\bk\|}\over (m+\|\bk\|)!} \; .
$$
The proof of \equ(Sigmaest) is also similar to 
that of \equ(Kestnorm). Define  $ G = G_1 G_2 $ 
where 
$$
G_1=  \big | \alpha  -
        p_{m-1}^2/2 + i t^{-1} \big|^{-k_{m-1}}
 \; \big|  \alpha  -
        p_m^2/2 + i t^{-1} \big|^{-k_m}\; ,  \quad 
G_2=  \prod_{j=0}^{m-2}  
\bigg|  \alpha  -
        p_j^2/2 + i t^{-1} \bigg|^{-(k_j+1)} \; .
$$
We shall use the measure $\nu$, $R(\alpha, p_{m-1}, p_m)$ and $Z(p_{m-1}, p_m)$
 as in the proof of the 
previous lemma. Then we can rewrite the left hand side of 
\equ(Sigmaest) as $ \int d\mu(\fp_m) \Big[ \int  \nu(d\a) \;  Z \; G
\Big]^{2-a} $.  We then use the 
H\"older inequality to arrive at 
$$\eqalign{ 
      &   \int |\Sigma (t, \fp, \bk)|^{2-a} d\mu(\fp_m) \cr
 & \le \int d\mu(\fp_m)  Z (p_{m-1}, p_m)^{2-a}
  \int_{-\infty}^\infty  \nu(d\a)  
G_1(\a, p_{m-1}, p_m)^{2-a} G_2 (\a, p_0, \cdots, p_{m-2})^{2-a} \; .
}
$$
The  integrations of $p_0 \cdots p_{m-2}$  in 
$\int d\mu(\fp_m)   G_2^{2-a}$ can be estimated the same way 
if we use
$$
        \sup_\a\sup_{p_j} \int {dp_{j-1} \over
        |\a - p_{j-1}^2/2 + it^{-1}|^{(k+1)(2-a)}
        <p_{j-1}-p_j>^{2d}} \leq C_a t^{(1-a) +k(2-a) }
\Eq(sup)
$$
We have thus proved that (cf. \equ(K2)):
$$\eqalign{ 
         \int d\mu(\fp_m)&  \Big(\Sigma (t, \fp,  \bk)\Big)^{2-a}
   \leq C_a  t^{(m-1)(1-a) + \|k \|(2-a) -k_{m-1}(2-a)-k_m(2-a) }  \cr
&      \times \int  {dp_{m-1}dp_m |\Phi(p_m)|^2 Z (p_{m-1}, p_m)  ^{2-a}
         \over 
<p_{m-1}>^{2d}<p_m>^{2d}   } \int \nu(d\a) G_1(\a, p_{m-1}, p_m)^{2-a} 
}
$$
{F}rom the definition of $\nu$ and the $Z$ estimate \equ(twodenom), 
the last integration is bounded by
$$\eqalign{ 
\le  & \int d\a	   \int dp_{m-1} dp_m {|\Phi(p_m)|^2\Big( 1 + \log_+ 
	\Big|   2 t \(p_{m-1}^2-p_m^2\)\Big|\Big)^{1-a}
         \over <p_{m-1}>^{2d}<p_m>^{2d}
	 |(p_{m-1}^2- p^2_m)/2 + it^{-1}|^{1-a}}\cr
& \times
\big| \alpha  -
        p_{m-1}^2/2 + i t^{-1} \big|^{-k_{m-1}(2-a)-1}
 \; \big|  \alpha  -
        p_m^2/2 + i t^{-1} \big|^{-k_m(2-a)-1}
}
$$
We now perform the $p_{m-1}$ integration. It is easy to check that 
it is bounded by $C_a t^{(1-a) +k_{m-1}(2-a) }$  uniformly for 
any values of $\a$ and $p_m$. Then we perform the $\alpha$ integration,
which is bounded by $C_a t^{k_m(2-a) }$. If $k_m=0$, we have to
borrow an extra decay in $\alpha$ from the $p_{m-1}$ cutoff to
make the $\a$ integration finite.
Finally, we use $\Phi\in L^2$ for the $p_m$ integration.
Putting all these bounds
together, we have proved the Lemma. 
\qed\enddemo

\bigskip

\subheading{ 3.2  $L^2$ Estimates}
\bigskip

The key estimate is the following Lemma. 

\proclaim{\Lemma (main1)} Recall the definition of $\psi_n$ in \equ(psiexp).
We have for any $n$ 
$$
       \bE \| \psi_{n}(t)\|^2 \leq (C\lambda^2 t)^n { 1 \over (n!)^{1/2}}
        \Bigl[ 1 + {(n!)^{3/2}(\log t)^{n+3} \over t^{1/2}}\Bigr] \; .
$$
Recall that $\lambda = \sqrt{\e}$, $t=T/\e$ and $T$ is fixed.
Let $n_0=n_0(\e)$ be given in \equ(n0) with some  $0<\gamma\ll 1$.
 It follows (cf.\equ(n0p)) that 
$$
 \lim_{M \to \infty} \lim_{\e \to 0} \bE \; \Big\| \sum_{n=M}^{n_0(\e)-1}
        \psi_n(T\e^{-1}) \; \Big\|^2 = 0 \; .
$$
\endproclaim

\demo{Proof of Lemma \equ(main1)} By definition of $\psi_n$ and 
$C_\pi:=C_\pi^{\delta(\xi)}$ defined in \equ(cpig), we have 
$$
        \bE \| \psi_{n}(t)\|^2 = U_{n,n} 
= \sum_{\pi \in \Pi_{n, n}} C_\pi \; .
\Eq(cpi)
$$ 
We shall estimate the contributions
from crossing, nested and simple pairings
separately. The proof of Lemma \equ(main1) follows 
from Lemmas \equ(cross),
\equ(nest) and \equ(simpleest). \qed \enddemo

\medskip
First we prove a weaker estimate:

\proclaim{\Lemma (apriori)}
For any $\pi\in \Pi_{n,n}$
$$
        |C_\pi| \leq (C\lambda^2 t)^n (\log t)^{n+2} \; .
$$
It follows immediately that
$$
        \bE \| \psi_n (t)\|^2 \leq n! (C\lambda^2t)^n (\log t)^{n+2} \; .
$$
\endproclaim

\demo{Proof of Lemma \equ(apriori)}
Since the number of pairings is $C^nn!$, the second bound follows
from the first one and \equ(cpi).
Recall the convention of relabelling $p$ in \equ(w). 
We can now use 
the definition of $C_\pi$ \equ(cpig)  to have 
$$
        |C_\pi|
        \leq \lambda^{2n}   \int d \fw_{2n+1} |K(t,\fp,n)||K(t,
        \fp',n)|\Delta_\pi^0( \fp, \fp') |F(\fp,n)||F( \fp',n)|\; .
$$
Define $\a_\ell := \a$ if $  \ell\le n$ and $\a_\ell:= \beta$ otherwise.
{F}rom the equation \equ(AB), we can solve the momentum 
$w_j$ for $j \in B$ as a function of $\fw_A$.
Together with the estimate of $K$ in \equ(Kest) we obtain
$$
      |C_\pi|
        \leq C \lambda^{2n}\int d\fw_A
        \int_{-\infty}^\infty d \a   d \beta     
	\prod_{0\le k \le 2n}
        { 1 \over | \a_k -w_k^2/2 + i /t | } |F(\fp,n)||F(\fp',n)| \; .
$$

Recall that the elements of $A$ are denoted by
$a_1=0<a_2 <  \ldots < a_{\bar n +1}$ and here $\bar n = n$.
{F}rom the definition of $F$ in \equ(F),
 \equ(normiteration) and \equ(middlesup), 
we have
$$
        |F(\fp,n)||F(\fp',n)|\leq {C^n |\Phi(w_0)||\Phi(w_{2\bar n+1})|
	\over <w_0>^{2d}
        <w_{k_1}>^{2d}<w_{k_2}>^{2d}}
        \prod_{\ell=2}^{n+1} {1\over <w_{a_\ell}
        - w_{a_{\ell-1}}>^{2d}}
$$
for any fixed $k_1 \not = k_2$ and $k_1, k_2 > 0$. 
Recall that $p_n=w_0$, $p_n'= w_{2\bar n+1}$,
$0\in A$, $2\bar n+1\in B$. Recall that $\Delta^0_\pi$ always
forces  the delta 
function $\delta(w_0-w_{2\bar n+1})$. Thus  we obtain for any $k > 0$, 
$$\eqalign{ 
    |C_\pi| \le & (C\lambda^2)^n
        \int_{-\infty}^\infty d \a d \beta  \int { |\Phi(w_0)|^2
	d\fw_A \over 
      <w_0>^{2d} <w_{k_1}>^{2d} <w_{k_2}>^{2d}}{1\over
        | \a-w_0^2/2 + i /t | | \beta-w_0^2/2 + i /t | } \cr
&        \times  \prod_{\ell = 2}^{n+1}
        { 1 \over | \a_{a_\ell} -w_{a_\ell}^2/2 + i /t |}
	\prod_{\ell = 2 }^{n+1}{1\over
        <w_{a_\ell} - w_{a_{\ell-1}}>^{2d}}
              \prod_{j \in B \atop j\not = 2\bar n+1}
        { 1 \over | \a_j -w_j^2/2 + i/t | } \; .
}
\Eq(sep)
$$

We  choose $k_1, k_2$ such that $\a_{k_1} = \a$ and $\a_{k_2} = \beta$.
 It is easy to check that  
$$
        \sup_{w}
        {1\over |\a_j - w^2/2+ it^{-1}| <w>^{2d}}
        \leq {Ct\over <\a_j>} \; .
\Eq(3.1)
$$
We use this estimate for $w=w_{k_2}$.
If we remove the  factor $<w>^{2d}$  in the denominator, we
only have to drop the $ <\a_j>$ in the denominator. Using this 
bound, we can estimate all the other factors depending on the momenta from $B$ 
in the expression  \equ(sep). We now estimate 
integrations of momenta in $A$. We can check the simple inequality 
$$
        \sup_{w_{a_{\ell-1}}} \int { dw_{a_\ell} 
        \over  <w_{a_\ell} >^{2d}}
        { 1 \over | \a_{a_\ell} -w_{a_\ell}^2/2 + i /t |
        <w_{a_\ell} - w_{a_{\ell -1}}>^{2d}} \leq  { C\log t \over 
< \a_{a_\ell} >} \; .
\Eq(3.2)
$$
Again, if we remove the denominator $<w_{a_\ell} >^{2d}$, 
the denominator $<\a_{a_\ell}>$ on the right hand side
has to be removed.  
We can now apply the last estimate to integration for $A$-momenta except $w_0$.
Hence   
$$
         |C_\pi| 
 \le (C\lambda^2t)^n (\log t)^n  \int  { |\Phi(w_0)|^2 dw_0 \over <w_0>^{2d}}
\int_{-\infty}^\infty 
         { d \a d \beta \over
        <\a> <\beta>   | \a-w_0^2/2 + i /t |
         | \beta-w_0^2/2 + i /t | } \; .
$$
The denominator $<\a> <\beta> $ comes from the choice of $k_1, k_2$ 
and the denominators on the right hand side of  \equ(3.1), \equ(3.2). 
Using 
$$
	\sup_w\int_{-\infty}^\infty {d\a \over <\a>|\a - w^2/2 + i/t|}
	\leq C\log t
\Eq(alphaint)
$$
we can bound the last integration above by $(C\log t)^2$.
Finally, by $\Phi\in L^2$, we obtain
$$
    |C_\pi| 
 \le (C\lambda^2t)^n (\log t)^{n+2}   \; . 
$$
\qed
\enddemo

The following remark will be useful later.
Suppose we consider only the product $\widetilde \Delta^0_\pi$
of a subset of delta functions from  $\Delta_\pi^0$ with rank $r$
according to the setup of part (II) of Lemma \equ(rank).
Then the method of the previous proof gives
$$
        \int d\fw_{2n+1} |K(t, \fp,n)||K(t, \fp',n)|
	\widetilde\Delta_\pi^0 ( \fp,\fp')
        G( \fp, n ;  \fp', n)|\widehat \psi_0(p_n)||\widehat\psi_0(p_n')|
	\leq (Ct)^{r-1} (\log t)^{2n-r+3}
$$
for any function $G$ with
$$
	 G( \fp, n ;  \fp', n) \leq \prod_{i=0}^{2n} {1\over 
	<w_i-w_{i+1}>^{20d}}\; .
$$
Moreover, the numbers of momenta within the
 two $K$-kernels do not have to be
the same. In full generality, the same proof as above gives
$$\eqalign{ 
         & \int d\fp_n d\fp_{n'}' |K(t, \fp, n)||K(t,  \fp', n')|
        \widetilde\Delta_\pi^0 ( \fp_n,\fp_{n'}')
        G( \fp, n ;\fp', n')|\widehat \psi_0(p_n)|
      |\widehat\psi_0(p_{n'}')| \cr
& 	\leq (Ct)^{r-1} (\log t)^{n+n'-r+3}
}
\Eq(fullgen)
$$
provided that  $\widetilde \Delta_\pi^0$
 can be written in the form of \equ(AB) with $|B|=r$ and 
$\delta(p_n-p_{n'}')$ is forced by $\widetilde \Delta_\pi^0$.

\bigskip

\subheading{ 3.3. Estimate of crossing pairings}

\bigskip

\proclaim{\Lemma (cross)}
If $\pi\in \Pi_{n,n}$ is crossing, then we have
$$
        |C_\pi|\leq t^{-1} (C\lambda^2t)^n  (\log t)^{n+3} \; .
\Eq(cross)
$$
Notice the extra factor $t^{-1}$ compared with 
the previous Lemma. 
\endproclaim

\demo{Proof of Lemma \equ(cross)}
First we present the proof for $d\ge3$. The case $d=2$ requires some technical
modicification which is discussed in Section 6.
Since $\pi$ is crossing, for some $j_0\in B$ with
 $j_0\not = 2\bar n+1$ we have
$$
        w_{j_0} = \pm w_{\ell_1} \pm w_{\ell_2} \pm \ldots \; ,
$$
where $\ell_1, \ell_2,\ldots \in A$ according to Lemma 
\equ(crossmatrix). Notice that Lemma \equ(rank) implies that
there are at least three momenta on the right hand side, 
hence we can assume that $\ell_1, \ell_2 \not = 0$. 

Similarly to \equ(sep) we have 
(recall that  the
delta function $\delta(w_0-w_{2\bar n+1})$ is forced)
$$\eqalign{ 
    |C_\pi| \le \,\,\,  &  \lambda^{2n}  \int_{-\infty}^\infty
         d \a d \beta  \int  d\mu(\fw_A)
        { 1 \over | \a-w_0^2/2 + i /t | | \beta-w_0^2/2 + i /t | } \cr
&         \times \[ \prod_{\ell \in A \atop \ell\not =0}
         { 1 \over | \a_\ell -w_\ell^2/2 + i /t | } \] \; 
         \bigg |  { 1 \over  \a_{j_0} - w_{j_0}^2/2 + i /t  } \bigg |_
       {w_{j_0} =(\pm w_{\ell_1} \pm w_{\ell_2} \pm \ldots)} \; \; Y_B \; ,
}
\Eq(sep1)
$$
where
$$
Y_B=          \prod_{j \in B\atop  j\not = j_0, 2\bar n+1}
        { 1 \over | \a_j -w_j^2/2 + i /t | } \; 
$$
and
$$
	d\mu(\fw_A) := {|\Phi(w_0)|^2d\fw_A\over <w_0>^{20d}}
	\prod_{j=1}^{n-1} {1\over <w_{a_j}-w_{a_{j+1}}>^{20d}}
$$
(notice that the argument of $\Phi$ is $w_0$ and not $w_n$).
We can bound $ | \a_j -w_j^2/2 + i /t |^{-1}$ by $t^{-1}$. Thus 
the  last factor $Y_B $ is bounded by  $t^{n-1}$. 
One can easily check the inequality for $d\ge 3$
$$\eqalign{ 
        \sup_{\g_i = \a,\beta}\sup_{ v }
        \int dw_{1}\int dw_{2} & <w_1>^{-2d} <w_2>^{-2d} 
        \Big| \g_{0} - \Big( \pm w_{1} 
\pm w_{2} + v \Big)^2/2 + i /t \Big|^{-1} \cr
&        \times{ 1 \over | \g_{1} - w_{1}^2/2 + i /t | }
        { 1 \over | \g_{2} - w_{2}^2/2 + i /t | }
         \le C(\log t)^3 \; .
}
\Eq(3denom)
$$
We can use this inequality to estimate the $w_{\ell_1}$
and $w_{\ell_2}$ integrations.  After this, the
integrations of $\a$ and $\beta$  and the other momenta   in \equ(sep1) 
are now  similar to the proof of the previous Lemma.
First we integrate all momenta but $w_0$. 
We gain an extra factor
$<\a>^{-1}<\beta >^{-1}$ (cf. \equ(3.1), \equ(3.2)), and we obtain
$$
 \eqalign{ |C_\pi| 
 	\le \, \, & (C\lambda^2t)^n t^{-1}(\log t)^{n+1}  
	\int  { |\Phi(w_0)|^2 dw_0 \over <w_0>^{2d}} \cr
	&\times \int_{-\infty}^\infty 
         { d \a d \beta \over
        <\a> <\beta>   | \a-w_0^2/2 + i /t |
         | \beta-w_0^2/2 + i /t | } \; .
}
\Eq(crossextra)
$$
Then we integrate $\a, \beta$ using \equ(alphaint).
The last integration is $dw_0$ and we have to use  $\Phi\in L^2$.
Each  integration gives at most a factor $\log t$.
The details are left to the reader.
Hence  \equ(sep1) is bounded by
$$
        |C_\pi|\le (C\lambda^2t)^nt^{-1} (\log t)^{n+3} \; ,
$$
which completes the proof for $d\ge 3$.
\qed
\enddemo

\bigskip

\subheading{ 3.4. Estimate of nested pairings}

\bigskip

\proclaim{\Lemma (nest)}
If $\pi \in \Pi_{n,n}$ is nested then
$$
        |C_\pi|\leq (C\lambda^2t)^nt^{-1}(\log t)^{n+2} \; .
\Eq(nest)
$$
\endproclaim

\demo{Proof of Lemma \equ(nest)}
A nested pairing (Fig. 2) contains two delta functions of
the type $\delta( -p_i + p_{i+1} - p_j + p_{j+1})$
and $\delta (-p_k + p_{k+1} - p_u + p_{u+1})$ with
$i < k < u < j<n$ (or the same type of delta functions
with all primed variables.  Without loss of generality we
can assume that a  nested pairing occurs in the nonprimed variables).
 Recall that this pair of delta functions
is called a {\it nest} with an {\it outer} and an {\it inner} delta
function (the first one is called outer, the second one is inner).

A nested pairing can have several inner delta function. Consider a 
nest which is minimal in the sense that none of its inner
delta function is an outer delta function in some other nest.
Since the pairing has no crossing,
all the inner delta functions in this nest are joining two
neighboring bullets. In other words, for all inner delta
function $\delta (-p_k + p_{k+1} - p_u + p_{u+1})$ we have $u= k+1$ (as in
Fig. 2)
and hence $p_k=p_{k+2}$ is forced.
Let $p_{c_1}, p_{c_2},\ldots p_{c_v}$ be those momenta with
$i+1=c_1< c_2 < \ldots <c_v = j$ which are forced to be
the same as $p_k=p_{k+2}$, i.e.
$$
        p_{c_1} = p_{c_2} =\ldots =p_k = p_{k+2} = \ldots
        =p_{c_v} 
\Eq(iddelta)
$$
(in particular $k$ and $k+2$ are among the $c_1, c_2,\ldots c_v$,
i.e. $v\ge 2$).
By Lemma \equ(forceid), these are those momentum lines in the
graph among $p_{i+1}, p_{i+2}, \ldots p_j$ 
 such that if we remove any of them along with $p_k$,
the graph becomes disconnected (and we include $p_k$ as well).
On Fig. 2 $v=4$, $c_1= i+1$, $c_2 = k$, $c_3 = k+2$, $c_4= j$.

It is obvious that $\Delta_\pi^0$ does not depend on $p_{k+1}$.
Since the only delta function  that contains $p_{k+1}$
is
$\delta(-p_k + p_{k+1} - p_{k+1}+p_{k+2})$.
We also claim that
using the notations \equ(iddelta)
$$
        \Delta_\pi^0 (\fw ) =
        \widetilde\Delta_\pi^0(\fp^*,\fp')
	\times \prod_{a=1}^{v-1}\delta(p_{c_a}-p_{c_{a+1}})
\Eq(deltadecomp)
$$
where $\fp^*: = \fp_{I^*}$ with
$$
        I^*:= I_n \setminus \{   c_1, c_2, \ldots c_v\} 
$$
and $\widetilde\Delta_\pi^0(\fp^*,\fp')$ is a product of
delta functions depending only on $\fp^*, \fp'$.
The point is that the momenta $\{ p_{c_1}, p_{c_2},\ldots \}$
 occur only in the identifying
delta functions as explicitly written in \equ(deltadecomp).
 The decomposition \equ(deltadecomp) is obvious from the graph
using  Lemma \equ(forceid). $\, [ \, $  The formal proof goes as follows.
 Since
$\pi$ is noncrossing, we know  from Lemma \equ(crossmatrix)
that $\Delta_\pi^0$ has a representation
where all delta functions are identifications.
Consider this representation.
If a momentum $p_{c_a}$
($1\leq a \leq v$) appears in a delta function, then it has the form
$\delta  (p_{c_a} - w_r)$ or $\delta (w_r-p_{c_a})$
for some $r$.
We claim that $r=c_b$ with some $1\leq b \leq v$, i.e.
$p_{c_a}$ appears only in the product of delta functions
identifying $p_{c_1}=p_{c_2} =\ldots = p_{c_v}$.
Clearly $i< r \leq j$, otherwise  the outer delta function
would prevent the graph becoming disconnected if we remove
the edge $e(w_r)$ and $e(p_{c_a})$. But then  $r$ must be one
of the $c_b$ as $\Delta^0_\pi$ does not depend on
other momenta with index between $i$ and $j$ by the minimality
of the nest. $\, ]$

Remark that $\widetilde \Delta^0_\pi(\fp^*, \fp')$ forces
$p_n=p_n'$. For, \equ(deltadecomp) decomposes $\Delta^0_\pi(\fw)$
into two factors having no common variables. Since $p_n, p_n'$
belong to the arguments of $\widetilde \Delta^0_\pi(\fp^*, \fp')$,
and $\Delta^0_\pi(\fw)$ forces $p_n=p_n'$, hence 
$\widetilde \Delta^0_\pi(\fp^*, \fp')$ does so, too.

We also need the following identity
$$
        K\Big(t, \fp_{I}, \underbrace{q,q, \ldots q}_{v \,\, times}\Big)
        =(-i)^v \int_0^t ds \,{s^{v-1}\over (v-1)!} e^{-is q^2/2}
        K(t-s, \fp, I) \; .
\Eq(iterKid)
$$
Recall the definition 
$$
        C_\pi = \lambda^{2n}\int d\fp_n d\fp_n'
        \overline{  K }(t, \fp, n)
        K(t, \fp', n)\Delta_\pi^0(\fp, \fp')
        \overline{F( \fp,n)}F(\fp',n) \; .
$$
Define the function 
$$
        E(\fp^*, \fp_n', p_k):= 
        \[ \overline{F( \fp,n)}F(\fp',n)\]_{p_{c_a} =       p_{k}, 
a=1 \cdots, v } \; , 
\Eq(Edef)
$$
which inherits the smoothness in $p_k$  and   decay 
  from $F$. 
Together with the identity \equ(iterKid) and 
the notation of $\fp^*$, we have 
$$
     C_\pi = \lambda^{2n}  \int d\fp^*\, dp_k
        \, d\fp_n' \int_0^t ds
        {s^{v-1} i^v\over (v-1)!}e^{is p_k^2/2}
	\overline{K(t-s, \fp, I^*)}
        K(t; \fp', n)\widetilde\Delta_\pi^0(\fp^*, \fp')
        E(\fp^*,\fp_n', p_k) \; .
$$

We can integrate $p_k$ by parts. We first rewrite 
the $d$-dimensional integration as 
$$
        \int s^{v-1}e^{is p_k^2/2} E(\fp^*, \fp_n', p_k)
        dp_k 
	= (2 \pi)^{-d/2} \int_0^\infty  s^{v-1}e^{is Q /2}  
      \( \int_{p_k^2=Q}  E(\fp^*, \fp_n', p_k ) d\omega \) 
	  Q^{{d\over 2}-1}dQ \; ,
$$
where $d\omega$ is the normalized 
surface measure on the sphere $p_k^2=Q$ in the
$p_k$ space. Define 
$$
        H(\fp^*,\fp_n', p_k): = 2 i Q^{-{d\over 2}+1}
	 {\partial\over \partial Q}
              \Bigg( Q^{{d\over 2}-1}
	 \int_{p_k^2=Q} E(\fp^*, \fp_n', p_k ) d\omega
        \Bigg)_{Q:= p_k^2}
$$
$$
= i(d-2)   p_k^{-2} 
        \int E(\fp^*, \fp_n', p_k ) d\omega
      + 2 i  {\partial\over \partial Q}
        \Bigg(  \int_{p_k^2=Q} E(\fp^*, \fp_n', p_k ) d\omega
        \Bigg)_{Q:= p_k^2} \; .
\Eq(Hdef)
$$
We can integrate $Q$ by parts to have for $v\ge 2$,
$$
        \int s^{v-1}e^{is p_k^2/2} E(\fp^*, \fp_n', p_k)
        dp_k = \int s^{v-2}e^{is p_k^2/2} H(\fp^*, \fp_n', p_k) 
	dp_k \; .
$$
Notice that $H$ inherits the decay from $F$. 
There is a singularity at $p_k =0$, 
integrable for  dimension $d\ge3$. Notice this 
singular term disappears in dimension $d = 2$.

We can now rewrite $C_\pi$ as 
$$
        C_\pi={\lambda^{2n} i^v  \over v-1}
         \int d\fp^*\, dp_k\, d\fp_n'  \int_0^tds
        {s^{v-2}\over (v-2)!}e^{is p_k^2/2} 
	\overline{K(t-s, \fp, I^*)}
        K(t, \fp', n)\widetilde \Delta_\pi^0(\fp^*, \fp')
        H(\fp^*,\fp_n', p_k) \; .
\Eq(cpirew)
$$
Using the  identity \equ(iterKid) to integrate $s$, we have
$$
        C_\pi= {i\lambda^{2n} \over v-1}
          \int d\fp^* \, dp_k \, d\fp_n' \Bigg(\prod_{a=1}^{v-2} dq_a \Bigg)
        \overline{K\Big(t,  \fp^*, p_k, q_1, q_2,\ldots q_{v-2} \Big)}
$$
$$
        \times K(t,  \fp', n)
        \widetilde\Delta_\pi^0(\fp^*, \fp') 
	\Big( \prod_{a=1}^{v-2}\delta (p_k - q_a)\Big)
        H(\fp^*, \fp_n', p_k)\; .
\Eq(repeat)
$$
where we have written  
$$
\overline{K\Big(t,  \fp^*, 
\underbrace{p_k, p_k, \ldots p_k}_{v-1 \,\, times} \Big)} 
=
\Bigg( \int \prod_{a=1}^{v-2} dq_a  \Bigg)
\overline{K\Big(t,  \fp^*, p_k, q_1, q_2,\ldots q_{v-2} \Big)}
\Big( \prod_{a=1}^{v-2}\delta (p_k - q_a)\Big)  \; .
$$
Notice that we now have momenta  $\fp^*, p_k, 
q_1, q_2,\ldots q_{v-2} $ and $\fp_n'$. The number of momenta 
$\fp_n'$ is $n+1$; that of  $\fp^*, p_k, 
q_1, q_2,\ldots q_{v-2}$  is  $n$. 
The new delta function in \equ(repeat) is
$$
        \widetilde\Delta_\pi^0(\fp^*, \fp') 
\Big( \prod_{a=1}^{v-2}\delta (p_k - q_a)\Big)
\Eq(newdelta)
$$
which has rank $n$ and it still forces $p_n=p_n'$.
The function $H$ depends on $\fp^*, p_k $ and $\fp_n'$. 
Define a function $H^*$ depending on  $\fp^*, p_k, 
q_1, q_2,\ldots q_{v-2} $ and $\fp_n'$ by 
$$
        H^*(\fp^*, p_k,  q_1, q_2, \ldots q_{v-2}, \fp_n')
        : = H(\fp^*, \fp_n', p_k) \prod_{a=1}^{v-2} 
         {1\over <p_k-q_a>^{20d}} \; .
$$
Here we have included the cutoff in large $q_a$. Due to the 
delta function $\Big( \prod_{a=1}^{v-2}\delta (p_k - q_a)\Big)$, 
we can replace $H$ in \equ(repeat) by
$H^*$ without changing its value. 
We are now in the same setup as in  Lemma \equ(apriori)
except that the rank of the delta functions \equ(newdelta)
is now $n$, reduced from $n+1$. 
We can apply the remark at the end of
the proof of Lemma \equ(apriori), \equ(fullgen), to have 
$$
        |C_\pi|\leq (C\lambda^2t)^nt^{-1}(\log t)^{n+2}\; .
$$
\qed
\enddemo

\bigskip
\subheading{ 3.5. Estimate of simple pairings}
\bigskip

\proclaim{\Lemma (simpleest)}
If $\pi \in \Pi_{n,n'}$ is simple, then
$$
        | C_\pi |\leq {(C\lambda^2t)^{\bar n}\over (\bar n!)^{1/2}} +
        O\Big( (C_a\lambda^2 t)^{\bar n} t^{-{a\over2}} \Big)
\Eq(simplepairest) 
$$
for any $0\leq a < 1$
(recall that $\bar n = (n+n')/2$).
\endproclaim

Before the proof, we need several notations.
The structure of the momenta in a  simple pairing 
can easily be described. We shall rename the variables
according to the pairing. 
Though for the $L^2$-norm we need to consider pairings $\pi\in\Pi_{n,n}$ only,
we set up the formulas for the more general case
$\pi\in\Pi_{n,n'}$  needed later on for  computing the Wigner
transform.

First notice that a simple pairing has the property that
all Type I delta functions are minimal, i.e., 
for any Type I delta function 
$\delta(-p_i + p_{i+1} - p_j + p_{j+1})$ in a simple pairing  
we have $j=i+1$; 
and similarly for Type I'.
Hence this delta function does not depend on $p_j\equiv p_{i+1}$.
Moreover, since each momentum appears exactly twice in the delta functions,
we see that $\Delta_\pi^0$ is independent of $p_j \equiv p_{i+1}$.

\proclaim{\Definition (external)}
In case of a simple pairing $\pi$, we call a momentum {\bf external}
if $\Delta_\pi^0$ depends on it. Otherwise it is called {\bf internal}.
We shall use the letter $r$ for external and $q$ for internal momenta.
\endproclaim

Let $\underline {\bk}$ be an array of double indices for the 
internal momenta
$$
      \underline {\bk}: = \Bigl( 01, 02, \ldots 0k_0, 11, 12, \ldots 1k_1,
        \ldots m k_m\Bigr)\; ,
$$ 
where $ \bk= (k_0, k_1, \ldots k_m)$ and $k_j\ge 0$ will
denote the number of the internal
momenta for  $j$. We shall denote $\|\bk\| = \sum k_j$
and $ | \bk|:= m$. These numbers will depend on the structure of 
the pairing $\pi$. Let $|\pi|: = m$. Define
$$
         \fr_{m}
        =  (r_{j })_{ 0\le  j \le m} , \quad \fr_{m, \widehat 0}
        =  (r_{j })_{ 1\le  j \le m}
$$
to be the external momenta.
Denote the internal momenta
by $\underline {\fq}=(q_{j \ell})_{ 0\le \ell < \infty,
0\le j < \infty }$ and denote
$$
        \underline{\fq}_{\bk}
        =(q_{j \ell})_{ 1\le \ell \le k_j,
        0\le j \le m } \; .
$$
The following
lemma about the structure of a simple pairing (see also Fig 3.)
will be used.

\proclaim{\Lemma (simplestr)}
If $\pi$ is simple, then it has $m= |\pi|$ Type II delta functions:
$\delta (-p_{i_j-1}+ p_{i_j} + p_{i'_j-1}' - p_{i'_j}')$ with $j=1, 2,\ldots m$
and $ i_0:= 0<i_1 <\ldots < i_m$, $i_0':= 0< i_1' < \ldots <i_m'$.
 For each $j=0, 1, \ldots , m$ there are
$k_j$  Type I delta functions in the form
$\delta (-p_u + p_{u+1} - p_{u+1}+ p_{u+2}) = \delta(-p_u+p_{u+2})$
with $i_j\leq u < i_{j+1}$.
We shall rename the momenta such that $r_j : = p_{i_j}$, and
 $q_{j\ell}:= p_{u+1}$ if $\delta (-p_u + p_{u+1} - p_{u+1}+ p_{u+2}) = 
\delta(-p_u+p_{u+2})$ is the $\ell^{th}$
Type I delta function with $i_j\leq u< i_{j+1}$. The notations are similar
for the Type I' delta functions. We claim that $\Delta_\pi^0$ is
equivalent to the product of the following delta functions
$$
       \delta( p_{i_j} - p_{i'_j}' ) \qquad j=0,1, \ldots , m \; ;
\Eq(typeII)
$$
$$
	\delta(p_{i_j} - p_{u+2})
\Eq(typeI)
$$
for each $u$ such that $q_{j\ell} = p_{u+1}$ for
$\ell = 1, 2, \ldots , k_j$, $j=0, 1, \ldots , m$
  (i.e. $p_u$ and $p_{u+2}$ are the
"external legs" belonging to the internal momentum $p_{u+1}$) ;
and
$$
       \delta( p_{i_j} - p_{u+2}')  
\Eq(typeI')
$$
for each $u$ such that $q_{j\ell}' = p_{u+1}'$,
$\ell = 1, 2, \ldots , k_j'$, $j=0, 1, \ldots , m$.
The total number of potentials is
$$
        2\bar n =n+n' =
        2\Big(|\bk|+ \|\bk\| + \|\bk'\| \Big)
$$
and we also have the relation 
$$
        n= m +2\|\bk\| , \qquad n'= m+2\|\bk'\| \; .
$$
The total number of delta functions in \equ(typeII), \equ(typeI), 
\equ(typeI') is $m+1 +\|\bk\| + \|\bk'\| = \bar n +1$.
\endproclaim

\demo{Proof of Lemma \equ(simplestr)}

For simplicity, we redraw the graph of a simple pairing
such  that the pairing lines corresponding to
 Type II deltafunctions are  straight lines inside $C$.
These lines do not cross each other.
An application of  Lemma \equ(forceid)
shows that the delta functions \equ(typeII), \equ(typeI) and \equ(typeI')
are forced by $\Delta^0_\pi$.  Each of these identifications contains at least
one momentum which does not appear elsewhere, hence the rank
of the product of these identifications is the same as the number of 
delta functions
in  \equ(typeII), \equ(typeI) and \equ(typeI').
Simple counting shows that these numbers are $m+1$, $\| \bk \|$
and $\| \bk' \|$, respectively, hence the rank 
 is $m+1+\|\bk\|+\|\bk'\| = \bar n +1$
which equals to the rank of $\Delta^0_\pi$ by Lemma \equ(rank).
 This shows that
$\Delta^0_\pi$ is equivalent to the product of
\equ(typeII), \equ(typeI) and \equ(typeI').
For example, the simple pairing on Fig. 3 has $m=3$, $\bk = (0,0,2, 0)$ ,
$\bk'= (0,1, 0,0)$, $n= 7$, $n' = 5$. 
\qed
\enddemo

\centerline{\epsffile{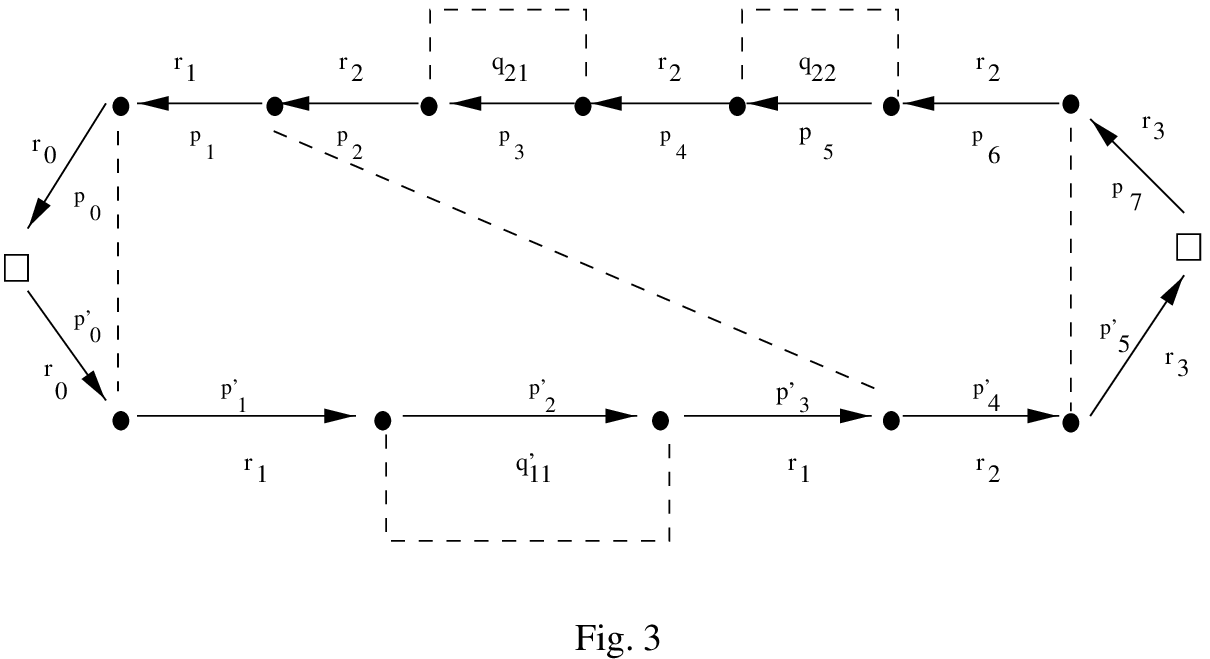}}

We now introduce some more notations associated with the new labeling.
Recall that for any set of momenta $\fp_I$
the  free evolution kernel is
$$
        K(t; \fp, I): = (-i)^{|I|-1}\int_0^t   \Bigl( \prod_{j\in I}
        \rd s_j\Bigr)  \delta(t-\sum_{j\in I}s_j)\prod_{j\in I}
        e^{-is_jp_j^2/2} \; . 
$$
For any disjoint sets $I, J$, we have the  semigroup property
$$
        K(t; \fp, I\cup J) = -i\int_0^t\int_0^t ds_1 ds_2 \delta(t-s_1-s_2)
        K(s_1; \fp, I)K(s_2; \fp, J)\; .
\Eq(semigroup)
$$
Define for $ \bk = (k_0, k_1, \ldots k_m)$
$$
        Q(t;   \fr, \underline {\fq} ;  {\bk}):
         = K(t; \underbrace{r_0, r_0,\ldots r_0}_{k_0+1}, q_{01},
        \ldots, q_{0k_0},    \ldots,
        \underbrace{r_{m}, r_{m}\ldots, r_{m}}_{k_{m}+1},
        q_{m1}, \ldots, q_{mk_m})\;  .
$$
{F}rom the semigroup equality  \equ(semigroup) we can rewrite 
$Q$ as 
$$
 Q(t;   \fr, \underline {\fq} ;  {\bk}) 
 = (-i)^{m}\int_0^t \Bigl( \prod_{j=0}^m
        \rd s_j\Bigr)
        \delta(t-\sum_{j=0}^m s_j)\prod_{j=0}^m K(s_j;
\underbrace{r_j, \ldots, r_j}_{k_j+1},  q_{j 1}, \cdots,
        q_{j k_j})\; .
$$
Clearly, $
Q(t;   \fr, \underline {\fq} ; \bk) = K(t, \fp, n)$.
If there is no internal momentum, we  denote it by  
$$
        Q^*(t;   \fr;  \bk): =
        K(t; \underbrace{r_0, r_0,\ldots r_0}_{k_0+1},
       \ldots,\underbrace{r_{m}, r_{m}\ldots, r_{m}}_{k_{m}+1})\; .
\Eq(multKstardef)
$$

The potential part in $F$ has 
contribution from the internal and external momenta.
The internal part is denoted by  
$$
        P(\fr; \underline{\fq} ; \bk) : = \prod_{j=0}^{|\bk|}
        R^2(r_{j}-q_{j1}) R^2(r_j-q_{j2})
        \ldots R^2(r_j- q_{jk_j}) \; ; 
\Eq(2L)
$$
the external  part  by
$$
        M(\fr, m) =  \prod_{j=0}^{m-1}  R(r_{j}-r_{j+1}) \; .
\Eq(Mdef)
$$
Therefore, we have
$$
F(\fp, n) = P( \fr;  \underline{\fq} ; {\bk})
	 M(\fr, |\bk|) \widehat\psi_0(r_m) \; 
$$
after renaming the momenta.
We can now rewrite $C_\pi$ \equ(cpig) for a simple pairing 
$\pi\in \Pi_{n,n'}$ as follows:
$$\eqalign{
      C_\pi 
        =& \lambda^{2\bar n} \int d\fr_m d\underline{\fq}_{\bk}
        d\underline{\fq'}_{\bk'}
        \overline{Q(t;   \fr, \underline {\fq} ; \bk)}
        Q(t;   \fr, \underline {\fq'} ; \bk')\cr
&\times         \overline{P( \fr;  \underline{\fq} ; {\bk})}
	\overline{ M(\fr, |\bk|)}
        P( \fr;  \underline{\fq'} ; {\bk'}) M(\fr, |\bk'|)
        |\widehat \psi_0(r_m)|^2 \; ,
}
\Eq(csim)
$$
where
$$
        d\underline{\fq}_{\bk}
        =\prod_{ 1\le \ell \le k_j , 0\le j \le m }  dq_{j \ell}
\qquad   d\fr_m
        =\prod_{ 0\le  j \le m}   dr_{j }  \; ,
$$
and similarly for $d\underline{\fq}_{\bk'}'$.

Recall the K-identity in \equ(Kid). This implies the identity for $Q$:
$$
       Q(t;   \fr, \underline {\fq} ; \bk)
 = i e^{t\eta/2} \int d\alpha \; e^{-i\alpha t}
        \prod_{j=0}^m \[ \left \{  { 1\over \alpha  -
        r_j^2/2 + i \eta} \right \}^{k_j+1}
        \prod_{\ell=1}^{k_j} { 1\over \alpha  -
        q_{j\ell}^2/2 + i \eta} \] \; .
\Eq(Qid)
$$
We now perform the  $q$ and $q'$ integration  in $C_\pi$ 
in the expression \equ(csim). 
Define  $\Theta_{\a,\eta}(r)$ and 
$G_\eta(\a, \beta, \fr, \fr', \bk, \bk')$ by 
$$
        \Theta_{\a,\eta}(r): = \int
        { R^2(r-q) \over \a - q^2/2+ i\eta} dq \; ,
\Eq(Thetadef)
$$
$$
        G_\eta(\a, \beta, \fr, \fr', \bk, \bk')
        := \overline{ M} (\fr, m) M (\fr', m) \prod_{j=0}^m
           \Big[\overline{\Theta}_{\a,\eta}(r_j)\Big]^{k_j}
        \Big[\Theta_{\beta,\eta}(r_j')\Big]^{k'_j}\; .
\Eq(Gdef)
$$
Notice that $\Theta$ depends on $r$ only through its norm as
$R$ is spherically symmetric.
We also need the notations later on: 
$$
        \Theta_{\a, 0^+}(r): =  \lim_{ \eta \to 0^+} \Theta_{\a,\eta}(r)\; ,
\Eq(Theta0)
$$
$$
        G_0(\a, \beta, \fr, \fr', \bk, \bk')
        := \lim_{\eta\to 0^+}
        G_\eta(\a, \beta, \fr, \fr', \bk, \bk') \; .
\Eq(G0)
$$
Then we have the identity  
$$\eqalign{ 
        C_\pi	= & \lambda^{2\bar n}e^{t\eta} \int d\fr_m
        \int_{-\infty}^\infty d\alpha  d\beta \; e^{i(\alpha-\beta)  t}
        \cr
&     \times \prod_{j=0}^m \[ { 1\over \alpha  -
        r_j^2/2 - i \eta} \]^{k_j+1}
         \[ { 1\over \beta  -
        r_j^2/2 + i \eta} \]^{k'_j+1}
         G_\eta(\a, \beta, \fr, \fr,\bk, \bk') 
|\widehat\psi_0 (r_m)|^2 \; .
}
\Eq(GL2)
$$

We shall need several properties of $\Theta_{\a,\eta}(r)$
(including the existence of the limit in \equ(Theta0)). These
are summarized in the following general lemma  about singular
integral operators. The proof is found in the appendix.

\proclaim{\Lemma (Ylemma)}
Let
$$
	\big( Y_zf \big)(r) : = \int {f(r-q)\over z-q^2} dq
$$
be a family of linear operators parametrized by
a complex parameter $z$ with $0<  \hbox{Im} \, z < 1/2$.
Let $\a : = \hbox{Re} \, z $ 
and $\eta: = \hbox{Im} \, z$.

(i) For $d\ge 3$ we have
$$
	\Big\|  \, Y_zf \,  \Big\|_\infty 
	\leq (\hbox{const.}) \,  \| f \|_{2d, 2d} \; ,
\Eq(Yest)
$$
and for $d=2$
$$
	\Big\|  \, Y_zf \,  \Big\|_\infty 
	\leq (\hbox{const.}) \, (1+[\log\a]_-)  \| f \|_{2d, 2d} \; ,
\Eq(Yest2)
$$
where $[x]_-=-\min \{ 0, x \}$, i.e. the coefficient is $1+|\log\a|$
for $\a<1$ and is 1 for $\a\ge 1$.
Here the constants may depend on $d$ but is independent of $\eta$.

(ii) If  $z'= \a' + i\eta'$ and $\eta \ge \eta'$, then for $d\ge 3$
$$
	\Big\|  \, Y_zf - Y_{z'}f \,  \Big\|_\infty
	\leq (\hbox{const.}) \,  |z-z'| \, \eta^{-1/2}
	 \| f \|_{2d, 2d}  \; ,
\Eq(Ycont)
$$
and for $d=2$
$$	
	\Big\|  \, Y_zf - Y_{z'}f \,  \Big\|_\infty
	\leq (\hbox{const.}) \,  |z-z'| \Big(\eta^{-1/2} + {1\over |z|}
	+ {1\over |z'|}\Big)
	 \| f \|_{2d, 2d}  \; .
\Eq(Ycont2)
$$

\endproclaim

{\it Remark:} Obviously
$$
	\Theta_{\a, \eta}(r) = 2\Big[Y_{2(\a + i\eta)}
	\big( R^2\big)\Big](r) \; ,
\Eq(obv)
$$
hence the existence of the limit \equ(Theta0) follows from \equ(Ycont).
Since 
$$
	\Big\| R^2(r- \cdot) - R^2(r' - \cdot )\Big\|_{2d, 2d}
	\leq C|r-r'| \; 
$$
with $C$ depending on the Schwarz norm of $R$, 
 we also have from \equ(Yest) that for  $d\ge3$
$$
	\Big| \Theta_{\a, \eta} (r) - \Theta_{\a, \eta} (r')\Big|	
	\leq C|r-r'| \; . 
\Eq(also)
$$
In $d=2$  the estimate \equ(also) is valid with an extra factor
$(1+[\log\a]_-)$.

\medskip
\demo{Proof of Lemma \equ(simpleest)}
\medskip

We first derive the following apriori estimate for $C_\pi$.
This estimate is better than Lemma \equ(apriori) by removing the 
factors $\log t$, but it still
does not recover the $1/n!$.

\proclaim{\Lemma (simpleapriori)}
If $\pi \in \Pi_{n, n'}$ is  a simple pairing, then
$$
	|C_\pi| \leq (C\lambda^2 t)^{\bar n} \; .
\Eq(simpleapriorieq)
$$
\endproclaim

\demo{Proof of Lemma \equ(simpleapriori)}
We present the proof for $d\ge 3$. The case $d=2$ requires some
technical modifications which are explained in Section 6.
It is clear from \equ(Yest) that 
$$
	| \Theta_{\a, \eta}(r)|\leq C
\Eq(unifTheta)
$$
uniformly in $\a, \eta$ and $r$.
 Hence, by \equ(assump),
$$
	\Big| G_\eta(\a, \beta, \fr, \fr,\bk, \bk')\Big|
	 |\widehat\psi_0 (r_m)|^2 d\fr_m
	\leq C^{\|\bk\| + \|\bk'\| + 2m}d\mu(\fr_m)
\Eq(measuredom)
$$
with a constant independent of $\eta, \a, \beta$. 
Recall the definition of $\Sigma (t, \fr,  \bk)$ in \equ(Sigma). 
Thus \equ(GL2) implies that ($\eta = t^{-1}$)
$$
	|C_\pi|\leq (C\lambda)^{2\bar n} e^{t\eta} \int d\mu(\fr_m)
		\Sigma (t, \fr, \bk)\Sigma(t, \fr,  \bk') \; .
$$
By the Schwarz inequality
$$
	\Sigma (t, \fr,  \bk)\Sigma(t, \fr, \bk')
	\leq t^{\|\bk\|-\|\bk'\|} \Sigma^2 (t, \fr,  \bk')
	+ t^{\|\bk'\|-\|\bk\|} \Sigma^2 (t, \fr, \bk) \; .
\Eq(sigmaschwarz)
$$
We conclude the proof of Lemma \equ(simpleapriori) from 
the estimate of $\Sigma (t, \fr,  \bk)$ in \equ(Sigmaest) with $a =0$. 
\qed
\enddemo

The estimate in \equ(simpleapriorieq) for a simple pairing 
is not optimal and there is 
an extra $1/n!$ missing. We now recover this factor partially.
 Notice that the $d\alpha d\beta$ integrations
in \equ(GL2) could  be performed 
more precisely if $G$ were independent of $\alpha, \beta$.
This dependence however is very  mild 
and can be removed. Thus the main term will have no $\alpha, \beta$
dependence and   their integrations can be done explicitly. 

Recall the definition  ${ \Theta}_{\alpha, 0^+}$
in \equ(Theta0). 
Define 
$$
        \Theta(v):= { \Theta}_{v^2/2, 0^+}(v)\; .
\Eq(Thetadef1)
$$

\proclaim{\Lemma (Clemma)} 
We have the following representation for $C_\pi$, $\pi$ simple;
$$\eqalign{ 
      C_\pi =  \lambda^{2\bar n}\int d\fr_m \overline{Q^*(t; \fr; \bk)}
        Q^*(t; \fr; \bk')
   |M(\fr, m)|^2  &
        \overline{\Theta(v_0)}^{\|\bk \|}
         \Theta(v_0 )^{\| \bk'\|} |\widehat\psi_0(r_m)|^2 \cr
      + O\Big((C_a\lambda^2t)^{\bar n}t^{- {a\over2}}\Big) \; 
}
\Eq(Kagain)
$$
for any $0\leq a < 1$.
\endproclaim

\demo{Proof of Lemma \equ(Clemma)}
Again, the proof is presented for $d\ge 3$. The case $d=2$ is
referred to Section 6.
We claim that  $G_\eta\, $ in the formula \equ(GL2) can be replaced by 
$|M(\fr, m)|^2
        \overline{\Theta(r_0)}^{\|\bk \|}
         \Theta(r_0 )^{\| \bk'\|} $ with small errors. 
Assuming this replacement in \equ(GL2), we can use the $Q$-identity \equ(Qid)
with no $\fq$ to integrate the $\a$ and $\beta$ in \equ(GL2). 
This proves the main term in \equ(Kagain). 

{F}rom the definition of $C_\pi$ in \equ(GL2), 
the error from the replacement is bounded by 
$$\eqalign{ 
        C\lambda^{2\bar n}\int d \fr_m \int_{-\infty}^\infty d\alpha
          d\beta &  \prod_{j=0}^m \Bigr |  { 1\over \alpha  -
        r_j^2/2 - i \eta}  \Bigr | ^{k_j+1}
        \Bigr |  { 1\over \beta  -
        r_j^2/2 + i \eta} \Bigr | ^{k'_j+1} \cr
&
        \times  \Bigg|G_\eta(\a, \beta, \fr, \fr, \bk, \bk')
         -  |M(\fr, m)|^2
        \overline{\Theta(r_0)}^{\|\bk \|}
         \Theta(r_0 )^{\| \bk'\|}  \Bigg| 
	|\widehat\psi_0(r_m)|^2 \; .
}
\Eq(schwerr)
$$
Recall the definition of $G$ in \equ(Gdef). 
Then the difference in \equ(schwerr) is equal to 
$$
G_\eta (\a, \beta, \fr, \fr, \bk, \bk') 
        -  |M(\fr, m)|^2
        \overline{\Theta(v_0)}^{\|\bk \|}
         \Theta(v_0 )^{\| \bk'\|} 
$$
$$
=     |M(\fr, m)|^2
\left \{ \prod_{j=0}^m
           \Big[\overline{\Theta}_{\a,\eta}(r_j)\Big]^{k_j}
        \Big[\Theta_{\beta,\eta}(r_j)\Big]^{k'_j}
-  \overline{ \Theta(r_0)}^{\|\bk \|}
         \Theta(r_0 )^{\| \bk'\|}  \right \}  \; .
\Eq(ck)
$$
Recall that $\Theta_{\a, \eta}(r)$ depends only on $|r|$, hence the
function $\Theta^*_{\a, \eta}(|r|): = \Theta_{\a, \eta}(r)$ is well defined.
The inequality \equ(also) is valid for $\Theta^*$ as well:
$$
	\Big| \Theta^*_{\a, \eta}(|r|) - \Theta^*_{\a, \eta}(|r'|)\Big|
	\leq C\big| |r|-|r'|\big| \; .
\Eq(also1)
$$
Now we have the following estimate 
$$
	\eqalign{ \Theta_{\a,\eta}(r_j) - 
	\Theta(r_0)  = & \; 
	\[ \Theta^*_{\a,\eta}(|r_j|) - \Theta_{\a,\eta}^*(\sqrt{ 2|\a|} )\]
	+ \[ \Theta_{\a,\eta}^*(\sqrt{ 2|\a|}) 
	- \Theta_{\a,\eta}^*( |r_0|)\] \cr
	& +\[\Theta_{\a,\eta}^*( |r_0|) -\Theta_{r_0^2/2,\eta}^*( |r_0|)\]
	  + \[ \Theta^*_{r_0^2/2,\eta}(|r_0|)- 
	\Theta_{r_0^2/2, 0^+}^*(|r_0|)\] \cr
	\leq& \;  C\Big( \Big| |r_j|-\sqrt{2|\alpha|}\Big| + 
	 \Big| |r_0|-\sqrt{2|\alpha|}\Big|\Big) 
	 + C\eta^{-1/2}\Big( |\alpha - r^2_0/2| + \eta\Big)\cr
	\leq & \; C\Big( |\a|^{-1/2} 
	+ \eta^{-1/2}\Big) \Big( \Big| \alpha - {r_j^2\over 2}\Big|	
	+ \Big| \alpha - {r_0^2\over 2}\Big| \Big) + 
	C\eta^{1/2} \; .
}
\Eq(telescope)
$$

Similar estimate is valid for $\Theta_{\beta, \eta}(r_j)$.
Recall that $\eta= 1/t$. Using the uniform bound on $\Theta$ 
and a telescopic sum to bound the curely
bracket  in \equ(ck) 
we can bound \equ(schwerr) by 
$$\eqalign{ 
	& C^{\bar n}\!\! \int d \mu(\fr_m) \int_{-\infty}^\infty \!\!
	d\alpha
          d\beta  \prod_{j=0}^m \Bigr |  { 1\over \alpha  -
        r_j^2/2 - i \eta}  \Bigr | ^{k_j+1}
        \Bigr |  { 1\over \beta  -
        r_j^2/2 + i \eta} \Bigr | ^{k'_j+1} \Big[ \eta^{-1/2} \! 
	+ |\a|^{-1/2} \!+ |\beta|^{-1/2}\! \Big] \cr
&    \times 
 \Bigg[ O\Big( \a - r_0^2/2\Big) + O\Big(
	\beta -r_0^2/2\Big) + \sum_j \left \{ 
O\Big( \a - r_j^2/2\Big) + O\Big(
	\beta -r_j^2/2\Big) + O(\eta) \right \} \Bigg] \; .
}
\Eq(finsch)
$$
Suppose first that the term $\big[
\eta^{-1/2} + |\alpha|^{-1/2} +|\beta|^{-1/2}\big]$ were not present. 
The error from $O(\a-r_j^2/2)$ can be bounded in a similar way
as $O(\a-r_0^2/2)$. 
So we shall focus only on $O(\a-r_0^2/2)$. 
The net effect of the term $O\Big( \a - r_0^2/2\Big)$ is that it
decreases $k_0$ by 1. By the proof of Lemma
\equ(simpleapriori), this gives an
extra $t^{-1}$ factor unless $k_0=0$. In this case
$$
	\Bigr |  { 1\over \alpha  -
        r_0^2/2 - i \eta}  \Bigr | \times O\Big( \a - r_0^2/2\Big)
$$
is bounded, and thus the $dr_0$ integration gives a factor $t^{k_0'}$.  
We use the argument from the proof of Lemma
\equ(simpleapriori) to bound the integration of the other momenta
$r_1, \cdots , r_m$. Again, we obtain an extra $t^{-1}$ factor.  
The estimate for the $ O\Big(\beta -r_0^2/2\Big)$
term is identical. Finally the $O(\eta)$ term  gives immediately
a factor  $t^{-1}$ from its choice. Hence the net effect of the
terms in the second square bracket is a factor $t^{-1}$ compared
with the main term.

Now we include the factor  $\big[
\eta^{-1/2} + |\alpha|^{-1/2} + |\beta|^{-1/2}\big]$  as well.
We claim that the net effect of this factor is estimated
by  $C_a^{\bar n}\eta^{-1+{a\over2}}$
for any $0\leq a<1$,
hence \equ(finsch) has an extra factor $t^{-1}\eta^{-1+{a\over2}} = 
t^{-{a\over2}}$
compared with the main term. This claim is clear for the
$\eta^{-1/2}$ term.
Now we consider the term $|\alpha|^{-1/2}$, the term with $|\beta|^{-1/2}$
is identical. We follow the argument in the previous paragraph.
When referring to Lemma \equ(simpleapriori), we need to consider
the quantity
$$
	\widetilde \Sigma (t, \fp, \bk) := \int_{-\infty}^\infty
	{d\a\over \sqrt{|\a|}}
	 \prod_{j=0}^m \Bigg| {1\over \a - p_j^2/2 - it^{-1}}\Bigg|^{k_j+1} .
$$
instead of $\Sigma (t, \fp, \bk)$.
We need the following modification of the estimate \equ(Sigmaest)
$$
	\int \big| \widetilde \Sigma (t, \fp, \bk) \big|^{2-a}
	d\mu(\fp_m) \leq t^{1-{a\over 2}}  \big( C_a t\big)^{(1-a)m
	+ (2-a)\| \bk\|} \; 
\Eq(mod)
$$
for any $0\leq a < 1$.
Combined with the trivial bound $\big| \widetilde \Sigma (t, \fp, \bk) \big|
\leq (Ct)^{m+\|\bk\|}$ we obtain
$$
	\int \big| \widetilde \Sigma (t, \fp, \bk) \big|^2
	d\mu(\fp_m) \leq t^{1-{a\over 2}}  \big( C_a t\big)^{m
	+ 2\| \bk\|} \; .
$$
The rest of the proof of Lemma \equ(simpleapriori) is unchanged.

The proof of \equ(mod) is similar to that of \equ(Sigmaest).
The only modification is to change $R(\a, p_{n-1}, p_n)$ in \equ(R) to
$$
	\widetilde R (\a, p_{n-1}, p_n):= \sqrt{|\a|}R (\a, p_{n-1}, p_n)
$$
and to change $Z(p_{n-1}, p_n)$  in \equ(Z) to
$$
	\widetilde Z(p_{n-1}, p_n) : = \int_{-\infty}^\infty
	{d\a\over \sqrt{|\a|}|\a - p_{n-1}^2/2 + i\eta||\beta
	-p_n^2/2 + i\eta|} \; .
$$
Simple calculation shows that the estimate \equ(twodenom) is valid for
$\widetilde Z$ with an extra $\eta^{-1/2}$ factor on the right hand side.
This shows that the term $|\a|^{-1/2}$ in \equ(finsch) is effectively
a factor $\big( \eta^{-1/2}\big)^{2-a}$ in the final result as we claimed.
This proves Lemma \equ(Clemma).
\qed\enddemo

To finish the proof of Lemma \equ(simpleest),
we now estimate the main term on the right hand side of \equ(Kagain). 
By the Schwarz inequality,
$$
\eqalign{
        \Big| \overline{Q^*(t; \fr; \bk)}
        Q^*(t; \fr; \bk')\Big| \leq & t^{\|\bk'\|-\|\bk\|}
	\Bigg[ {(m+\|\bk \|)!\over (m+\|\bk'\|)!}
	\Bigg]^{1/2}
	|\overline{Q^*(t; \fr; \bk)}|^2 \cr
       & +t^{\|\bk\|-\|\bk'\|}\Bigg[ {(m+\|\bk' \|)!\over (m+\|\bk\|)!}
	\Bigg]^{1/2}
	|Q^*(t; \fr; \bk')|^2,
}
$$
We now use the  estimate \equ(Qestcomb) with $a = 1/2$  
and \equ(measuredom) to have 
$$
        |C_\pi|
        \leq {(C \lambda^2 t)^{\bar n}
	 \over \Big[(m+\|\bk\|)!(m+\|\bk'\|)!\Big]^{1/2}}
        + O\Big((C_a\lambda^2t)^{\bar n}t^{-{a\over2}}\Big).
$$
Finally, using that $m+\|\bk\| \ge n/2$, $m +\|\bk' \|\ge n'/2$
 and $n! \leq C^{\bar n} [(n'/2)!][(n/2)!]$, we obtain the
statement of Lemma \equ(simpleest). \qed
\enddemo  %demo of simpleest

\bigskip

\subheading{ 4. Wigner transform of the main term}
\numsec=4\numfor=1\numtheo=1

\bigskip

As it is explained in Step 2.1, we have to compute the Wigner
transform of the main term
$$
        \psi^{main}_{t, M} : = \sum_{n=0}^{M-1}\psi_n(t).
$$
Recall $\widehat J_\e(x, v) =\e^{-d} \widehat J(\xi\e^{-1}, v)$. 
The goal of this Section
is to compute 
$$
         \bE\int
        \e^{-d}\overline{\widehat J(\xi\e^{-1}, v)}
         \widehat W_{\psi^{main}_{T/\e,M}}
        (\xi, v) d\xi dv
        = \bE\int\overline{  \widehat J_\e(\xi, v)}\; \overline{\widehat
        \psi_{T/\e, M}^{main}} 
        \Big(v-{\xi\over 2}\Big)
        \widehat\psi_{T/\e, M}^{main}\Big(v+{\xi\over 2}\Big)d\xi dv
$$
in the limit $\lim\limits_{M\to\infty}\lim\limits_{\e\to0}$. By definition 
of $U^\Gamma$ in \equ(UG) and $C_\pi^\Gamma$ in \equ(cpigG), we have 
$$
	\bE\int
        \e^{-d}\overline{\widehat J(\xi\e^{-1}, v)}
         \widehat W_{\psi^{main}_{T/\e,M}}
        (\xi, v) d\xi dv
	= \sum_{n, n'=0}^{M-1} U^{\widehat J_\e}_{n,n'}
	= \sum_{n,n'=0}^{M-1} \sum_{\pi\in \Pi_{n,n'}} C_\pi^{\widehat J_\e}.
\Eq(one)
$$
If we put $ \widehat J_\e(\xi, v)$ to be a delta function in $\xi$,
$C_\pi^{\widehat J_\e}$ becomes $C_\pi$ computed in Section 3. 
The conclusions were that  $C_\pi$ is of lower order if $\pi$ is 
crossing or nested and it can be written as \equ(Kagain) 
if $\pi$ is  simple. Furthermore, for simple pairing it is bounded by 
 \equ(simplepairest) and \equ(simpleapriorieq).
All these estimates   can be extended to 
$C_\pi^{\widehat J_\e}$ as well, uniformly in $\e$.
We now state them and 
comment on the  modifications needed for the proofs. Since $J$ is fixed, we
can introduce the simplifying notation
$$
	C_\pi^\e: =C_\pi^{\widehat J_\e},\qquad\qquad
	U^\e_{n,n'}: = U^{\widehat J_\e}_{n,n'} \; .
$$

\proclaim{\Lemma (wigerror)} The contributions from the 
crossing and  nested pairings are negligible in the sense that 
$$
        U_{n,n'}^{\e} = \sum_{\pi \in \Pi_{n, n'}\atop \pi \,\, simple}
        C_\pi^{\e}
        + O\Big( \bar n!(C\lambda^2t)^{\bar n} t^{-1} 
	(\log t)^{\bar n +3}\Big)\; .
$$
In particular for any $M$ fixed we have 
$$
	\lim_{\e\to0} \sum_{n,n'=0}^{M-1} U_{n,n'}^{\e}
	= \lim_{\e\to0}\sum_{n,n'=0}^{M-1}
	  \sum_{\pi \in \Pi_{n, n'}\atop \pi \,\, simple}
        C_\pi^{\e} \; .
\Eq(firsttrunc)
$$
Since $t=T\e^{-1}$, $\lambda =\sqrt{\e}$.
\endproclaim

\demo{Proof of Lemma \equ(wigerror)}
The proofs of the apriori estimate Lemma \equ(apriori)
and Lemma \equ(cross) can be carried out with only minor  changes.
We only need to  replace   the term
$$
        {1\over |\alpha  - w_0^2/2 + it^{-1}||\beta - w_0^2/2 + it^{-1}|}
$$
in \equ(sep) and \equ(sep1) by 
$$
        {1\over |\alpha  - {1\over 2}(\widetilde w_0-{\xi\over 2})^2
         + it^{-1}||\beta - {1\over 2}(\widetilde w_0+
	{\xi\over 2})^2 + it^{-1}|} \; 
$$
with $\widetilde w_0 : = w_0 +\xi/2$.
The rest is the same. 
The proof for the nested pairings is essentially unchanged (cf. Lemma 
\equ(nest)).
\qed
\enddemo

We can generalize \equ(GL2) for a simple pairing  $\pi\in\Pi_{n,n'}$ 
to include the test function $J$: 
$$\eqalign{ 
         C_\pi^{\e}
        = &     \lambda^{n+n'}  e^{t\eta} \int d \xi  \int d\fv_m
        \overline{\widehat J_\e(\xi, v_0)}\int_{-\infty}^\infty d\alpha
         d\beta e^{i(\alpha-\beta)  t} \cr
&        \times \prod_{j=0}^m \[ { 1\over \alpha  -
        {1\over 2}( v_j-{\xi\over 2})^2 - i \eta} \]^{k_j+1}
         \[ { 1\over \beta  -
         {1\over 2}(v_j+{\xi\over 2})^2 + i \eta} \]^{k'_j+1} \cr
&         \times G_\eta\Big(\a,
        \beta, \fv-{\xi\over 2}, \fv+{\xi\over 2},\bk, \bk'\Big)
        \overline{ \widehat \psi}_0
        \Big(v_m -{ \xi\over 2}\Big) \widehat 
       \psi_0\Big(v_m + {\xi\over 2}\Big) \; ,
}
\Eq(GL2xi)
$$
where
$$
        (\fv\pm\xi/2)_j= v_j\pm\xi/2 \; . 
$$
Recall also that
$$
       n= 2 \|\bk\|+m ,  \; n'= 2\|\bk'\|+ m,  \quad   |\bk|= |\bk'|=m \; .
$$
The  apriori estimate for $C_\pi$, Lemma \equ(simpleapriori),  
can be extended to $C_\pi^{\e}$: 

\proclaim{\Lemma (wignerapriorilemma)}
For any simple pairing $\pi\in \Pi_{n,n'}$
$$
	|C_\pi^{\e}|\leq (C\lambda^2 t)^{\bar n} \; .
\Eq(wignerapriori)
$$
\endproclaim

\demo{Proof of Lemma \equ(wignerapriorilemma)}
The proof is almost identical to that of  Lemma \equ(simpleapriori).
We can  bound $|C_\pi^{\e}|$ by 
$$
\eqalign{
	|C_\pi^{\e}|\leq  &(C\lambda^2)^{\bar n}
	\int d\xi d\fv_m\; \bigg|\widehat J_\e(\xi, v_0)\bigg |\; 
	\Sigma\Big(t; \fv-{\xi\over 2}, \bk\Big)\Sigma \Big( t; 
	\fv+ {\xi\over 2}, \bk'\Big) \cr
	&\times \sup_{\a, \beta}
 	\Bigg| G_\eta\Big(\a,
        \beta, \fv-{\xi\over 2}, \fv+{\xi\over 2},\bk, \bk'\Big)\Bigg|
 	\Bigg|  \overline{ \widehat \psi}_0
        \Big(v_m -{ \xi\over 2}\Big) \widehat 
       \psi_0\Big(v_m + {\xi\over 2}\Big)\Bigg|\; .
}
$$

Recalling the definition of $\Phi$ \equ(Phi), we have by a Schwarz
inequality
similar to \equ(sigmaschwarz)
$$
	\Sigma\Big(t; \fv-{\xi\over 2}, \bk\Big)\Sigma \Big( t; 
	\fv+ {\xi\over 2}, \bk'\Big)\Bigg| \overline{ \widehat \psi_0
        \Big(v_m -{ \xi\over 2}\Big)} \widehat \psi_0
	\Big(v_m + {\xi\over 2}\Big)
	\Bigg| 
$$
$$
	\leq 
	 t^{\|\bk'\| -\|\bk\|}{\Sigma^2\Big(t; \fv-{\xi\over 2}, \bk\Big)
	\Phi^2\Big( v_m- {\xi\over 2}\Big) \over  <v_m - \xi/2>^{20d}}
	+  t^{\|\bk\| -\|\bk'\|}{\Sigma^2 \Big( t; 
	\fv+ {\xi\over 2}, \bk'\Big)
	\Phi^2\Big( v_m+ {\xi\over 2}\Big) \over  <v_m + \xi/2>^{20d}} \; .
\Eq(facts)
$$
By definition of $G_\eta $ in \equ(Gdef), we have the simple bound
$$
	\sup_{\a, \beta}\Big|  G_\eta\Big(\a,
        \beta, \fv-{\xi\over 2}, \fv+{\xi\over 2},\bk, \bk'\Big)\Big|
	\leq \prod_{j=0}^{m-1} {C\over \Big\langle \Big(v_j\pm {\xi\over 2}
	\Big) -\Big( v_{j+1} \pm {\xi\over 2}\Big)\Big\rangle^{20d}} \; .
$$
Recall the definition of the measure $\mu$ in \equ(mu). 
{F}rom the last two bounds,  we have the following  bound
similar to \equ(measuredom) in the proof of Lemma \equ(simpleapriori):  
$$
	\sup_{\a,\beta}
	\Sigma\Big(t; \fv-{\xi\over 2}, \bk\Big)\Sigma \Big( t; 
	\fv+ {\xi\over 2}, \bk'\Big)\Big|  G_\eta\Big(\a,
        \beta, \fv-{\xi\over 2}, \fv+{\xi\over 2},\bk, \bk'\Big)\Big|	
	\Bigg| \overline{ \widehat \psi_0
        \Big(v_m -{ \xi\over 2}\Big) }\widehat \psi_0
	\Big(v_m + {\xi\over 2}\Big)
	\Bigg| d\fv_m 
$$
$$
	\leq C^m \Bigg[
	t^{\|\bk'\| -\|\bk\|}
	\Sigma^2\Big(t; \fv-{\xi\over 2}, \bk\Big)	
	d\mu\Big(\fv_m-{\xi\over 2}\Big)
	+ t^{\|\bk\| -\|\bk'\|}
	\Sigma^2\Big(t; \fv+{\xi\over 2}, \bk'\Big)
	d\mu\Big(\fv_m+{\xi\over 2}\Big)
	\Bigg]
  	\; .
\Eq(wigmeasuredom)
$$
Hence we can bound $|C_\pi^{\e}|$ by 
$$
\eqalign{
	|C_\pi^{\e}|\leq & (C\lambda^2)^{\bar n} t^{\|\bk'\| -\|\bk\|}
	\int d\xi  d\mu\Big(\fv_m-{\xi\over 2}\Big)
	\; \bigg|\widehat J_\e(\xi, v_0)\bigg |\; 
	\Sigma^2\Big(t; \fv-{\xi\over 2}, \bk\Big)\cr
	&+ (C\lambda^2)^{\bar n} t^{\|\bk\| -\|\bk'\|}
	\int d\xi d\mu\Big(\fv_m+{\xi\over 2}\Big)\;
	 \bigg|\widehat J_\e(\xi, v_0)\bigg |\; 
	\Sigma^2 \Big( t; 
	\fv+ {\xi\over 2}, \bk'\Big) \; .
}
$$
Using that $\int d\xi\sup_v\big|\widehat J_\e(\xi, v)\big | $
is bounded uniformly in $\e$, we only have to perform the $\fv$
integration. After a change of variables $\fv \to \fv \pm \xi/2$
we can use  \equ(Sigmaest) with $a=0$.
  This concludes the proof of 
\equ(wignerapriori).
\qed\enddemo

We have the following Lemma parallel to the Lemma \equ(Clemma):

\proclaim{\Lemma (Wlemma)} 
We have the following representation for $C_\pi^\e$ for 
a simple pairing $\pi\in \Pi_{n,n'}$:
$$
         C_\pi^{\e} = C_\pi^{\e, main}
        + O\Big((C_a\lambda^2t)^{\bar n} t^{-{a\over2}} \Big)
\Eq(CM)
$$
$(0\leq a < 1$) with
$$
	 C_\pi^{\e, main}: =\int d \xi\int d \fv_m \,
	\overline{\widehat J_\e(\xi, v_0)}
        Y(\xi,\fv, \bk, \bk') \; ,
\Eq(CM1)
$$
where
$$\eqalign{ 
        Y(\xi, \fv, \bk, \bk')
        = &  \l^{ 2m+2 \|\bk\|+2 \|\bk'\| }
	 \overline{\Theta(v_0)}^{\|\bk \|}
         \Theta(v_0 )^{\| \bk'\|} \; |M(\fv, m) |^2 \cr
&         \times \overline{ Q^* \Big(t; \fv-{\xi\over 2}; \bk\Big)}
          Q^* \Big(t; \fv+{\xi\over 2}; \bk'\Big)
        \widehat W_{\psi_0 }(\xi, v_m) \; .
}
$$
\endproclaim

\demo{Proof of Lemma \equ(Wlemma)} The proof is parallel to 
that of Lemma \equ(Clemma). We shall only note the difference
in $d\ge 3$.
The term similar to \equ(ck) is now bounded by 
$$
        G_\eta\Big(\a, \beta, \fv-{\xi\over 2}, \fv+{\xi\over 2},
        \bk, \bk'\Big) \; -
        \;  |M(\fv, m)|^2\prod_{j=0}^m
        \overline{\Theta(v_0)}^{k_j}
         \Theta(v_0 )^{k_j'}
\Eq(4.9)
$$
$$
        \leq C^m \big[ \eta^{-1/2} + |\a|^{-1/2} + |\beta|^{-1/2}\big]
	\Bigg[  O(\xi)+ O(\eta)+
	\sum_{j=0}^m O\( \a-{v_j^2\over 2}\)
         +\sum_{j=0}^m O\( \beta-{v_j^2\over 2}  \)\Bigg]
	|M(\fv, m)|^2.
$$
Recall again the simple bound \equ(wb) for the pairing between Wigner transform and
a test function. 
We can now follow the proof  of Lemma \equ(Clemma) 
to bound the last three terms in the second square bracket in \equ(4.9). 
Since $\widehat J$ decays  sufficiently fast in $\xi$, we have  
$$
	\int d\xi \sup_v |\widehat J_\e(\xi, v)| \; |\xi|
	= \int d\xi \e^{-d} \sup_v |\widehat J (\xi\e^{-1}, v)||\xi| \leq \e
	= C t^{-1} \; .
$$
This provides the bound needed for the new error term $O(\xi)$.
\qed \enddemo

We have all the ingredients for the following Lemma which is parallel 
to Lemma \equ(simpleest). The proof is the same 
and we omit it. 

\proclaim{\Lemma (mainapr)} We have the following estimate 
on $C_\pi^{\e, main}$
for a simple $\pi \in \Pi_{n,n'}$ uniformly in $\e$:
$$
	|C_\pi^{\e, main}|\leq {(C\lambda^2 t)^{\bar n}\over
	 (\bar n!)^{1/2}}\; .
\Eq(mainapriori)
$$
\endproclaim

\medskip

Combining Lemma \equ(wigerror) and Lemma \equ(Wlemma),
 we have thus proved that 
for any $M$ fixed 
$$
	\lim_{\e\to0} \sum_{n,n'=0}^{M-1} U_{n,n'}^{\e}
	=  \lim_{\e\to0}\sum_{n,n'=0}^{M-1}
	\sum_{\pi\in \Pi_{n,n'}\atop\pi\,\, simple}
	 C_\pi^{\e, main} \; .
\Eq(sectrunc)
$$
We now sum up $C_\pi^{\e, main}$ for all pairings with a fixed
number of external momenta $m+1$. Recall that $m$ depends on $\pi$
and we use the notation $m=|\pi|$.
Since we have a cutoff in $M$, 
not all of these pairings (with a fixed
number of external momenta $m+1$) appeared  in \equ(sectrunc). 
But the number of simple
pairings in $\Pi_{n,n'}$ is at most $C^{\bar n}$. 
Hence for any $M$ fixed, we have  from Lemma \equ(mainapr),
$$
	\Bigg| \sum_{m=0}^{M-1}\sum_{\pi \, : \, |\pi|=m
	\atop \pi \,\, simple} C_\pi^{\e, main}  -
	\sum_{n,n'=0}^{M-1}\sum_{\pi\in \Pi_{n,n'}\atop \pi \,\, simple} 
	C_\pi^{\e, main} \Bigg|
	\leq {(C\lambda^2 t)^M\over (M!)^{1/8}} \; .
$$ 
We can thus extend the summation to 
infinity with negligible errors (recall that $\lambda^2 t = T$ which is fixed).
The conclusion of Lemma \equ(wigerror), \equ(Wlemma) and \equ(mainapr) is 
$$
	\lim_{M\to\infty}\lim_{\e\to0} \sum_{n,n'=0}^{M-1} U_{n,n'}^{\e}
	=  \lim_{M\to\infty}\lim_{\e\to0}\sum_{m=0}^{M-1}
	\sum_{\pi \, ; \, |\pi|=m\atop\pi\,\, simple}
	 C_\pi^{\e, main} \; .
\Eq(conclusion)
$$
{F}rom the definition of $ C_\pi^{\e, main} $ in \equ(CM1),
we have for any $m$ fixed 
$$
	\sum_{\pi \, ; \, |\pi|=m\atop\pi\,\, simple}
	 C_\pi^{\e, main} 
	= \sum_{\bk, \bk' \atop |\bk|=|\bk'|=m}
	\int d\xi\int d\fv_m \overline{\widehat J_\e (\xi, v_0)}
	Y(\xi, \fv , \bk, \bk') \; .
\Eq(twostar)
$$
We now sum up $\bk$, $\bk'$.
For any $m$ and $a$, there is an  identity
$$
      \sum_{ \bk \, : \, |\bk| = m}
          a^{\|\bk\|} Q^*(t;   \fr;  {\bk})
	= e^{-ita} K(t; \fr, m) \; .
\Eq(onestar)
$$
We now give a proof of this identity. 
By the semigroup property \equ(semigroup)
and the definition \equ(multKstardef) of $ Q^*$, we have
$$
	\sum_{ \bk \, : \, |\bk| = m}
          a^{\|\bk\|} Q^*(t;   \fr;  {\bk})
	= (-i)^m \int_0^t \Big(\prod_{j=0}^m ds_j\Big)
	\delta \Big( t-\sum_{j=0}^m s_j\Big)
	\prod_{j=0}^m  \sum_{k_j=0}^\infty a^{k_j}
	K(s_j; \underbrace{r_j, r_j,\ldots, r_j}_{k_j+1})  \; .
\Eq(4.3)
$$
We have the simple identity 
$$
	\sum_{k_j=0}^\infty a^{k_j}
	K(s_j; \underbrace{r_j, r_j,\ldots, r_j}_{k_j+1})  = 
    \sum_{k_j=0}^\infty {(-is_j a)^{k_j}\over k_j!}
	e^{-is_jr_j^2/2}  
\; = \; 	e^{-is_j a}\; e^{-is_jr_j^2/2}  \; .
$$
Using this identity in \equ(4.3) and recalling the definition of
$K$, we have proved \equ(onestar). 
We now sum over $\bk$ and $\bk'$ and apply the identity 
\equ(onestar) for $a= \lambda^2
\Theta(v_0)$:
$$%\eqalign{ 
        \!\!\! \sum_{ \bk, \bk'\atop|\bk|= |\bk'|=m} \!\!\!\!
          Y(\xi, \fv, \bk, \bk') 
        =  \l^{ 2m}
        e^{i t\l^2  [ \ov{ \Theta}(v_0) -\Theta(v_0)] }
        \overline{ K} \Big(t; \fv-{\xi\over 2}; m\Big)
          K \Big(t; \fv+{\xi\over 2}; m\Big) \; |M(\fv, m) |^2
        \widehat W_{\psi_0 }(\xi, v_m) \; .
\Eq(Ysum)
$$

{F}rom \equ(twostar) and  \equ(Ysum), we have a new representation 
of $\sum_\pi \,C_\pi^{\e, main} $: 
$$
\eqalign{
	\sum_{\pi \, : \, |\pi|=m\atop \pi \,\, simple} 
C_\pi^{\e, main} & = 
	   \l^{ 2m}
        e^{i t\l^2  [ \ov{ \Theta}(v_0) -\Theta(v_0)] }
	 \int  d\xi \int d \fv_m   \e^{-d} 
	\overline{\widehat J( \xi \e^{-1}, v_0)}\cr 
       &  \times\overline{ K} \Big(t; \fv-{\xi\over 2}; m\Big)
          K \Big(t; \fv+{\xi\over 2}; m\Big) \; |M(\fv, m) |^2
        \widehat W_{\psi_0 }(\xi, v_m) \; .
}
$$
By definition of $K$,
$$
        \overline{K \Big(t; \fv-{\xi\over 2}; m\Big)}   {K}
        \Big(t; \fv+{\xi\over 2}; m\Big)
        =   \int_0^\infty d s_0 ds_1 \ldots ds_m \;
        \delta\(t-\sum_{j=0}^m s_j\)
$$
$$
        \times \int_0^\infty
         d s_0' d s_1' \ldots ds'_m \;
        \delta\(t-\sum_{j=0}^m s'_j\)\;
        \prod_{j=0}^m \exp {\Bigg\{ {i \over 2}\Big[
         s_j  \( v_j-{\xi\over 2}\)^2 - s'_j \(v_j+{\xi\over 2}
        \)^2 \Big ] \Bigg\}} \; .
$$
Let $a_j = (s_j + s'_j)/2$,
$b_j = (s_j - s'_j)/2$. Thus 
$$\eqalign{ 
        \overline{K  \Big(t; \fv-{\xi\over 2} ;m\Big)  } {K}
        \Big(t; \fv+{\xi\over 2}; m\Big)
        = & 2^{m}
    \prod_{j=0}^m   \int_0^t d a_j 
	 \; \;   \delta \(t-\sum_{j=0}^m  a_j \)
        e^{ -i \xi \cdot \sum_{j=0}^m a_j v_j }  \cr
        & \times \prod_{j=0}^m  \int_{-a_j}^{a_j}db_j\; \; 
        \delta \(\sum_{j=0}^m b_j\)
        e^{ i \sum_{j=0}^m b_j v_j^2 } \; . 
}
$$
The last integral can be rewritten as
$$
           \int_{-\infty}^{\infty} d \bb \; \chi_\ba (\bb) \;
	\;  e^{  i \sum_{j=0}^{m-1} b_j (v_j^2- v_m^2) } \; ,
$$
where
$$
	   \chi_\ba(\bb) :=   \chi\( - a_0 \le \sum_{j=0 }^{m-1} 
        b_j \le a_0 \)
         \prod_{j=0}^{m-1}  \chi \( -a_j \le b_j \le a_j \) \; .
$$
Here $\chi$ is the characteristic function and
$\bb = (b_0, b_1, b_2,\ldots b_{m-1})$, $\ba = (a_0, a_1, \ldots a_m)$.

We change the variables
to the macroscopic variables: 
$\a = \e a,  \zeta= \e^{-1} \xi  $, $t = \e^{-1}T$
with $\e= \lambda^2$.  Notice that $i [ \ov{ \Theta}(v_0) -\Theta(v_0)] = 
2\; \hbox{Im} \;  \Theta(v_0)$.
We can now rewrite $\sum_\pi \,C_\pi^{\e, main} $  as 
$$
	\sum_{\pi \, : \, |\pi|=m\atop \pi \,\, simple} 
C_\pi^{\e, main}= 2^{m} \int  d \fv_m
	 |M(\fv, m) |^2 e^{2T\hbox{Im}\Theta(v_0)}
	 \Bigg(\prod_{j=0}^{m}   \int_0^T d \a_j \Bigg)
        \delta \( T-\sum_{j=0}^{m } \a_j  \) 
$$
$$
        \times  \int_{-\infty}^{\infty} d \bb \; \chi_{\e^{-1}\balpha} (\bb) \;
        e^{  i \sum_{j=0}^{m-1} b_j (v_j^2- v_m^2)}
	\int d\zeta 	\;  \overline{\widehat J(\zeta, v_0)}\; 
	e^{ -i\zeta\cdot\sum_{j=0}^{m} \a_j  v_j }
	\widehat  W_{\psi_0 }(\e \zeta, v_m) \; .
$$
Notice that the $\l^{2m}$  factor in \equ(Ysum) 
is  cancelled by $\e^{-m}$ 
from the change of variables.
The last integral 
is the Fourier transform of a product
and  thus  it is equal to
$$
\int dX \overline { J(X, v_0)}
W_{\psi_0 }^\e\[  \( X-\sum_{j=0}^m  \a_j v_j\), v_m \] 
= \int dY \overline { J}\Big(Y+ \sum_{j=0}^m  \a_j v_j, v_0\Big)
W_{\psi_0 }^\e(Y, v_m)
\; .
$$
{F}rom the definition of the initial data
 $\psi_0=\psi_0^\e$ in the main Theorem, 
$$
W_{\psi_0^\e}^\e(Y, v_m) =\e^{-d} W_{\psi_0^\e}(Y/\e, v_m) 
\to  F_0(Y, v_m)
\Eq(wlim)
$$
weakly as $\e\to0$.
For any fixed $m$ and $\balpha$
$$
        \int_{- \infty}^{\infty}d \bb  \; \chi_{\e^{-1}\a}(\bb)\;
        \prod_{j=0}^{m-1} e^{i  b_j (v_j^2- v_m^2)}
        \to \prod_{j=0}^{m-1} \Big( 2\pi\delta(v_j^2- v_m^2)\Big)
$$
weakly as $\e\to0$. In particular
$$
	\lim_{\e\to 0}\int dv_0\ldots dv_{m-1} |M(\fv, m)|^2
	 \overline { J}\Big(Y+ \sum_{j=0}^m  \a_j v_j, v_0\Big)
	\int_{- \infty}^{\infty}d \bb  
	\; \chi_{\e^{-1}\a}(\bb)\;
        \prod_{j=0}^{m-1} e^{i  b_j (v_j^2- v_m^2)}
$$
$$
        = \int dv_0\ldots dv_{m-1}|M(\fv, m)|^2
	\overline { J}\Big(Y+ \sum_{j=0}^m  \a_j v_j, v_0\Big)
	\prod_{j=0}^{m-1} \Big( 2\pi\delta(v_j^2- v_m^2)\Big)
\Eq(deltalim)
$$
since $J, R\in {\Cal S}$. This convergence is in 
${\Cal S}(\bR^d_Y\times \bR_{v_m}^d)$. Hence we can multiply 
the limits \equ(wlim) and  \equ(deltalim).

After changing the variable $Y$ back  to $X$ 
we have thus proved that
$$
 	\lim_{\e\to0}
	\sum_{\pi\, : \, |\pi|=m\atop \pi \,\, simple} C_\pi^{\e, main}
	= \int  d Xd \fv_m
 	 \overline{J( X, v_0)} e^{2T\hbox{Im} \Theta(v_0)}
\Eq(three)
$$
$$
  	\times\Big(\prod_{j=0}^m   \int_0^T d \a_j\Big)
	   \delta \(T-\sum_{j=0}^m  \a_j \)
	\prod_{j=0}^{m-1} \sigma(v_j,  v_{j-1})
         F_0 \(X - \sum_{j=0}^{m} \a_j
        v_j, v_m \) \; ,
$$
where
$$
	\s(U, V): = 4\pi R^2(U-V) \d(U^2-V^2) \; .
$$
Combining \equ(one), \equ(conclusion) and \equ(three) we obtain 
$$
         \lim_{M \to \infty}   \lim_{\e \to 0}
	\bE \int \overline{\widehat J_\e (\xi, v)}\widehat
	W_{\psi_{T/\e, M}^{main}}(\xi, v) d\xi dv
$$
$$
	 =  \sum_{m=0}^\infty \int dX \int dV \overline {J (X, V)}
	\int dV_1 \ldots dV_m \int_0 d\btau_m
	\; \delta\Big( T -\sum \tau_j\Big)
$$
$$
	\times e^{2 T Im \Theta(V)}
         \s (V, V_1 )\s (V_1, V_2)
       \ldots  \s (V_{m-1}, V_m )
      F_0 \(X - \sum_{j=0}^{m} \tau_j
        V_j, V_m \)   \; ,
\Eq(Yfinal)
$$
with $V=V_0$.
Since this holds  for all smooth functions $J$, 
$$
        \lim_{M\to\infty}\lim_{\e\to0}
	\bE   W^\e_{\psi_{T/\e, M}^{main}}(X, V)  = F_T (X, V)
$$
weakly with
$$
 \eqalign{        F_T(X, V): = & \sum_{m=0}^\infty
         e^{2T Im\Theta(V)}
          \int dV_1 \,  dV_2 \, \ldots dV_m
        \int_0 d\btau_m \, \,
        \delta\Big(T-\sum_{j=0}^m \tau_j\Big)\cr
      &  \times  \s (V, V_1 )\s (V_1, V_2)
       \ldots  \s (V_{m-1}, V_m )
      F_0 \Big( \; X - \sum_{j=0}^{m} \tau_j
        V_j, V_m \; \Big)   \; .
}
$$
We can check that the function $F_T(X, V)$ satisfies the equation
$$
        \partial_T F_T(X, V)
 + V\cdot \nabla_X F_T(X, V)
 =    \int  \[ \s (U, V)  F_T(X, U)\] dU  + 2 \; 
\hbox{Im}  \; \Theta(V)  F_T(X, V) \; .
\Eq(4.5) 
$$
Recall $\sigma(V)$ is the total differential
cross section given by 
$$
        \sigma(V) := \int \sigma (U,V) dU \; .
$$
By radial symmetry,  $\sigma(V) = \sigma (|V|)$.
Recall the definition of $\Theta$ from \equ(Thetadef) 
and \equ(Thetadef1). It is easy to check that 
$$
        \hbox{Im} \;  \Theta(V)
        = -2 \pi \int_{\bR^d}dq
         R^2(V-q)  \delta(V^2 - q^2)
          =-\frac{\sigma(V)}{ 2} \; .
$$
This is actually the second order
equality of  the optical theorem. From the last relation
and the  equation \equ(4.5), we have proved that the 
Wigner transform of the main term converges weakly to the solution 
of the linear Boltzmann equation with the scattering kernel $\s$.

\bigskip

\subheading{ 5. Partial Time Integration 
and Estimate of $\Psi_{n_0}(t)$}
\numsec=5\numfor=1\numtheo=1

\bigskip

We shall estimate the $L^2$ norm of $\Psi_{n_0}(t)$ in this section
(recall \equ(decomp) for the definition).
Our basic idea is to subdivide the time integration into smaller
time intervals of size $t/\kappa(\e)$ with $\kappa(\e)$ to be chosen in 
\equ(ka). We then use the Duhamel formula to estimate 
the evolution in each time interval. In other words, instead of 
using the  Duhamel formula to control directly the wave function 
$\Psi_{n_0}(t)$, we first subdivide the time interval
and  then use the Duhamel
formula in each time interval. We shall see that this procedure
improves the error estimates. 

We now define several time dependent wave functions below. 
The wave functions  have two variables: $(t, p_0)$ for the  time and 
momentum. We
shall omit one or both variables  if they 
are not relevant in the formulas. The choice of $\kappa = \kappa(\e)$
and $n_0= n_0(\e)$ depend on $\e$, but we will omit this argument
in most cases.
Let $\theta_j = j t/ \kappa  $ for
$j=0, 1, \ldots  \kappa  $. Rewrite $\Psi _{n_0}(t)$ 
in \equ(decomp) as 
$$
        \Psi _{n_0}(t) = (-i)\int_0^t   ds \,\,   e^{-i(t-s) H  }
        {\widetilde \psi}_{n_0}(s)
        = (-i)\sum_{j=0}^{ \kappa  -1} e^{-i(t-\theta_{j+1})H}
        \int_{\theta_j}^{\theta_{j+1}}  e^{-i(\theta_{j+1}- s)H}
        {\widetilde\psi}_{n_0 }(s) d s .
$$
We now use the  Duhamel formula to expand the unitary
group $e^{-i(\theta_{j+1}-s)H} $.
Recall the definition of $L$   in \equ(L). Define 
a kernel 
$$
        \widetilde D(t; \fp, m):=K(t; \fp, m-1)
        L(\fp, m)
$$
and the corresponding operator 
$$
        \Bigl( \widetilde\cD_t(m) \widehat\psi\Bigr)(p_0) : =
        \int d\fp_{m, \widehat 0} \widetilde D(t; \fp, m)
       \widehat \psi(p_{m}) \; .
$$
For $n\ge m$ and $s > \theta$, 
define the wavefunction with the $(m+1)$-th collision after
 time $\theta$ and $n$ total collisions  up to time $s$ by 
$$
        \widehat \psi _{n, m,\theta}(s)
        = \int_\theta^s d\s \widetilde{D}_{s-\s}(n-m)  \widehat
	  {\psi}_{m}(\s)\; .
$$
We also define the amputated versions of these functions in momentum
space:  for $n>m$
$$
        { \widetilde\psi} _{ n, m, \theta}(s, p_0)
        : = \l\int dp \,\widehat V(p_0-p)
         \widehat \psi _{n-1, m, \theta}(s, p) \; . 
$$
The  Duhamel formula for $e^{-i(\theta_{j+1}-s)H} $ gives 
$$
         \Psi _{n_0} = U_1+U_2
\Eq(psidec) 
$$
where
$$
        U_1(t) = \sum_{n_0\leq n< 4n_0}\sum_{j=0}^{ \kappa  -1}
        e^{-i(t-\theta_{j+1})H}
        \psi_{n, n_0, \theta_j}(\theta_{j+1}) \; ,
\Eq(U1)
$$
$$
        U_2(t) = (-i) \sum_{j=0}^{ \kappa -1}
        e^{-i(t-\theta_{j+1})H}
        \int_{\theta_j}^{\theta_{j+1}}
         e^{-i(\theta_{j+1}-s)H}
        {\widetilde\psi}_{4n_0, n_0, \theta_j }(s)ds \; .
\Eq(U2)
$$

{F}rom the unitarity of $ e^{-i(t-\theta_{j+1})H}$
and the triangle inequality, we can bound $U_1$ 
by 
$$
        \pa U_1 \pa_2^2  \le C \; n_0 \; 
          \kappa  \sum_{n_0\leq n< 4n_0}  \sum_{j=0}^{ \kappa  -1}
        \pa \psi_{ n, n_0, \theta_j}(\theta_{j+1}) \pa^2_2 \; .
\Eq(U1e)
$$
This is a discrete version of the following inequality based on
the unitarity and triangle inequality. Its proof is obvious. 

\proclaim {\Lemma (unitarity)}
For any function $h_s$ we have
$$
         \Big\| \int_0^t e^{-i(t- s) H} h_{s} \; ds  \Big\|_2^2
        \le t  \int_0^t  \pa h_{s}  \pa_2^2 ds \; .
$$
\endproclaim

{F}rom this Lemma, we can bound $U_2 $ by 
$$
         \pa U_2(t) \pa_2^2 \le t \sum_{j=0}^{ \kappa -1}
        \int_{\theta_j}^{\theta_{j+1}}
        \pa \widetilde   \psi_{4n_0, n_0, \theta_j }(s) \pa_2^2 ds
        \le t^2 \sup_j \sup_{\t_j\le s \le \t_{j+1}}
        \pa \widetilde   \psi_{4n_0, n_0, \theta_j }(s) \pa_2^2 \; .
\Eq(U2e)
$$
The $L^2$ norms $\pa \psi_{ n, n_0, \theta_j}(\theta_{j+1}) \pa^2_2$
and $\pa \widetilde   \psi_{4n_0, n_0, \theta_j }(s) \pa_2^2$
in the last two equations are estimated in the following Lemma
which we use for $\tau = \theta_{j+1}$ and $\tau =s $.

\proclaim {\Lemma (parttime)}
For $0\leq \tau -\t \leq t/ \kappa  $, $\,\,\theta \leq t- t/\kappa$,
 $ \, \kappa  \ge 2$,  $\,  n_0 \le n $ and $0\leq a < 1$ we have
$$
        \bE \| \psi _{n,n_0, \theta}(\tau)\|^2 \leq
         { (C\lambda^2 t)^{n}
        \over (n!)^{1/2}}\Bigl[ (\log t)^4 + {(n!)^{3/2} 
	(C_a\log t)^{n+5 }\over t^{a/2}}\Bigr]
\Eq(first)
$$
(we will use this for $n_0\leq n < 4n_0$).
For $ 3n_0 +2\leq n  $ (we use it for $n=4n_0$)
$$
        \bE \| \psi _{n,n_0, \theta}(\tau)\|^2 \leq
        { n!  (C\lambda^2 t)^{n}
        (\log t)^{n+4 }\over \kappa^{n_0}}
\Eq(third)
$$
and 
$$
        \bE \|\widetilde  \psi _{n,n_0, \theta}(\tau)\|^2 \leq
       { n!  (C\lambda^2 t)^{n} (\log t)^{n+4 }
	\over \kappa^{n_0}} \; .
\Eq(fourth)
$$
\endproclaim

\noindent 
{\it Remark}: These estimates are not optimal. 
The true exponent of $ \kappa $ in \equ(third), 
instead of $n_0$,  
is $n-n_0$  corresponding to $n-n_0$ collisions in a  
time interval of length $t/\kappa$. 
In the last estimate \equ(fourth), 
we can get an extra ${1\over t}$ due to the definition of
$\widetilde\psi$ starting with a potential. 
But we do not need these stronger results.

We now use this Lemma to bound $U_1$ and $U_2$.
Recall the scaling of coupling constant $\l=\sqrt{\e}$ and $t= T/\e$. 
Notice that $\l^2 t$ is bounded by $T$, independent
of $\e$. Hence we can bound $U_1$ and $U_2$ by 
$$\eqalign{
        \pa U_1(t) \pa_2^2  \,\,\, \le & \; \; 
        { C^{n_0} n_0^2  \; \kappa^2  |\log \e|^4   
        \over \sqrt {n_0! }} 
         +  { C_a^{n_0}  \;  n_0^2  \; \kappa^2    
         \; |\log \e |^{4n_0+5 }  \e^{{a\over 2}}  \; (4n_0)!  } \cr    
         \pa U_2(t) \pa_2^2 \,\,\, 
        \le \;  & \; 
         {  \;  (4n_0)!  \; C ^{n_0} |\log \e|^{4n_0+4 } \over  
      \e^2\kappa^{n_0}} \; .
}
\Eq(u)
$$
Though these estimates look complicated, they are 
from a few simple observations. The estimate on $U_1$
has two terms: the first one is from the simple pairings; the second 
from the non-simple pairings. 
The last term for the non-simple pairings
has the factor $(4n_0)!$ from the number of pairings and 
  $ \e^{a/2}$  from the phase cancellation
of the non-simple pairings.
  The first
term from the simple pairings has neither $(4n_0)!$
nor the factor $ \e^{a/2}$. It gains at least a factor $\sqrt {n_0!}$
in the denominator from the time integration. The true
size of this factor should be $n_0!$ and we
lost half of it in the estimate. 
Since we use 
the unitarity in estimating $U_1$, 
we in principle lose a factor $t$. But $U_1$
is fully expanded in a subinterval of length $t /\kappa$. 
This reduces the factor from $t$ to $\kappa^2$. 
In the estimate of $U_2$, the key factor is the denominator 
$\kappa^{n_0}$, coming from putting $3n_0$ collisions in a
time interval $t/\kappa$. The true size of 
this factor is  $\kappa^{3n_0}$, which we obtain only partially.

We choose 
$$
        n_0  = n_0(\e):=    { \gamma  | \log \e | \over \log |\log \e | }, \quad
       \kappa=  \kappa(\e)  :=  |\log \e|^{1/\g^2}
\Eq(ka)
$$
with a sufficiently small positive $\gamma \ll 1$.
Notice this choice gives
$$
      \lim_{\e \to 0}  \e^{\alpha}  \big|\log\e\big|^{n_0(\e)} =0,  \quad
	\lim_{\e \to 0}  \e^{\alpha}  \big[ n_0(\e)\big]! =  0, \quad 
 \lim_{\e \to 0} \e^{\beta} \kappa(\e)^{n_0(\e)} = \infty
$$
for any $\alpha > \gamma$ and $\beta < \g^{-1}$.
Thus the  three terms on the right hand sides of \equ(u) 
vanish in the limit $\e \to 0$. 
The rest of this section is devoted to prove Lemma 5.2. 

\demo{Proof of Lemma \equ(parttime)}
Denote 
$$
        I_{n-n_0,n}: = \Big\{ n-n_0, n-n_0+1,  \ldots , n\Big\} \; .
$$
Define the free evolution operator with constraint 
given by the parameters $n_0, \t$ as  
$$
        K^\#(t, \t; \fp, n, n_0): = 
	\int_\t^t K(t-s, \fp, n-n_0-1)K(s,\fp, I_{n-n_0,n}) ds \; .
\Eq(Ksharf)
$$
The sign $\#$ refers to the partial time constraint. 
We can rewrite $\psi_{n, n_0, \t}(\tau)$ as
$$
        \widehat\psi_{n, n_0, \t}(\tau, p_0) 
= \int  K^\#(\tau, \t; \fp, n, n_0)
        L(\fp, n)\widehat\psi_0(p_n) d \fp_{n, \widehat 0} \; .
$$
The pairing structure in the Gaussian expectation  
$\bE \|\psi_{n, n_0, \t}(\tau)\|^2$ depends only on the
potential part and is independent of the $K$-kernels.
Thus the generated deltafunctions
are given by  $\Pi_{n,n}$ from Section 3 and 
are independent of $n_0$. Hence we can write (cf. \equ(cpig))
$$\eqalign{ 
        \bE \|\psi_{n, n_0, \t}(\tau)\|^2 = & 
\sum_{\pi\in\Pi_{n,n}} C_\pi^\# \cr
        C^\#_\pi := & \lambda^{2n} \int d\fp_n d\fp_n'
	 \overline{K^\#(\tau, \t; \fp, n, n_0)}
        K^\#(\tau, \t;\fp', n, n_0)
	 \Delta^0_\pi(\fp, \fp') \overline{F(\fp, n)}
        F(\fp', n) \; .
}
\Eq(cstar)
$$
For the amputated function we just omit $p_0$ from the argument of $K^\#$:
$$\eqalign{
        \bE \|\widetilde\psi_{n, n_0, \t}(\tau)\|^2 
	= & \sum_\pi \widetilde C_\pi^\# \cr
        \widetilde C^\#_\pi := &  \lambda^{2n} \int d\fp_n d\fp_n'
	 \overline{
        \widetilde K^\#(\tau, \t; \fp, n, n_0)}
        \widetilde K^\#(\tau, \t;\fp', n, n_0)
	 \Delta^0_\pi(\fp, \fp') \overline{F(\fp, n)}
        F(\fp', n)
}
$$
where
$$
        \widetilde K^\#(\tau, \t; \fp, n, n_0):
        = \int_\t^\tau
	 K(\tau-s, \fp, I_{1, n-n_0-1})K(s,\fp, I_{n-n_0,n}) ds \; .
$$
 Since the number of simple pairings is  at most $C^n$, 
Lemma \equ(parttime)  follows from Lemmas 
\equ(partapriori),
\equ(partcross), \equ(partnested) and \equ(partsimple)  below. 
\qed\enddemo

We first prove the following apriori estimate. This 
implies in particular 
the estimates \equ(third) and \equ(fourth) of Lemma \equ(parttime). 
The estimate obtained in the following Lemma is not optimal
and the remark
after Lemma  \equ(parttime) applies  as well. 

\proclaim{\Lemma (partapriori)}
For any $\pi\in\Pi_{n,n}$ we have for $n \ge n_0$
$$
        |C^\#_\pi|\leq (C\lambda^2 t)^n  (\log t)^{n+4}
$$
and for $n \ge 3n_0+2 $
$$
        | C^\#_\pi|\leq {(C\lambda^2 t)^n  (\log t)^{n+4}
	\over \kappa^{n_0}}
\Eq(squeeze)
$$
$$
        |\widetilde C^\#_\pi|\leq{ (C\lambda^2 t)^n
         (\log t)^{n+4} \over \kappa^{n_0}} \; .
\Eq(tildesq)
$$
\endproclaim

\demo{Proof  of Lemma \equ(partapriori)}
Recall the K-identity in Lemma \equ(K-id2). We can extend it
to the following identity for $  K^\#(\tau, \t; \fp, n, n_0)$.
Notice that we now need two integrations for two time constraints: 
$$\eqalign{
        K^\#(\tau, \t; \fp, n, n_0) =
        - \int_\t^\tau ds & 
        \; e^{(\tau-s)\eta}\; e^{s \widetilde\eta}\int_{-\infty}^\infty
        d\a d\widetilde \a \; e^{-i\a (\tau-s)} \; e^{-i\widetilde\a s} \cr
& \times \prod_{j=0}^{n-n_0-1}
        {1\over \a - p_j^2/2 + i\eta}\prod_{j\in I_{n-n_0, n}}
        {1\over \widetilde\a - p_j^2/2 + i\widetilde\eta} \; ,
}
\Eq(Krep)
$$
where we choose $\eta:= (\tau-\t)^{-1}$, $\widetilde\eta := t^{-1}$. 
We can integrate $s$ to have 
$$\eqalign{
        K^\#(\tau, \t; \fp, n, n_0)
        = i& \int_{-\infty}^\infty
         d\a d\widetilde \a \; {e^{-i\tau\widetilde\a +\tau\widetilde\eta}
         - e^{-i\a(\tau-\t)
	-i\t\widetilde\a +(\tau- \t)\eta +\t\widetilde\eta}
        \over \a-\widetilde\a + i(\eta-\widetilde\eta)} \cr
& \times \prod_{j=0}^{n-n_0-1}
        {1\over \a - p_j^2/2 + i\eta}\prod_{j\in I_{n-n_0, n}}
        {1\over \widetilde\a - p_j^2/2 + i\widetilde\eta} \; .
}
$$
This yields the following estimate using that
$\tau\widetilde\eta \leq 1$, $(\tau-\t)\eta\leq 1$, $\t\widetilde\eta\leq 1$:
$$
\eqalign{
        |K^\#(\tau, \t; & \fp, n, n_0)| \cr
	&\leq \int_{-\infty}^\infty
         d\a d\widetilde \a {1\over
        | \a-\widetilde\a + i(\eta-\widetilde\eta)|}\prod_{j=0}^{n-n_0-1}
        {1\over |\a - p_j^2/2 + i\eta|}\prod_{j\in I_{n-n_0, n}}
        {1\over |\widetilde\a - p_j^2/2 + i\widetilde\eta|} \; .
}
\Eq(Kestapr)
$$
Define 
$$\alpha_k := \cases  \widetilde\a & \hbox{ if }  k\leq n-n_0-1 \cr 
 \a  &  \hbox{ if }
  n-n_0\leq k\leq n \cr  
 \beta & 
 \hbox{ if }  n+1\leq k\leq n+n_0+1 \cr
\widetilde\beta &  \hbox{ if }   n+n_0+2\le k  
\endcases
$$
$$
\eta_k := \cases \widetilde \eta & \hbox{ if }  k\leq n-n_0-1 \cr 
  \eta  &  \hbox{ if }
  n-n_0\leq k\leq n \cr  
 \eta & 
 \hbox{ if }  n+1\leq k\leq n+n_0+1 \cr
\widetilde\eta &  \hbox{ if }   n+n_0+2\le k  \; .
\endcases
$$
Hence we can bound $C^\#_\pi$ by 
$$
        |C^\#_\pi|\leq \lambda^{2n}
        \int d\fw_A \int_{-\infty}^\infty d\a d\widetilde\a d\beta  d\widetilde\beta {1\over
        | \a-\widetilde\a + i(\eta-\widetilde\eta)|} {1\over
        | \beta-\widetilde\beta + i(\eta-\widetilde\eta)|}
$$
$$
        \times\prod_{0\leq k\leq 2n} {1\over |\a_k - w_k^2/2 + i\eta_k|}
        |F(\fp, n)||F(\fp', n)| \; .
$$
If $j\in B$, we consider $w_j$  as a function of $\fw_A$ given by 
\equ(AB). Thus we can estimate $|C^\#_\pi|$ by the following 
bound  similar to \equ(sep):
$$\eqalign{ 
        |C^\#_\pi|\leq & \lambda^{2n}  \int_{-\infty}^\infty
         d\a d\widetilde\a d\beta d\widetilde\beta
        {1\over | \a-\widetilde\a + i(\eta-\widetilde\eta)|
        | \beta-\widetilde\beta + i(\eta-\widetilde\eta)|} \cr
&        \times \int d\mu(\fw_A)
        {1\over|\a- w_0^2/2 + i\eta||\beta -w_0^2/2+i\eta|}
        \prod_{\ell\in A\atop \ell\not = 0}
        {1\over |\a_\ell - w_\ell^2/2 + i\eta_\ell|} \cr
&         \times  \prod_{j\in B\atop j\not = 2n+1}
        {1\over |\a_j - w_j^2/2 + i\eta_j|}
}
\Eq(sq1)
$$  
Notice that $\eta,\widetilde\eta\ge t^{-1}$
and $|\eta -\widetilde\eta|\ge t^{-1}$ (since $\tau-\t \leq  {t\over \kappa}
\leq {t\over 2}$). 
Following the proof of
Lemma  3.4, we can bound  $|C^\#_\pi| $ by 
$$
 |C^\#_\pi|       \leq(C\lambda^2 t)^n  (\log t)^{n+4} \; .
$$

We now prove the  stronger estimate \equ(squeeze)
for $n \ge 3n_0+2$. Notice that we can bound the last product
 in \equ(sq1) by 
$$
        \prod_{j\in B\atop j\not = 2n+1}
        {1\over |\a_j - w_j^2/2 + i\eta_j|} \leq 
       \prod_{j\in B\atop j\not = 2n+1} |\eta_j|^{-1}  \; .
$$
{F}rom the definition of $B$ in 
\equ(AB), we have  $|B|= n+1$. The total number of $\widetilde\eta$'s 
among $\eta_1, \eta_2, \ldots
\eta_{2n}$ is $2n_0+2$. Hence among the $\eta_k$ in  $B$, 
there are at least $n-2n_0-2 (\ge n_0)$ of them with 
$\eta_k = (\tau -\t)^{-1}$ . Hence
$$
\prod_{j\in B\atop j\not = 2n+1} |\eta_j|^{-1}  \le 
   t^n \Big(  {\tau -\t\over t}
        \Big)^{n_0} \leq {t^n\over   \kappa^{n_0}} \; .
$$
This proves \equ(squeeze). 

Finally, to obtain  \equ(tildesq),  we just notice that the factor
$|\a- w_0^2/2 + i\eta|^{-1}|\beta -w_0^2/2+i\eta|^{-1}$ is not
present in the formula for $\widetilde C^\#_\pi$
similar to \equ(sq1). The rest of the proof is the same. 
If we keep track of this omission carefully, 
then there is one less time integration
and we obtain  an extra $t^{-1}$ in the estimate of 
$\widetilde\psi_{n, n_0, \t}(t)$. But
we do not need it here.
This finishes the proof of Lemma \equ(partapriori). 
\qed\enddemo

\proclaim{\Lemma (partcross)}
If $\pi$ is crossing then
$$
        |C^\#_\pi|\leq (C\lambda^2t)^n t^{-1} (\log t)^{n+5} \; .
$$
\endproclaim

\demo{Proof of Lemma \equ(partcross)}
The proof is essentially  the same as in Lemma \equ(crossing).
We only have to use 
$\eta,\widetilde\eta\ge t^{-1}$
and $|\eta -\widetilde\eta|\ge t^{-1}$. We shall not 
repeat the argument here. 
\qed\enddemo

\proclaim{\Lemma (partnested)}
 For any nested $\pi\in \Pi_{n,n}$, the following bound holds:
$$
        |C^\#_\pi|\leq
         (C\lambda^2 t)^n t^{-1}(\log t)^{n+4} \; .
$$
\endproclaim

\demo{Proof of Lemma \equ(partnested)}
Recall from the proof of Lemma \equ(nest) that a nest consists of an inner
and an outer delta function: $\delta (-p_i+p_{i+1} - p_j + p_{j+1})$
and $\delta(-p_{k} +p_{k+1} - p_u + p_{u+1})$, $i< k<u<j$, 
and we can assume that the
inner delta function is minimal: $u=k+1$. 
There can be several inner deltas 
 belonging to the same outer delta. We can consider a nest where all
the inner deltas are minimal. Such a nest is called minimal nest.
Recall the time threshold $s$ in \equ(Ksharf).
If the outer delta is entirely before the time 
$s$ or entirely after $s$, 
we can follow the proof of Lemma \equ(nest). Instead of using 
the apriori
estimate Lemma \equ(apriori) in the proof of Lemma \equ(nest), 
we now use Lemma \equ(partapriori). We thus obtain
the estimate $(C\lambda^2t)^n t^{-1} (\log t)^{n+3}$.

\bigskip\bigskip
\centerline{\epsffile{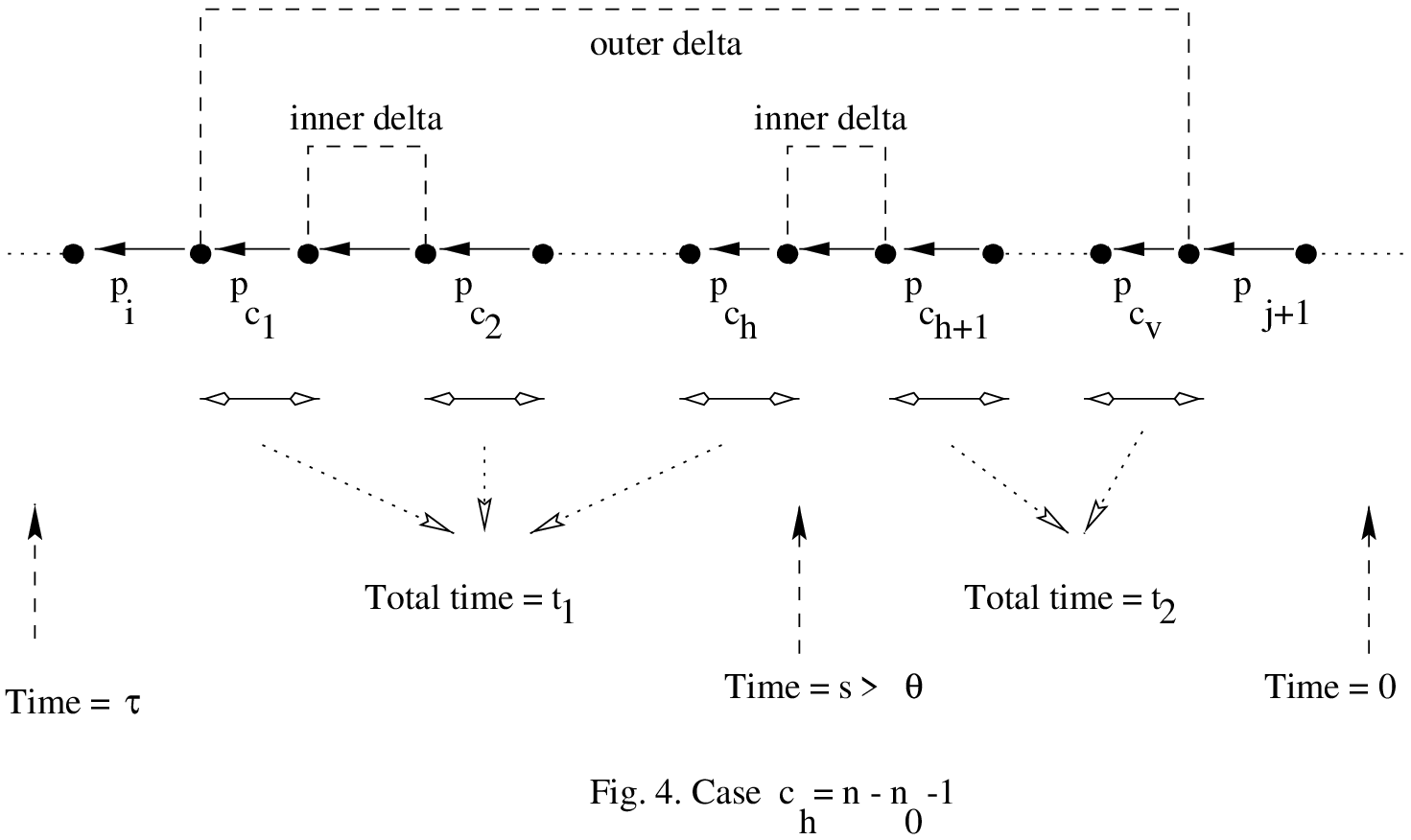}}

The only case left to consider  is when the outer delta of a 
minimal nest bridges over the time
constraint $s$, i.e. if $i\leq n-n_0-1$ but $j\ge n-n_0$.
Recall that we have $u= k+1$ for 
all inner delta
function $\delta (-p_k + p_{k+1} - p_u + p_{u+1})$ (as in
Fig. 2). 
Hence by Lemma \equ(forceid), $p_k=p_{k+2}$.
Let $p_{c_1}, p_{c_2},\ldots p_{c_v}$ be those momenta with
$i+1=c_1< c_2 < \ldots <c_v = j$ which are 
the same as $p_k=p_{k+2}$, i.e.
$$
        p_{c_1} = p_{c_2} =\ldots =p_k = p_{k+2} = \ldots
        =p_{c_v} \; .
$$
Recall 
$I^*:= I_n\setminus\{c_1, \ldots , c_v\}$ and 
$\widetilde\Delta_\pi^0(\fp^*, \fp)$ (see \equ(deltadecomp)).
Suppose that $p_{c_h}$ is the first momentum after time $s$, i.e.
$1\leq h\leq v-1$ is defined by $c_h\leq n-n_0-1 < n-n_0
\leq c_{h+1}$.
Recall the identity \equ(iterKid). We apply this identity
to the two $K$ kernels defining $K^\#$ (see \equ(Ksharf))
with $q$ replaced by   $p_k$. Thus  we obtain
$$\eqalign{ 
 K^\#(\tau, \t; \fp, n,n_0) =& \int_\t^\tau ds \int_0^{\tau-s} dt_1
        K\Big(\tau -s-t_1, \fp, I^*\cap I_{n-n_0-1}\Big)
        {t_1^{h-1}\over (h-1)!} e^{-it_1p_k^2/2} \cr
        & \qquad \times 
        \int_0^s dt_2 K\Big( s -t_2, \fp, I^*\cap I_{n-n_0, n}\Big)
        {t_2^{v-h-1}\over (v-h-1)!} e^{-it_2p_k^2/2} \; .
}
$$
Recall the definition of 
$E(\fp^*, \fp', p_k)$  \equ(Edef). We thus  have the identity 
$$
\eqalign{
        C^\#_\pi = &\lambda^{2n}\int d\fp^* dp_k d\fp_n'
	K^\#(\tau , \t; \fp', n,n_0) \widetilde\Delta_\pi^0 (\fp^*, \fp')
        E(\fp^a*, \fp', p_k) \cr
        & \times \int_\t^\tau ds\int_0^{\tau-s} dt_1
        \overline{K  \Big(\tau-s-t_1, \fp, I^*\cap I_{n-n_0-1}\Big)
        {t_1^{h-1}\over (h-1)!} e^{-it_1p_k^2/2}}\cr
        &\times 
        \int_0^s dt_2 \; 
	\overline{K   \; \Big( s -t_2, \fp, I^*\cap I_{n-n_0, n}\Big)
        {t_2^{v-h-1}\over (v-h-1)!} e^{-it_2p_k^2/2}} \; .
}
$$ 
Following the proof of Lemma \equ(nest), we integrate $p_k$ 
by parts. We thus obtain an identity for $C^\#_\pi$ in terms of 
$H(\fp^*, \fp', p_k)$  (see \equ(Hdef)), like \equ(cpirew), but with 
an extra factor $(t_1 + t_2)^{-1}$. Rewrite this factor 
by the simple identity 
$$
        {1\over t_1+t_2} = \int_0^\infty  e^{-(t_1+t_2)\mu} 
	d\mu\; . 
$$
We obtain the following expression of $C^\#_\pi$:
$$\eqalign{ 
        C^\#_\pi =&  \lambda^{2n}\int d\fp^* dp_k d\fp_n' 
         \int_0^\infty d\mu \,\, 
        K^\#(\tau , \t; \fp', n,n_0) \widetilde\Delta_\pi^0 (\fp^*, \fp')
        H(\fp^*, \fp', p_k) \cr
&         \times\int_\t^\tau ds\int_0^{\tau-s} dt_1
       \overline{  K  \Big(\tau -s-t_1, \fp, I^*\cap I_{n-n_0-1}\Big)
        {t_1^{h-1}\over (h-1)!} e^{-it_1(p_k^2/2- i\mu)}} \cr
&        \times 
        \int_0^s dt_2 \overline{  K \Big( s -t_2, \fp,
         I^*\cap I_{n-n_0, n}\Big)
        {t_2^{v-h-1}\over (v-h-1)!} e^{-it_2(p_k^2/2- i\mu)}} \; .\cr
}
\Eq(5.8)
$$
We can perform the $dt_1 dt_2$ integrations
so that the last two lines are equal to 
$$\eqalign{ 
        \int_\t^\tau ds \; & \overline{ K^*   \Big( \tau -s,
        \fp^2_{I^*\cap I_{n-n_0-1}},  
         \underbrace{p_k^2 - 2i\mu, \ldots ,
        p_k^2 - 2i\mu}_{h \,\, \hbox{times}}\Big) } \cr
&  \qquad \qquad \qquad       \times\overline{K^* \Big(s,
        \fp^2_{I^*\cap I_{n-n_0, n}},
        \underbrace{p_k^2 - 2i\mu, \ldots ,
        p_k^2 - 2i\mu}_{v-h \,\, \hbox{times}}\Big)} \; ,
}
$$
where $K^*$ is defined in \equ(Kstardef) and
 $\fp_{I^*\cap I_{n-n_0-1}} = 
\{ p_j\, : \, j\in I^*, j\in I_{n-n_0-1}\}$.
If the lower bound  $\t$ in  the $s$ integration 
were replaced by $0$, we could use the semigroup identity of $K$ 
in \equ(semigroup) to obtain once more  $K^*$, i.e.,
it is equal to 
$$
\overline{K^*   \Big( \tau ,
        \fp^2_{I^*\cap I_{n-n_0-1}},  
         \underbrace{p_k^2 - 2i\mu, \ldots ,
        p_k^2 - 2i\mu}_{v \,\, \hbox{times}}\Big)} \; .
$$
The lower bound $\t$ complicates slightly the estimate.

We can now follow the proof of the 
apriori estimates 
Lemma \equ(apriori) or  Lemma \equ(partapriori) to bound 
$ |C^\#_\pi|$  by 
$$
        |C^\#_\pi|\leq \lambda^{2n}\int_{-\infty}^\infty d\mu\int d\mu(\fw_A)
        \int_0^\infty  d\a d\widetilde\a
        d\beta d\widetilde\beta
	\,\,  {1\over |\a-\widetilde\a + i(\eta-\widetilde 
	\eta)|}
        {1\over |\beta-\widetilde\beta + i(\eta -\widetilde \eta)|}
$$
$$
        \times H(\fp^*, \fp', p_k)\Bigg|
	{1\over \a - p_k^2/2 + i(\eta +\mu)}\Bigg|^h
        \Bigg|{1\over \widetilde\a - p_k^2/2 + 
	i(\widetilde\eta + \mu)}\Bigg|^{v-h}
        \!\!\!\prod_{0\leq k \leq 2n\atop k\not = c_j, \, 1\leq j\leq v}
       \!\!\! {1\over |\a_k - w_k^2/2 + i\eta_k|} \; .
$$
Here $\eta: = (\tau -\t)^{-1}$ and $\widetilde\eta : = t^{-1}$.
The definition of $\a_k$ and $\eta_k$ are the same as in the proof of
Lemma \equ(partapriori).
For each $\mu$ fixed, we can estimate
the right hand side of the last inequality as in 
Lemma \equ(partapriori).  
The only difference is the regularization 
of $p_k^2$  by $i(\mu+ \eta)$ instead of $i\eta$. 
The   dependence on 
$\mu$ is rather easy to track. Integrating this $\mu$ dependent
estimate over $\mu$, 
we have 
$$
         |C^\#_\pi|
   \leq (C\lambda^2 t)^n t^{-1}(\log t)^{n+4} \; .
$$
The proves the Lemma \equ(partnested). \qed\enddemo

\proclaim{\Lemma (partsimple)}
For any simple $\pi \in \Pi_{n,n}$
$$
        |C^\#_\pi|\leq {(C\lambda^2 t)^n\over (n!)^{1/2}}(\log t)^4
        +O\Big( (C_a\lambda^2 t)^n t^{-{a\over2}} (\log t)^{n+2}\Big)
$$
with $0\leq a < 1$.
\endproclaim

The proof of Lemma \equ(partsimple) is similar to that
of Lemma \equ(simpleest). 
Instead of the K-identity \equ(Kid), we now use 
the representation \equ(Krep) for $K^\#$. This gives a representation
of the free evolution with time constraint. 
The rest of the proof for 
Lemma \equ(partsimple) is similar to Lemma \equ(simpleest).

\newpage

\subheading{ 6. The two dimensional case}
\numsec=6\numfor=1\numtheo=1

\bigskip

We now prove Theorem 1.1 for $d=2$. This requires to extend all
the arguments in section 3, 4 and 5 to $d=2$. 
The key Lemmas in these sections are Lemmas
\equ(cross)-\equ(Clemma), Lemmas 4.2, 4.3 and Lemma 5.2. 
Since the proofs in Sect. 4 and 5
are taken from arguments in Sect. 3, once we extend  the proofs 
in Sect. 3 to $d=2$,
the modifications needed in Sect. 4 and 5 are obvious
and will not repeated here. 
So we shall only
prove Lemmas \equ(cross)-\equ(Clemma) for $d =2$. 
We first note that the estimates in Lemma \equ(Ylemma), 
needed in the proof 
of Lemma  \equ(simpleapriori), 
are worse for  $d=2$ than for  $d\ge 3$. 
The trouble comes from the small
momentum regime. The function $\Theta (v)$ has a $\log |v|$
singularity for small $v$ in the real part of $\Theta$; 
the imaginary
part is bounded. This means that the constant $C$
in Lemma \equ(simpleapriori) (which partly comes from
\equ(unifTheta))  blows up logarithmically 
if the zero  momenta set has a positive probability  with respect
to  the initial wavefunction. 
On the other hand, the Boltzmann equation (1.4) involves 
only  $\hbox{Im}\, \Theta$,  which
remains bounded. Hence the solution to (1.4) depends weakly continuously 
on its initial data, to be made precise later on. 
We now use this regularity property of the Boltzmann equation (1.4)
and  the unitarity of the Schr\"odinger evolution 
to cutoff the small momenta regime of the initial wave function.

Let $\t >0$ be the cutoff threshold
and let $\chi^\t$ denote the smooth  radial  cutoff function 
supported outside  the ball of radius $\t$ centered at  the origin
with  $\chi^\t(r)\equiv 1$
for $|r|\ge 2\t$.
We decompose the initial wave function into 
$$
	\psi_0^\e = \psi_0^{\e, \t} +\big(\psi_0^\e -\psi_0^{\e, \t}\big)
\Eq(decompos)
$$
where 
$$
	\widehat\psi_0^{\e, \t}(r): = \widehat\psi_0^\e(r)\chi^\t(r)
$$
We can compute the Wigner distribution of $\psi_0^{\e, \t}$
easily. The proof is rather simple and will be postponed.

\proclaim{\Lemma (cutoff)}
For any $\t>0$  let $F^\t_0(X, V) = |h(X)|^2 \delta (V - \nabla S(X))
\big[\chi^\t (V)\big]^2$. 
Then the following weak limit exists and equals to $F^\t_0(X, V)$:
$$
	\lim_{\e\to 0} W^\e_{\psi_0^{\e, \t}} (X, V) \; 
= F^\t_0(X, V) \; .
$$
\endproclaim

Let $\psi^{\e, \t}_{t} = e^{-itH}\psi^{\e, \t}_0$
be the solution to the Schr\"odinger
equation \equ(scheq) with initial condition $\psi_0^{\e, \t}$.
Let $F_T^\t$ be the solution to the Boltzmann equation \equ(Beq) with
the initial condition $F_0^\t$ and the collision kernel given in the 
main theorem (1.1). The following key Lemma is a cutoff version of 
Theorem 1.1.

\proclaim{\Lemma (2dthm)}
Let $d = 2$. With the notations above, for any $\t>0$ and $T>0$
$$
	\lim_{\e\to0} \bE W^\e_{\psi^{\e,\t}_{T/\e,\omega}} (X, V)
	= F_T^\t (X, V)
\Eq(2dthmeq)
$$
weakly in $\cS(\bR^{2d})$.
\endproclaim

The proof of this Lemma will be given later on. 
We now  take the $ {\t\to0}$ limit in \equ(2dthmeq) to prove Theorem 1.1
for $d=2$.
Recall the continuity property \equ(optim) of the Wigner transform 
with respect to the $L^2$ norm of the wave function. We use \equ(optim)
with respect to  the decomposition 
$$
	\psi_t^\e = \psi_t^{\e, \t} + e^{-itH}\big(
	\psi_0^\e -\psi_0^{\e, \t}\big) .
\Eq(decompose)
$$
By the unitarity of the Schr\"odinger evolution 
we have $\|\psi_t^\e\|_2=1$ and 
$$
	\Big\|  e^{-itH}\big(\psi_0^\e -\psi_0^{\e, \t}\big)
	\Big\|_2 = 
	\Big\| \psi_0^\e -\psi_0^{\e, \t}\Big\|_2 \; .
$$
Clearly, 
$$
	 \lim_{\t\to0}\sup_{\e\leq 1}
	\Big\| \psi_0^\e -\psi_0^{\e, \t}\Big\|_2 = 0 \; .
$$
Thus from \equ(decompose), we have in particular $\|\psi_t^{\e,\t}\|$ is 
uniformly bounded.
For any $\e$ and $\t$ we have (cf. \equ(optim))
$$
	\Big| \bE \, \,\langle J, W^\e_{\psi^\e_{T/\e}}\rangle
	- \bE \, \,\langle J, W^\e_{\psi^{\e, \t}_{T/\e}}\rangle\Big|
	\leq C(J) \sqrt{\bE \big\| \psi_t^{\e, \t}\big\|_2^2 \,\, 
	\bE \big\|  e^{-itH}\big(\psi_0^\e -\psi_0^{\e, \t}\big)
	\big\|_2^2}
$$
with a $J$-dependent constant $C(J)$,
hence
$$
	\lim_{\t\to0}\sup_{\e\le 1}
	\Big| \bE \,\, \langle J, W^\e_{\psi^\e_{T/\e}}\rangle
	- \bE \,\, \langle J, W^\e_{\psi^{\e, \t}_{T/\e}}\rangle \Big| =0 \; .
\Eq(wigcont)
$$
{F}rom \equ(2dthmeq) and \equ(wigcont) we have that 
$$
	\lim_{\e\to0} \bE  W^\e_{\psi^\e_{T/\e}}(X, V)
	= \lim_{\t\to0} F_T^\t(X, V)
$$
weakly.

Notice  that  the Boltzmann evolution (1.4) is weakly continuous 
in the following sense: Suppose that we have a sequence of initial data
$F_0^\t(X, V)$ such that
$$
	\lim_{\t\to0} F_0^\t(X, V) = F_0(X, V) 
$$
Let $F_T^\t(X, V)$ denote the solution to the Boltzmann equation 
with initial data $F_0^\t(X, V)$. Then 
$$
	 \lim_{\t\to0} F_T^\t(X, V) = F_T(X, V)
$$
weakly. This can be checked easily since (1.4) is a linear equation
and the collision kernel is bounded  in $d=2$.  
In our case, 
$$
F_0^\t(X, V) =	|h(X)|^2\delta
	(V-\nabla S(X)) \big[ \chi^\theta (V)\big]^2
	\to |h(X)|^2\delta
	(V-\nabla S(X)) := F_0(X, V) 
$$
weakly as $\t\to0$ provided that 
the set $\{ (X, V) \; : \; V=0\}$
has zero measure with respect to the probability measure $F_0(X, V)
dX dV$. This concludes Theorem 1.1 for $d =2$.

It remains to prove  Lemmas 6.1 and 6.2. The proof of  Lemma 6.1 is just a simple exercise
in real analysis.

\demo{Proof of Lemma \equ(cutoff)}
For any smooth testfunction $J$, from the definition of the Wigner 
distribution (1.6) we have 
$$
\eqalign{
	&  \int dX dV \overline{J(X, V)} 
	 W^\e_{\psi_0^{\e, \t}} (X, V) \cr 
& 	=  \;  \e^{-d} \int d\zeta dv 
	\overline{\widehat J(\zeta, v)}
	\chi^\t \big(v-{\zeta\e\over 2}\big) \chi^\t
	\big(v+{\zeta\e\over 2}\big) \int e^{i\big[ {(X-Y)\cdot v - (S(X)-S(Y))\over \e}
	- \zeta\cdot{X+Y\over 2}\big]} \overline{h(X)}h(Y) dX dY \cr
	  \; .
}
\Eq(basicneed)
$$
By Taylor expansion there exist bounded smooth functions 
$F^\t$ and $G^\t$ such that 
$$
	\chi^\t \big(v-{\zeta\e\over 2}\big) \chi^\t
	\big(v+{\zeta\e\over 2}\big) = \big[ \chi^\t (v)\big]^2
	- \e^2\zeta^2 F^\t(v) + \e^3\zeta^3 G^\t(v, \zeta \e)
$$
Since $d=2$,  we need to compute 
$$
\eqalign{ \lim_{\e\to0} \,\, & \int d\zeta dv \overline{
	\widehat J(\zeta, v)}\Big[ \; 
	\e^{-2}\big[  \chi^\t (v)\big]^2+ \zeta^2
	 F^\t(v) + \e\zeta^3 G^\t(v,\zeta\e) \Big]\cr
	&\times\int e^{i\big[ {(X-Y)\cdot v - (S(X)-S(Y))\over \e}
	- \zeta\cdot{X+Y\over 2}\big]}\overline{h(X)}h(Y) dX dY \; .
}
$$
The term with $G^\t$ has an extra $\e$, so for this term one can estimate
all integrands by absolute values and it gives a negligible
contribution. For the other two terms, we need the following fact: 
for any Schwarz functions $H, I$
$$
\eqalign{ \lim_{\e\to0}  \e^{-2}\int d\zeta dv \overline{
\widehat I(\zeta, v)}H(v)
	&\int e^{i\big[ {(X-Y)\cdot v - (S(X)-S(Y))\over \e}
	- \zeta\cdot{X+Y\over 2}\big]}\overline{h(X)}h(Y) dX dY \cr
	&= \int dX |h(X)|^2\overline{ I(X, \nabla S(X))} H(\nabla S(X)) \; .
}
\Eq(H)
$$
Applying this statement to $H= F^\t$ and $I=-\Delta_X J$, we have
$$
\lim_{\e\to0} \,\, \int d\zeta dv \overline{
	\widehat J(\zeta, v)} \zeta^2
	 F^\t(v)  \int e^{i\big[ {(X-Y)\cdot v - (S(X)-S(Y))\over \e}
	- \zeta\cdot{X+Y\over 2}\big]}\overline{h(X)}h(Y) dX dY 
\; = 0
$$
since  the factor $\e^{-2}$ is missing in this expression. 
Applying this statement to  $H = \big[ \chi^\t \big]^2$ and $I=J$, 
we  have 
$$\eqalign{ 
&	 \lim_{\e\to0} \,\,\, \e^{-2} \int d\zeta dv 
	\overline{\widehat J(\zeta, v)}
	\big[  \chi^\t (v)\big]^2  \int e^{i\big[ {(X-Y)\cdot v - (S(X)-S(Y))\over \e}
	- \zeta\cdot{X+Y\over 2}\big]} \overline{h(X)}h(Y) dX dY \cr
&	= \int dX |h(X)|^2 \overline{J(X, \nabla S(X))}
	 \big[\chi^\t(\nabla S(X))\big]^2 \; .
}
$$
This proves  Lemma \equ(cutoff).
 
It remains to prove \equ(H). We first perform the $d\zeta$ integral
and then change the variables $A= (X+Y)/2$ and  $B=X-Y$. Thus 
the LHS of \equ(H) is equal to 
$$
 \lim_{\e\to0}  \e^{-2}\int dA \,\, dB \,\, dv \overline{J(A, v)}
	H(v) e^{i \e^{-1}\big[ B\cdot v - S\big(A+{B\over 2}\big) +
	S\big(A-{B\over 2}\big)\big]} \; \overline{ h\big( A 
	+ {B\over2}\big)} h\big( A -{B\over 2}\big)
\Eq(6.5)
$$
>From the Taylor expansion, the exponent can be written as
$$
i \e^{-1}\big[ B\cdot v - S\big(A+{B\over 2}\big) +
	S\big(A-{B\over 2}\big)\big]
= 
i \e^{-1}\big[ B\cdot ( v-\nabla S(A)) + B^2K(A, B)\big] \; , 
$$
where $K(A, B)$ is some bounded  smooth function. 
Changing the variable $B = \e^{-1} \beta$, we have that 
\equ(6.5) is  equal to 
$$\eqalign{ 
	&  \lim_{\e\to0}  \int dA \,\, d\beta \,\, dv \overline{J(A, v)}
	H(v) e^{i \beta\cdot ( v-\nabla S(A)) + i\e\beta^2K(A, \e\beta)}
	 \overline{ h\big( A 
	+ {\e\beta\over2}\big)} h\big( A -{\e\beta\over 2}\big) \cr
&	=  \int dA \,\, d\beta \,\, dv \overline{J(A, v)}
	H(v) e^{i \beta\cdot ( v-\nabla S(A)) } |h(A)|^2
}
$$
The last term is the right hand side  of \equ(H). 
This completes the proof of Lemma \equ(cutoff).
\qed

\enddemo

We now prove Lemma 6.2. As remarked in the begining of this section, 
we shall only
prove Lemmas \equ(cross)-\equ(Clemma) for $d =2$ with the initial 
wave function supported  
outside a ball of radius $\t$ in the Fourier space.
Lemma 3.6 can be proved in exactly the same
way.  The key steps to prove Lemma \equ(simpleest) are Lemmas 
\equ(simpleapriori) and   \equ(Clemma);
the rest of the arguments to prove Lemma \equ(simpleest) hold 
for $d =2$ as well.  
Thus we only have to prove Lemmas \equ(cross), \equ(simpleapriori)
and \equ(Clemma). 
We shall omit all superscripts $\t$ from the
notations for the rest of this section. 
All constants will depend on $\t$. 
Certainly, the support of  $\widehat\psi_0$ is now 
outside a ball of radius $\t$ in the Fourier space.

\medskip
\demo{Proof of Lemma \equ(cross) for  $d=2$ with cutoff initial function}
Following the original proof, we notice that the estimate
\equ(3denom) needs to be modified in $d=2$.
It is easy to see that \equ(3denom) remains true with an extra
$$
	{1\over |\a| + t^{-1}} + {1\over |\beta|+ t^{-1}}
$$
factor on the right hand side.Thus the same factor appears
in \equ(crossextra). Since  the initial wave function is supported 
outside a ball of radius $\t$, 
 we can restrict the $w_0$ 
integration in \equ(crossextra) to $|w_0|\ge\t$.  This cutoff ensures
that the two singularities, $\a\sim 0$ and $\a\sim w_0^2/2$ are
separated. Thus the
$d\a$ integration gives only an extra $\log t$ factor. 
The same argument applies to the $d\beta$ integration.
This proves Lemma \equ(cross) for  $d=2$ with cutoff initial wave function. 
\qed\enddemo

We now prove Lemma 
\equ(simpleapriori) and   \equ(Clemma) for $d=2$.

\demo{Proof of Lemma \equ(simpleapriori) for $d=2$ with cutoff initial
function $\psi_0^{\e, \t}$}
We first note that  the uniform bound \equ(unifTheta) is replaced in $d=2$ 
by 
$$
	|\Theta_{\a, \eta}(r)|\leq C (1 + [\log\a]_-).
$$
Hence \equ(measuredom) is modified by the factor $(1 + [\log\a]_-)^{\|\bk\|
+m}(1 + [\log\beta]_-)^{\|\bk'\|+m}$ on the right hand side, i.e.,
$$
	\Big| G_\eta(\a, \beta, \fr, \fr,\bk, \bk')\Big|
	 |\widehat\psi_0 (r_m)|^2 d\fr_m
	\leq C^{\|\bk\| + \|\bk'\| + 2m}
(1 + [\log\a]_-)^{\|\bk\|
+m}(1 + [\log\beta]_-)^{\|\bk'\|+m} d\mu(\fr_m)
\Eq(measuredom6)
$$
Thus from  \equ(GL2) 
$$
\eqalign{|C_\pi|\leq  &  (C\lambda)^{2\bar n}e^{t\eta} \int_{|r_m| \ge \t} 
 d\mu(\fr_m)
        \int_{-\infty}^\infty d\alpha  d\beta \cr
     &\prod_{j=0}^m \Bigg| { 1\over \alpha  -
        r_j^2/2 - i \eta} \Bigg|^{k_j+1}
         \Bigg| { 1\over \beta  -
        r_j^2/2 + i \eta} \Bigg|^{k'_j+1}
	(1 + [\log\a]_-)^{\|\bk\|+m}(1 + [\log\beta]_-)^{\|\bk'\|+m} \; .
}
\Eq(GL22d)
$$
The goal is to
show that \equ(GL22d) is bounded by $(C\lambda^2 t)^{\bar n} $.
The constant may depend on $\theta$.
Let 
$$
	\cB: = \Bigg\{ (\a, \beta) \; : \; |\a|\ge {\t^2\over 16} \; ,
	|\beta|\ge {\t^2\over 16}\; \Bigg\}
$$
and denote by $\cB^c$ the complement of $\cB$. Define 
$$
\eqalign{
{\Cal F}_{\cB} := &  (C\lambda)^{2\bar n}e^{t\eta} \int_{|r_m| \ge \t} 
 d\mu(\fr_m)
        \int_{ (\a, \beta) \in \cB} d\alpha \;  d\beta   \cr
     &\prod_{j=0}^m \Bigg| { 1\over \alpha  -
        r_j^2/2 - i \eta} \Bigg|^{k_j+1}
         \Bigg| { 1\over \beta  -
        r_j^2/2 + i \eta} \Bigg|^{k'_j+1}
	(1 + [\log\a]_-)^{\|\bk\|+m}(1 + [\log\beta]_-)^{\|\bk'\|+m} \; 
}
\Eq(GL22d1)
$$
and similarly define ${\Cal F}_{\cB^c} $. We have the following extension of Lemma 3.11.
This proves Lemma 3.11 for $d=2$. 
\qed
\enddemo

\proclaim{\Lemma (roughlemma)}
For any initial function and any $\pi\in \Pi_{n,n'}$ with $n+n'\leq 2n_0(\e)$ we have
$$
	|{\Cal F}_{\cB}|\leq  (C\lambda^2 t)^{\bar n} \; .
\Eq(rough)
$$
$$
	|{\Cal F}_{\cB^c}|\leq  t^{-1/2} (C\lambda^2 t)^{\bar n} \; .
\Eq(rough2)
$$
\endproclaim

\demo{Proof of Lemma \equ(roughlemma)}
We follow the proof of Lemma \equ(simpleapriori). 
Since $ (\a, \beta) \in \cB$,  
the logarithmic factors $(1 + [\log\a]_-)^{\|\bk\|+m}(1 + [\log\beta]_-)^{\|\bk'\|+m}$ 
is bounded by $C^{2\bar n}$. Thus 
we can directly follow the proof of Lemma \equ(simpleapriori)
to obtain a bound $(C\lambda^2t)^{\bar n}$ for ${\Cal F}_{\cB}$. This proves 
\equ(rough).

We now prove \equ(rough2). 
We first define the quantity
$$
	\widetilde \Sigma (t, \fp, \bk) := \int_{-\infty}^\infty
	d\a(1 + [\log\a]_-)^{\|\bk\|+m}
	 \prod_{j=0}^m \Bigg| {1\over \a - p_j^2/2 - it^{-1}}\Bigg|^{k_j+1} .
$$
which is a modification of $\Sigma$ defined in (3.26). The 
estimate \equ(Sigmaest) has the version for $\tilde \Sigma $:  
$$
	\int \big| \widetilde \Sigma (t, \fp, \bk) \big|^2
	d\mu(\fp_m) \leq  t^{{1\over 4}}  \big( Ct\big)^{m
	+ 2\| \bk\|} \; .
\Eq(mod2d)
$$
To prove it,  we
replace $R(\a, p_{m-1}, p_m)$ in \equ(R) by 
$$
	\widetilde R (\a, p_{m-1}, p_m):= (1+[\log\a]_-)^{m+\|\bk\|}
	R (\a, p_{m-1}, p_m)
$$
and replace  $Z(p_{m-1}, p_m)$  in \equ(Z) by 
$$
	\widetilde Z(p_{m-1}, p_m) : = \int_{-\infty}^\infty
	{(1+[\log\a]_-)^{m+\|\bk\|} \over |\a - p_{m-1}^2/2 + i\eta||\a
	-p_m^2/2 + i\eta|} d\a \; .
\Eq(modZ)
$$
Recall that $m+\|\bk\| \leq n\leq 2\bar n \leq 2n_0(\e)$ with 
$ n_0(\e)$  defined in (2.2) and satisfying 
\equ(n0p).
Simple calculation shows that  the estimate \equ(twodenom) is valid for
$\widetilde Z$ with an extra $t^{1/4}$ factor on the right hand side. 
Here we have to use \equ(n0p). 
This proves  \equ(mod2d).

Recall $(\a, \beta) \in \cB^c$ for ${\Cal F}_{\cB^c}$, hence
$|\a|\leq \t^2/8$ or 
$|\beta|\leq \t^2/8$. Suppose we have $|\a|\leq \t^2/8$. 
Since we only integrate $|r_m| \ge \t$ in \equ(GL22d)
the denominator 
$$
|\a - r_m^2 + i\eta| \ge \t^2/8
$$
Notice that this reduces $k_m$ by $1$. 
We now follow the proof of Lemma \equ(simpleapriori)
but use \equ(mod2d) and the remark above. Since $k_m$
is reduced by $1$, we 
we obtain an extra $t^{-1}$. Together with the factor
$t^{1/4}$ in \equ(mod2d), we have an overall factor $t^{-3/4} \leq t^{-1/2}$. 
This proves  \equ(rough2). 
\qed
\enddemo

\demo{Proof of Lemma  \equ(Clemma) for $d=2$ with cutoff initial function}
 We only remark the
modifications compared with the $d\ge 3$ case. Our goal is to estimate
\equ(schwerr) for $d=2$. Notice that 
$ \Big| G_\eta(\a, \beta, \fr, \fr,\bk, \bk')\Big|
	 |\widehat\psi_0 (r_m)|^2 d\fr_m$ satisfies 
\equ(measuredom6). Similar inequality holds if we replace 
$G_\eta(\a, \beta, \fr, \fr,\bk, \bk')$ by 
$|M(\fr, m)|^2
        \overline{\Theta(r_0)}^{\|\bk \|}
         \Theta(r_0 )^{\| \bk'\|} $. 
Thus from \equ(rough2) in
Lemma \equ(roughlemma) the contribution of the set $(\a, \beta)\in 
\cB^c$ to \equ(schwerr) is negligible.
We only have to consider 
the case $(\a, \beta) \in \cB$. 
Using  Lemma \equ(Ylemma), we have to modify 
\equ(also1) to
$$
	\Big| \Theta^*_{\a, \eta}(|r|) - \Theta^*_{\a, \eta}(|r'|)\Big|
	\leq C\big| |r|-|r'|\big| (1+[\log\a]_-)\; .
\Eq(also12d)
$$
Thus \equ(telescope) becomes 
$$
	\eqalign{ \Theta_{\a,\eta}(r_j) - 
	\Theta(r_0)  \leq &  \quad C |\a|^{-1/2}
	(1+[\log\a]_-)\Big( \Big| \alpha - {r_j^2\over 2}\Big|
	+ \Big| \alpha - {r_0^2\over 2}\Big| \Big) \cr
	& + C\Big( \eta^{-1/2} + {1\over |\a| + \eta} + {1\over |r_0|}
	\Big)\Big( |\alpha - r^2_0/2| + \eta\Big)  \; .\cr
}
\Eq(telescope2d)
$$
Hence  
we can bound \equ(schwerr) on the set $(\a, \beta)\in \cB$ by 
$$\eqalign{ 
	 & C^m \int d \mu(\fr_m) \int_{(\a, \beta)\in \cB} d\alpha
          d\beta  \prod_{j=0}^m \Bigr |  { 1\over \alpha  -
        r_j^2/2 - i \eta}  \Bigr | ^{k_j+1}
        \Bigr |  { 1\over \beta  -
        r_j^2/2 + i \eta} \Bigr | ^{k'_j+1} \cr
	&\times (1+[\log\a]_-)^{m+\|\bk\|}(1+[\log\beta]_-)^{m+\|\bk'\|}\cr
	&\times\Big[ \eta^{-1/2}
	+ {1\over |\a|^{1/2}} +{1\over |\a|+\eta}
	 + {1\over |\beta|^{1/2}} + {1\over |\beta |+\eta}
	+ {1\over |r_0|}\Big] \cr
&    \times 
 \Bigg[ O\Big( \a - r_0^2/2\Big) + O\Big(
	\beta -r_0^2/2\Big) + \sum_j \left \{ 
O\Big( \a - r_j^2/2\Big) + O\Big(
	\beta -r_j^2/2\Big) + O(\eta) \right \} \Bigg] \cr
}
\Eq(finsch2d)
$$
As in the proof for  $d\ge 3$, 
we gain an extra factor
$t^{-1}$ {f}rom the terms in the second square bracket in \equ(finsch2d). 
Again, we can estimate the logarithmic factors 
$(1 + [\log\a]_-)^{\|\bk\|+m}(1 + [\log\beta]_-)^{\|\bk'\|+m}$ 
in $\cB$  by $C^{\bar n}$. 
Thus   we can estimate \equ(finsch2d) 
following the proof for $d\ge 3$. This concludes the proof
of  Lemma \equ(Clemma) for $d=2$.
\qed
\enddemo

\newpage

\subheading{Appendix}

\bigskip

\demo{Proof of Lemma \equ(rank) }
{\it Part (i)}. Using the $\fw$ variables,
all Type (I), (I'), (II) deltafunctions can be written in
a unified way as $\delta (w_i -  w_{i+1}+ w_j-w_{j+1})$
with $0\leq i < j \leq 2\bar n$, and the Type III$(\xi)$ delta
function is $\delta (w_n - w_{ n +1} +\xi)$.
Notice that for this proof it suffices to consider the case $\xi=0$,
the extension to  nonzero $\xi$ is obvious.

{F}rom these
deltafunctions one can form a $(\bar n +1)\times ( 2\bar n + 2)$
matrix ${\Cal M}$ in a natural way; the columns correspond 
to variables $w_0, w_1, \ldots w_{2\bar n+1}$, each row
corresponds to a deltafunction and it contains $0, +1, -1$
according to the coefficients in the linear combination of $w$'s given in the
deltafunction. 
 The product of all the deltafunctions gives
the kernel of $ {\Cal M}$, as
a subspace of $\bR^{2\bar n +2}$ (coordinatized by the variables
$w_0, w_1, \ldots w_{2\bar n +1}$).

The proof of  Lemma  \equ(rank) is a simple Gaussian elimination
performed on the matrix ${\Cal M}$. We shall bring the
matrix into a form $\Big[ {\Cal M}_0 \, \Big| \, -I  \,\Big]$,
where $I$ is the $(\bar n +1)\times (\bar n + 1)$ identity matrix
and  ${\Cal M}_0$ is the matrix in \equ(AB).
The columns in the $-I$ part of that matrix correspond to
the variables indexed by the set $B$. Using the $-I$ matrix, 
this means that the rows of ${\Cal M}_0$ can also be associated with
the elements of $B$.
 From this elimination, in particular,
it follows that the rank of ${\Cal M}$ is $\bar n +1$.

Since we have to show some particular properties of ${\Cal M}_0$, 
a simple existence argument is not sufficient. It is
 simpler to give an explicit construction.

Each delta function 
$\delta (w_i -  w_{i+1}+ w_j-w_{j+1})$,
 in its original (non simplified) form,
 contains four momentum variables, indexed by $i < i+1 \leq j < j+1$.
Two of these appear with plus sign, two with minus sign
and the two middle ones can cancel each other. In this
case, the corresponding row of ${\Cal M}$ contain only two
nonzero entries.
The variable with the biggest index
$j+1$ always comes with a minus sign, and this index
is called {\it leading}.
Furthermore, the leading index of the Type III delta function
$\delta (w_n-w_{ n +1})$ is
$n +1$ by definition. Obviously, the leading indices are different for
different delta functions.
Define $B$ to  be the set of all leading indices, $|B|=\bar n +1$,
 and let $A$ be the complementary set of indices. It is clear from the
construction that $0\in A$ and $ 2\bar n +1\in B$.

\noindent
We  show that by permuting rows and 
by adding certain rows to some others in ${\Cal M}$, we can
achieve that 

\medskip

(i) all the columns, corresponding to $B$-elements, 
have only a single $-1$ nonzero entry;

(ii) all entries of the matrix are $0, -1$ or $1$;

(iii) the leading $-1$ is last  nonzero element in each row ;

(iv) the last row  is $(1, 0, 0, \ldots, 0, -1)$.

\medskip

It is obvious that with these row operations, the sum of the elements
in each row remains 0.
Hence, after rearranging the rows and columns, this will yield the form 
$\Big[{\Cal M}_0   \, \Big| \, -I \Big]$, with a matrix ${\Cal M}_0$
described in Lemma \equ(rank).

By looking at the description of the deltafunctions we observe
that in the matrix ${\Cal M}$
 the first column  contains only a $+1$
nonzero entry, the last column contains a single $-1$
and all other columns are either completely empty or
they contain exactly one $+1$ and one $-1$. 
In particular, the sum of all rows is $(1, 0, 0, \ldots, 0, -1)$.
The $B$-columns, i.e. the $b_1$-th, $b_2$-th etc. ones, are never empty.
For each $b\in B$, $b\neq  2\bar n +1$, let $R_b^+$ (and
$R_b^-$) to be the row which contains  a $+1$ ($-1$, respectively)
 in the $b^{th}$ place. 

We  permute the rows of ${\Cal M}$ such that $b\mapsto R_b^-$
be an increasing function. We put the row containing the leading $-1$
in the last column into the last row of the permuted matrix.
Since the leading $-1$'s are the last nonzero elements in
each row,
 we see that $R_b^- < R_b^+$ for each $b\in B$, $b\neq  2\bar n +1$.
For each such $b$, we add
the row $R_b^-$ to the row $R_b^+$ and we do it in the increasing order of
the $b$'s. More precisely, if $b_1 < b_2 < \ldots < b_{\bar n }$
are the elements of $B$ which differ from $ 2\bar n +1$, then
 first we add the row $R_{b_1}^-$ to $R_{b_1}^+$, then the
row $R_{b_2}^-$ to $R_{b_2}^+$, etc. We name these steps according
to their column index as $b_1$-step, $b_2$-step, etc.

In this way, each leading $-1$, except the one in the last row,
is added exactly once to cancel
 the $+1$ in its own column below it.  Hence after these operations,
the $B$-columns will contain a single $-1$.
 The column
which originally contained a single $-1$ is not affected.
Notice that no leading $-1$ is cancelled. This is a consequence of the
following two facts:

(i) we always add  a row
to some other row below;

(ii)  during the elimination, the elements above a leading $-1$ in its column
 are always zero.

\noindent 
The conclusion is that after these operations, the resulting matrix
has a single $-1$ in each $B$-column. 
Furthermore, these $-1$'s are the last nonzero elements in their rows
and the first nonzero elements in their column.

The last row of the resulting matrix is $(1, 0, 0, \ldots, 0, -1)$.
For,  along the elimination each row but the last one has been
added exactly once to some other row below. Hence, at the end,
the last row will be the sum of all the rows of the original matrix.

Finally, we have to show that the only possible entries are 
$0, -1$ and $ +1$. Suppose this is not the case. Since the entries
 originally 
were $0, -1, +1$, we can consider the first moment during the elimination
procedure,
when a $\pm 2$ was created. For definiteveness, assume  that 
it was in the $b$-step (when row $R_{b}^-$ was added
to row $R_b^+$), that an entry $+2$ appeared at the first time,
and suppose that it was in the $m^{th}$ column
(the proof for the case $-2$ is identical).
In other words, right before the $b$-step
both rows $R_{b}^-$ and $R_b^+$ contained $+1$ at the $m^{th}$ place.
 Certainly $m< b$ as $R_b^-$ has zero elements in the columns with
higher index than  $b$. 

During the elimination,
we call a row {\it untouched} if it has not been added yet to any other
row (but some other row could have been added to it).
In the original matrix ${\Cal M}$
there was at most  one $+1$ in the $m^{th}$ column and suppose
it was  in the row $R$.
Due to the algorithm,  both $+1$'s at the $m^{th}$ place at the $b$-step
must originate from this single $+1$.

Clearly $R\leq R_b^-$, since every row is added at most once to
some other row, always below, hence the $+1$ originally
in the row $R$ could not appear in a row above it.
For the same reason, $R\neq R_b^-$, since it is exactly the $b$-step
 when we add the $R_b^-$ row. Before this step, the $R_b^-$ row
was untouched and the $+1$ in the
this row has not been moved out. So the $+1$ in the $R_b^+$ row
could not have been created if $R=R_b^-$.
Hence, right before the $b$-step
there are two untouched rows (namely $R_b^+$ and $R_b^-$), both contain
a $+1$ at the $m^{th}$ place which were not there in the original
matrix ${\Cal M}$.

We show that this is impossible.
Notice that there could  several $+1$'s appear in the same column
along the elimination. But at every moment, there is at most one
of them which sits in an untouched row.
This is seen by induction on the step number. Whenever a new $+1$
is created in a column, it must come from another $+1$ 
in the same column, and this latter $+1$ sits in a row which has 
just  been touched. Hence the number of $+1$'s in untouched rows
never increases. Originally there was at most one $+1$ in the
$m^{th}$ column, so there could not be two of them before the $b$-step.

The proof of part (ii) is very similar to the case $r= \bar n +1$
and the details are left to the reader.
This completes the proof of Lemma \equ(rank).

\qed
\enddemo

\bigskip\bigskip

\demo{Proof of Lemma \equ(forceid)}
First we show that the identifications described in the theorem can be forced.
Fix two edges $e(w_i)$ and $e(w_j)$ and suppose that after removing them,
the graph $G=G_\pi$ splits into two components:
 $G\setminus \{e(w_i), e(w_j)\}= G_1 \cup G_2$,
$G_1\cap G_2=\emptyset$ (as graphs). We used that $G$ contained 
the circle graph $C$ with $V(G)=V(C)$, hence $G$ is two-connected,
so the number of components after removal of two edges is at most two.
Choose the indices such that the right square be in $V(G_2)$.

Notice that $G_i$ $(i=1,2)$ cannot be just a single bullet (it can 
be a square),
since each bullet has three edges adjacent to them. This corresponds to the
fact that $w_i=w_{i+1}$ type momentum identifications are not possible
(apart from the momenta at the squares).
$G_i$ certainly can be a square, but this leads to the trivial
deltafunction $\delta (w_{2\bar n +1}-w_0)$ or $\delta(w_n-w_{n+1})$
which are forced as we have seen. So from now on we can exclude this case.

Let $C\setminus \{e(w_i), e(w_j)\}= A_1 \cup A_2$ (as graphs), where
$V(A_i) = V(G_i)$, i.e. the arc (directed path) $A_i$ is a 
subgraph of $G_i$. Obviously $A_1\cap A_2=\emptyset $ (as graphs),
and since $G_1\cap G_2=\emptyset$, it is clear that all the pairing edges
in the original $G$ must be either entirely  within  $G_1$
or in $G_2$. On Fig. 5 $i=2$,  $j=8$, $A_1$ is the directed arc with edges
$e(w_3), e(w_4),\ldots, e(w_7)$ and $A_2$ is $e(w_9), e(w_{10}),\ldots 
e(w_{2\bar n +1}), e(w_0), e(w_1)$.

Consider $G_1$. All pairing lines starting from any point of $V(G_1)$
 must be joined to another bullet from $V(G_1)$.
Let us add up the arguments of all
 deltafunctions (Type I, I' and II) which correspond
to pairing lines within $G_1$, and if $G_1$ also contains the left
square, then add the argument of $\delta(w_n- w_{n +1})$ as well.
 Let us call this collection
of deltafunctions $\Delta$. Notice that in all deltafunctions the
sign is decided by the arrow: the sign is plus if the arrow is
incoming (into the vertex, associated with the deltafunction) and
minus otherwise.

In $\Delta$ every momentum belonging
to $E(G_1)$  appears exactly twice: once
with plus once with minus sign. 
For, notice that every vertex (except the left square)
has a unique delta function associated to
it; the one whose pairing line meets this vertex. 
The Type III delta function is associated to the  left square.
Consider now the momentum
 $w_m$ with  $e(w_m) \in E(G_1)$. This connects two
vertices of $V(G_1)$.
 Both delta functions
corresponding to these vertices are present in $\Delta$.
Hence the momentum $w_m$ appears exactly twice in the collection $\Delta$,
and since it corresponded to an arrow, which is outgoing from one
vertex and incoming into the other, the sign of these two appearances
are opposite. Hence when summing up the arguments of 
all the deltafunctions in $\Delta$, all these momenta cancel. 

 After summation, 
only those  momenta remain which appear in $\Delta$, but are not
part of $E(G_1)$. Since $\Delta$ consisted of delta functions associated
to vertices in $V(G_1)$, the remaining momenta are those which
are associated to edges in $E(G)\setminus E(G_1)$ and which edges are
adjacent to $V(G_1)$, i.e. exactly $e(w_i)$ and $e(w_j)$.
One of them must be incoming into $G_1$, the other is outgoing,
since at every vertex we have one incoming and one outgoing arrow, so
the same is true for any subset of $V(G)$, in particular for $V(G_1)$.
This completes the proof that the given deltas force $w_i=w_j$.
On Fig. 5, the arguments of the  three delta functions,
$\delta(w_2-w_3+w_5-w_6)$,
 $\delta(w_3-w_4+w_4-w_5)$, $\delta(w_6-w_7+w_7-w_8)$,
add up to $w_2-w_8$.

\bigskip\bigskip
\centerline{\epsffile{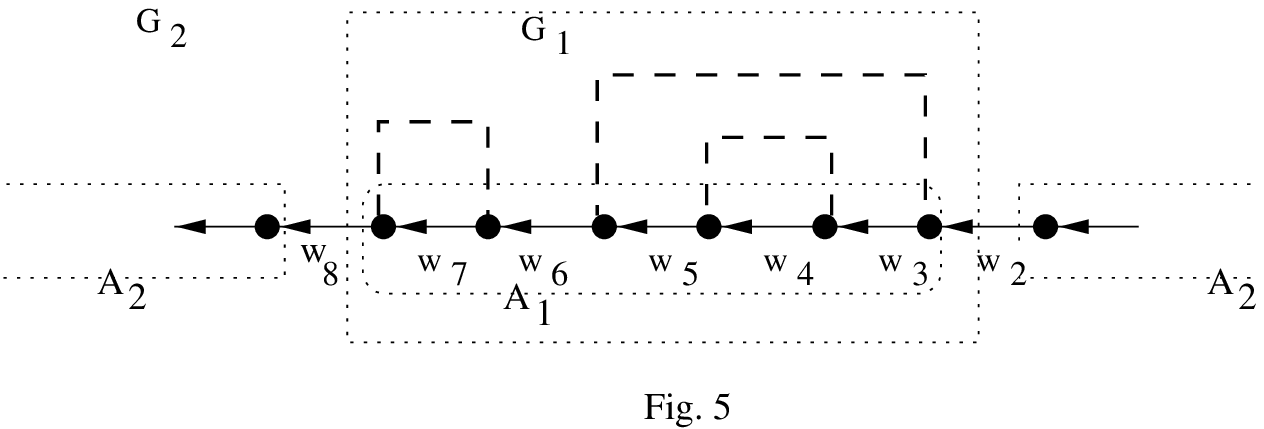}}

For the  proof of the other direction, we have to show that
if some linear combination of the arguments of the delta functions 
Type I, I', II,
forces $w_i =w_j$, then $G\setminus \{ e(w_i), e(w_j)\}$
has two components (see Fig. 6).

Since $C$ is a circle, $C\setminus \{ e(w_i), e(w_j)\} = A_1\cup
A_2$ where $A_1\cap A_2=\emptyset$ are two arcs.
 Hence the two possible components
of $G$, after the removal,
 must have the same vertex sets as the arcs: $V(A_1)=V(G_1)$ and $V(A_2)=V(G_2)$.
 So we have
to prove that if $w_i=w_j$ is forced, then there could be no
pairing line between $V(A_1)$ and $V(A_2)$ (which could save
the connectedness of $G \setminus \{ e(w_i), e(w_j)\}$).

Suppose that $G\setminus \{ e(w_i), e(w_j)\}$ is connected.
Then there is a pairing line $L$ which connects a vertex in $V(A_2)$,
say $P$, with a vertex in $V(A_1)$, say $Q$.
Of course $P$ and $Q$ are
bullets (not squares; as there is no pairing line coming out of a square).
Consider $C_L: = C\cup L$, more precisely, $C_L$ is a graph such that
$V(C_L)=V(C)$ and $E(C_L)= E(C) \cup \{ L\}$.  This is a circle
with an extra loop $L$. Since $L$ connects $A_1$ and $A_2$, there is a 
circle $\overline{ C}$ 
in $C_L$ which contains the edge $e(w_i)$, but not
 $e(w_j)$ and $\overline{ C}$ naturally contains $L$.
$\overline{C}$ contains an arc $A$ from $C$ 
built from momentum lines (the arc $A$ between $P$ and $Q$),
and clearly $\overline{C} = A\cup L$.

\bigskip\bigskip
\centerline{\epsffile{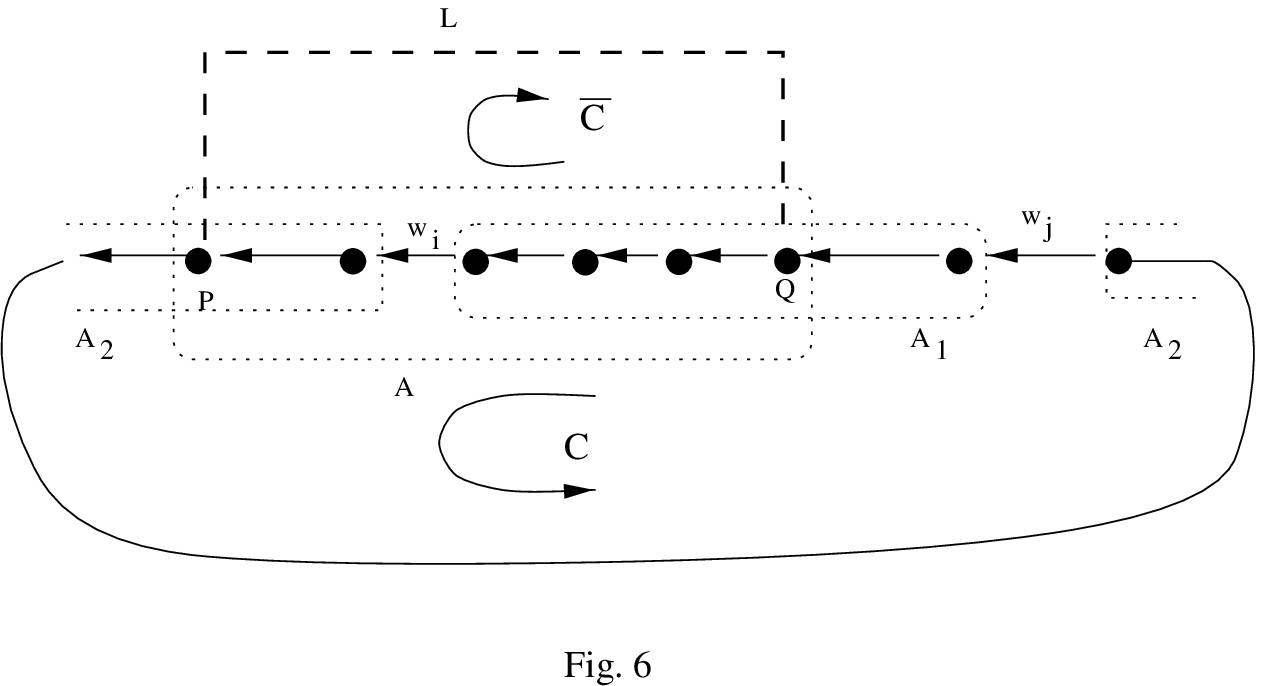}}

Now we  add a nonzero momentum
$q$ to all the momenta associated with the momentum lines in $A$
and leave the momenta associated 
with the complement arc $A^c: = C\setminus A$,
 containing $e(w_j)$, untouched. We claim that
this shift in the momenta is compatible with all the deltafunctions
Type I, I', II, III
but is not compatible with $\delta(w_i -w_j)$, hence the
latter cannot be forced by our deltas.

It is clear that this shift is not compatible with $\delta(w_i -w_j)$,
since after the shift $w_i \to w_i + q$, while $w_j$ is
unchanged.

To see that this momentum shift  is compatible with all our deltas, first
note that if a deltafunction (from Type I, I', II) is not the one
associated with $P$ or $Q$,
 then it remains valid after the shift. The reason is
that this deltafunction joins two bullets, and for each bullet there
is an incoming and an outgoing arrow. Depending on whether the bullet
belongs to $V(A)$ or not,  these two momenta get shifted or not 
simultaneously. But the structure of the deltafunctions is such that
the incoming and the outgoing momenta come with different signs,
hence either both are shifted or none of them,  the deltafunction
remains valid.
The Type III deltafunction clearly remains valid.

Finally we have to check the validity of the delta function connecting
$P$ and $Q$. But in the argument of this delta function
there are two momenta which are in $A$ and two which are in $A^c$,
and the ones which are in $A$ come with different signs.
So adding $q$ to the momenta in $A$ does not change this delta function
either. The proof is completed.

\qed
\enddemo

\demo{Proof of Lemma \equ(crossmatrix)}
In this proof
we use the notation $w_0, w_1, \ldots, w_{2\bar n +1}$.
In the matrix $\Big[ {\Cal M}_0 \, \Big| \, -I \Big]$, the first
$\bar n+1$ columns are associated with the elements of $A$, the remaining
columns with the elements of $B$. Hence, by the identity matrix,
the rows  of ${\Cal M}_0$ can be associated with  the elements of $B$ as well.

Suppose first that $\pi$ is crossing, i.e. it contains
two delta functions 
$\delta (w_i - w_{i+1} + w_j - w_{j+1})$ and 
$\delta (w_k - w_{k+1} + w_\ell - w_{\ell +1})$
such that $i < k < j < \ell$. By definition,  $j+1\in B$, i.e. $j+1= b_\nu$
with some $1\leq \nu\leq \bar n+1$
(see the proof of Lemma \equ(rank)).
Suppose that in ${\Cal M}_0$,
the $\nu$-th row (associated with $j+1\in B$)
 contains only one nonzero element.
Then $\Delta_0^\pi(\fw)$ forces the identification $w_{j+1} = w_u$
for some $u\in A$. Moreover, $u<j+1$ by Lemma \equ(rank).
Using Lemma \equ(forceid), we obtain that the removal of
the edges $w_u$ and $w_{j+1}$ disconnects $G_\pi$. But
this is impossible; the two dashed lines corresponding to these
two delta functions keep the graph connected. This can
easily be checked
 by distinguishing three cases: $u\leq i$, $i+1\leq u \leq k$
and $k+1\leq u < j$. Fig 7. shows these three possibilities for $w_u$.

\bigskip\bigskip
\centerline{\epsffile{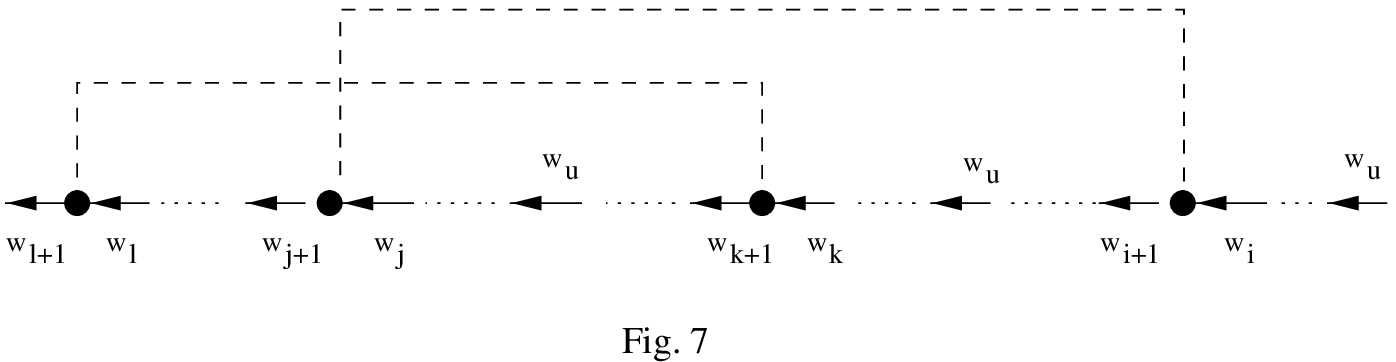}}

For the converse proof, assume that $\pi$ is noncrossing.
The set $B$ is uniquely given by $\pi$. 
We need the following auxiliary lemma, whose proof is postponed.

\proclaim{\Lemma (aux)} If $\pi$ is noncrossing, then
$\Delta^0_\pi(\fw)$ is equivalent to a product
of identification delta functions of the form
$$
	\prod_{b\in B} \delta (w_b - w_{a(b)})
\Eq(proddelta)
$$
where $a(b)\in A$. (The sets $A, B$ are defined in Lemma \equ(rank).)
\endproclaim

{F}rom the elements $a(b)$ one can form a matrix $\widetilde {\Cal M}_0$
as follows.
The $\nu$-th row of $\widetilde {\Cal M}_0$
 is associated with the element  $b_\nu\in B$,
 and the $\mu$-th column with $a_\mu\in A$,
such that the $\mu$-th entry of the $\nu$-th row is 1 for
$a(b_\nu)=a_\mu$, and  all other
entries are zero. We claim that $\widetilde {\Cal M}_0$ is the same 
as ${\Cal M}_0$ obtained in Lemma \equ(rank). If it were not the
case, then for some $b_\nu\in B$,
 the original $\Delta^0_\pi(\fw)$ would force both 
$\delta (w_{b_\nu} - w_{a_\mu})$
 and $\delta \Big(w_{b_\nu} - \sum_{\alpha \in A^*} (\pm
w_{\alpha})\Big)$ with some $A^*\subset A$, $A^*\neq \{ a_\mu \}$.
But then  $\Delta^0_\pi(\fw)$ would force the nontrivial delta function
$\delta\Big(w_{a_\mu} -  \sum_{\alpha \in A^*} (\pm
w_{\alpha})\Big)$ involving momenta only from $A$. This
delta function obviously
cannot be forced by $ \prod_{\nu=1}^{\bar n+1} \delta (w_{b_\nu} -
 w_{a(b_\nu)})$
but this latter product is equivalent to  $\Delta^0_\pi(\fw)$,
which contradiction proves $\widetilde {\Cal M}_0= {\Cal M}_0$, hence
the converse statement of the Lemma. \qed

\medskip

{\it Proof of Lemma \equ(aux).}
It is enough to show that for each $b\in B$ there is an $a(b)\in A$
such that $\delta (w_b - w_{a(b)})$ is forced by $\Delta^0_\pi(\fw)$.
For, by comparing the ranks of the product \equ(proddelta)
and $\Delta^0_\pi(\fw)$, we see that these are equivalent.

Since $b\in B$, we know that a pairing line ends at the beginning
of the arrow $w_b$. Consider the arrow at the beginning of this pairing line
(see Fig. 8).
 In other words, suppose that this pairing line corresponded to
 $\delta (w_{k_1} - w_{k_1+1} + w_{b-1}-w_b)$ 
with $k_1< b-1$, and consider $w_{k_1}$.
Since $\pi$ is noncrossing, the removal of $w_{k_1} , w_b$ disconnects
the graph, hence $\delta (w_{k_1} - w_b)$ is forced by $\Delta^0_\pi(\fw)$
using Lemma \equ(forceid). If $k_1\in A$, then we define $a(b): =k_1$.
If $k_1\in B$, then we continue the procedure; there is
a pairing line ending at $w_{k_1}$ and starting at, say, $w_{k_2}$.
I.e. there is a delta function 
 $\delta (w_{k_2} - w_{k_2 +1} + w_{k_1-1} - w_{k_1})$, $k_2< k_1-1$,
 and we see
that $\delta (w_{k_2} - w_{k_1})$ is forced, hence $\delta (w_{k_2} - w_b)$
is forced as well. Again, if $k_2\in A$, then $a(b):=k_2$,
otherwise we continue the procedure to find $k_3$ etc.
Since $0\in A$, we see that at some point 
this procedure terminates, yielding the value $a(b )\in A$
(on Fig. 8, $b, k_1, k_2\in B$, $k_3\in A$).
\qed

\enddemo

\bigskip\bigskip
\centerline{\epsffile{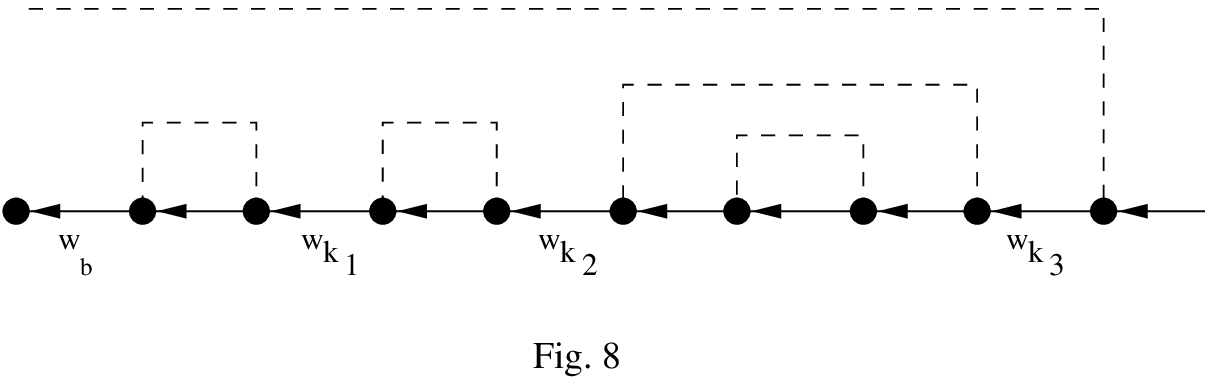}}

\bigskip

\demo{Proof of Lemma \equ(Ylemma)}
We shall consider  $r=0$, the general case is similar.
Let $Y(z) : = \big( Y_zf\big)(0)$. Define
$$
	f^*(Q): = \int_{S^{d-1}} f(-Q^{1/2}\omega) d\omega
$$
where $Q\ge 0$ and
the integration is on the unit sphere $S^{d-1}$ with respect to the
normalized surface measure. Clearly $f^*$ is bounded, smooth away 
from 0, it decays fast at infinity and 
its derivative has a $Q^{-1/2}$ singularity at $Q=0$.  We have 
$$
	Y(z)= \int_0^\infty {f^*(Q)\over z- Q} Q^{{d\over 2}-1} dQ .
\Eq(Ydef)
$$

(i) For the proof of \equ(Yest) we assume that $\a = \hbox{Re} \, z \ge 0$;
the case $\a < 0$ is similar but easier. We split the integration as follows
$$
	Y(z) = \int_J {f^*(Q)\over z- Q} Q^{{d\over 2}-1} dQ 
	+\int_{J^c}{f^*(Q)\over z- Q} Q^{{d\over 2}-1} dQ \; ,
$$
where $J:= \big[ \max (0, \a-1), \a +1 \big]$ and $J^c= [0,\infty)\setminus J$.
On the set $J^c$ we use the estimate $|z-Q|\ge 1$, hence this integral
is bounded by $\hbox{(const.)}\|f\|_{2d, 2d}$.
On the set $J$ we use the identity 
$$
	 \int_J {f^*(Q)\over z- Q} Q^{{d\over 2}-1} dQ
	 =  \int_J {g(Q)- g(\a)\over \a - Q + i\eta}
	dQ + g(\a)  \int_J
	{1\over \a - Q + i\eta} dQ
\Eq(Yexp)
$$
with
$$
	g(Q) : = f^*(Q)Q^{{d\over 2}-1} \; ,
$$
The function $g'(Q)$ inherits the $Q^{-1/2}$ singularity at the
origin from the derivative of $f^*$. We easily obtain for $|\a - Q|\leq 1$ that
$$
	g(Q)- g(\a)\leq \hbox{(const.)}\| f\|_{2d, 2d}
	\big| \sqrt{Q} - \sqrt{\a}\big| \; .
$$
Hence the first term  in \equ(Yexp) is estimated
$$
	\Bigg| \int_J {g(Q)- g(\a)\over \a - Q + i\eta}
	dQ\Bigg|
	\leq \hbox{(const.)}
	\|f\|_{2d, 2d}  \int_J {1 \over \sqrt{\a} + \sqrt{Q}}
	dQ \leq \hbox{(const.)} \|f\|_{2d, 2d} .
$$
The second term in \equ(Yexp) is essentially explicit:
$$
	 g(\a)  \int_J
	{1\over \a - Q + i\eta} dQ
	=  g(\a)  \int_J {(\a - Q) + i\eta\over (\a - Q)^2 + \eta^2}dQ
$$
The imaginary part of the integrand is uniformly bounded.
The integral
$$
	\int_{\max (0, \a-1)}^{\a +1} {\a - Q \over (\a - Q)^2 + \eta^2}
	 dQ
$$
is zero if $\a \ge 1$ by symmetry. For $\a< 1$
$$
	\int_{\max (0, \a-1)}^{\a +1} {\a - Q \over (\a - Q)^2 + \eta^2}
	 dQ = \int_{2\a}^{\a +1} {\a - Q \over (\a - Q)^2 + \eta^2}
	 dQ \leq \hbox{(const.)} |\log\a|
$$
Considering that $g(\a)\sim \a^{{d\over 2}-1}$ for small $\a$,
we see that the second term in
\equ(Yexp) is bounded for $d\ge 3$,
 and has a logarithmic singularity for $d=2$.
 This concludes the proof of part (i).

\medskip

(ii) For the proof of \equ(Ycont) and \equ(Ycont2),
notice that
$$
	{1\over z-Q} = -{d\over dQ} \log (z-Q) \; .
$$
Since $\hbox{Im} z >0$ and $Q$ is real, we always evaluate the 
logarithm on the upper half plane. So we can choose any branch 
of the logarithm and fix it. 
Integration by parts from \equ(Ydef) gives
$$
	Y(z) = -f^*(0)\log z + \int_0^\infty \log (z-Q) F(Q) dQ
$$
if $d=2$, and
$$
	Y(z) = \int_0^\infty \log (z-Q) F(Q) dQ
$$
if $d\ge 3$,
with
$$
	F(Q): = {d\over dQ}\Big( f^*(Q) Q^{{d\over 2}-1}\Big) \; .
$$
Notice that $F(Q)$ is smooth away from the origin, it has a strong
decay at infinity and it has a $Q^{-1/2}$ singularity at $Q=0$.
Hence
$$
	\Big|  Y(z)- Y(z')\Big| \leq  
	\|f^*\|_\infty \Big|\log z - \log z'\Big|+
	 \Bigg| \int_0^\infty \Big( \log (z-Q) -\log (z'-Q)\Big)
	F(Q) dQ\Bigg|
$$
and the first term is present only in $d=2$. For the first term
we use the following estimate which can  easily be checked
$$
	\Big|\log z - \log z'\Big| \leq (\hbox{const.})|z-z'|
	\Big( {1\over |z|}+{1\over |z'|}\Big) \; .
\Eq(ex)
$$
We rewrite  the second term 
$$
\eqalign{ \Bigg| \int_0^\infty \Big( \log (z-Q) -\log (z'-Q)\Big)
	F(Q) dQ\Bigg| &=  \Bigg|\int_{\Gamma(z, z')} d\xi
	\int_0^\infty {F(Q)\over \xi-Q} dQ\Bigg|\cr
	&\leq \int_{\Gamma(z, z')} d |\xi| \int_0^\infty {|F(Q)|\over
	|\xi-Q|} dQ
}
$$
where $\Gamma(z, z')$ is any path in the upper half plane
that connects $z, z'$.
 Here $d|\xi|$ is the arclength measure.
 It is an easy exercise to show that for $\hbox{Im}\, \xi>0$
$$
	\int_0^\infty{|F(Q)|\over |\xi- Q|} dQ\leq \hbox{(const.)}
	 |\xi |^{-{1\over 2}}\| f\|_{2d, 2d}
\Eq(fixxi)
$$
using the properties of $F$ listed above.
We choose the path $\Gamma(z, z')$ that goes from $z=\a + i\eta$ 
to $\a' + i\eta$ then to $z'=\a' + i\eta'$ along straight line
segments (recall that $\eta'\le \eta \le 1/2$). The integral
of the right hand side of \equ(fixxi) along this path
gives
$$
	\int_{\Gamma(z, z')} d |\xi| \int_0^\infty {|F(Q)|\over
	|\xi-Q|} dQ 
	\leq \hbox{(const.)}
	 |z-z'| \, |\eta|^{-1/2}\| f\|_{2d, 2d}
$$
Combining these estimates we obtain \equ(Ycont) and \equ(Ycont2).
\qed
\enddemo

{\it Acknowledgement.} We would like to thank H. Spohn
for his several comments and discussions on this project.
Part of this work was done during the time when 
L. E. visited  the Erwin Schr\"odinger Institute in Vienna
and  when both authors visited the Center of Theoretical Sciences in Taiwan.
We thank them  for the hospitality and 
the support of this work.

\bigskip

\beginsection References.

\item{\rtag {BBS}} C, Boldrighini, L. Bunimovich, Y.Sinai: On the Boltzmann 
equation for the Lorentz gas, J. Stat. Phys. {\bf 32}, 477-501, (1983).

\item{\rtag {DGL}}
D. Durr, S. Goldstein and J. Lebowitz:
 Asymptotic motion of a classical particle in a random 
potential in two dimensions: Landau model. Commun. Math. Phys. {\bf 113}, 
209-230 (1987).

\item{\rtag{EY1}} L. Erd\H os and H.-T. Yau: {\it Linear Boltzmann
equation as scaling limit of quantum Lorentz gas},
Advances in Differential
Equations and Mathematical Physics. Contemporary Mathematics {\bf 217},
137-155 (1998).

\item{\rtag{EP}}
R. Esposito, M. Pulvirenti and A. Teta: {\it The Boltzmann equation
for a one-dimensional quantum Lorentz gas}, Preprint (1998).

\item{\rtag {Ga}} G.  Gallavotti: Rigorous theory of the Boltzmann equation 
in the Lorentz gas. Nota inteerna n. 358, Univ. di Roma (1970).

\item{\rtag {HLW}} T. G. Ho, L. J. Landau and A. J. Wilkins: {\sl On the
weak coupling limit for a Fermi gas in a random potential}, Rev.  Math. Phys. 
{\bf 5},
 209-298 (1992).

\item{\rtag {KP}}
H.  Kesten, G. Papanicolaou: A limit theorem for stochastic acceleration, 
Commun. Math. Phys. {\bf 78}, 19-63 (1980).

\item{\rtag {S}} H. Spohn: Derivation of the transport equation for 
electorns moving through random impurities, J. Stat. Phys., {\bf 17}, 
385-412 (1977). 

\item{\rtag {Sp2}} H. Spohn: The Lorentz process converges to a random flight 
process. Commun. Math. Phys. {\bf 60}, 277-290 (1978).

\enddocument
\enddocument